\UseRawInputEncoding

\documentclass[accepted=2025-12-02,a4paper,onecolumn,superscriptaddress]{quantumarticle}
\pdfoutput=1

\usepackage{doi}
\usepackage{booktabs} 
\usepackage[numbers,sort&compress]{natbib}
\usepackage{amsmath}
\usepackage{microtype} 
\usepackage{tabularx}
\usepackage{multirow} 

\usepackage{graphicx}

\usepackage{amsfonts}
\usepackage{color}
\usepackage{mathtools}
\usepackage{nicematrix}

\usepackage{thmtools} 
\usepackage{thm-restate}
\usepackage{amsthm}
\usepackage{hyperref}

\usepackage{verbatim} 
\hypersetup{
	colorlinks = true,
	linkcolor =blue,
	citecolor=blue, 
	urlcolor=blue 
}

\usepackage{ulem} 

\usepackage{braket}
\usepackage{physics}
\usepackage{bm}

\usepackage{float}
\usepackage[capitalize]{cleveref}

\usepackage{tikz}
\usetikzlibrary{shapes.geometric, arrows} \usetikzlibrary{arrows.meta}
\usetikzlibrary{positioning}
\usetikzlibrary{calc}
\tikzstyle{process} = [rectangle, rounded corners, minimum width=3cm, minimum height=1cm, text centered, draw=black, align= center]
\tikzstyle{virtualprocess} = [rectangle, rounded corners, minimum width=3cm, minimum height=1cm, text centered, draw=black, dashed, align=center]
\tikzstyle{emptyprocess} = [rectangle, minimum width=3cm, minimum height=1cm, text centered, align=center]
\tikzstyle{point} = [circle, inner sep=0pt, minimum size=3pt, fill=black]

\tikzstyle{arrow} = [thick, ->, >=stealth]
\tikzstyle{dashedarrow} = [thick, ->, >=stealth, dashed]

\tikzstyle{connect} = [ thick,<->, >=stealth]

\tikzstyle{channel} = [rectangle, line width=0.7mm, minimum width=5cm, minimum height=3cm, text centered, draw=black, align = center]
\tikzset{
	pics/detector/.style args={#1,#2,#3}{
		code={
			\draw[line width=0.7mm] (0,-2) -- (0,2); 
			\draw[line width=0.7mm] (0,2) -- ++(1,0); 
			\draw[line width=0.7mm] (0,-2) -- ++(1,0);
			\draw[line width=0.7mm] (1,2) arc (90:-90:2); 
			\node[#3, align=center] (#1) at (1.3,0) {\scalebox{0.9}{\huge #2}}; 
		}
	}
}

\newcommand{\etadet}{\eta_{\text{det} }}
\newcommand{\dcprob}{d_{\text{det} }}
\newcommand{\etachar}{\Delta_{\eta}}
\newcommand{\dcchar}{\Delta_{\dc}}

\newcommand{\nint}{3}
\newcommand{\allk}{\vec{k}}
\newcommand{\tempevent}{\widetilde{\Omega}}
\newcommand{\constraints}{\mathcal{S}_\text{constraints}}

\newcommand{\Bound}{\mathcal{B}}
\newcommand{\Boundbasiczero}{\Bound_{0,0}}
\newcommand{\Boundbasicdelta}{\Bound_{\deltaone,\deltatwo}}

\newcommand{\Bounddecoymin}[1]{\Bound^\text{decoy}_{\text{min}-#1}}
\newcommand{\Bounddecoymax}[1]{\Bound^\text{decoy}_{\text{max}-#1}}

\newcommand{\ntot}{N_\text{tot}}

\newcommand{\dist}{d}
\newcommand{\inquotes}[1]{``#1"}

\newcommand{\half}{\frac{1}{2}}
\newcommand{\multiNorm}[1]{%
	\bigg\Vert #1 \bigg\Vert%
}
\newcommand{\mode}{\mathbf{d}}
\newcommand{\proj}{\mathrm{\Pi}}

\newcommand{\binfunction}[3]{F(#1,#2,#3)}

\newcommand{\fserf}{f_\text{serf}}
\newcommand{\num}{\bm{N}}

\newcommand{\qualityfactor}{c_q}
\newcommand{\con}{\text{con}}
\newcommand{\serfbound}{\kappa}
\newcommand{\accset}{ \mathcal{F}}

\newcommand{\id}{\mathrm{I}}

\newcommand{\etaChan}{\eta_{\text{ch}}}

\newcommand{\dc}{\text{dc}} 
\newcommand{\misAngle}{\theta} 
\newcommand{\tempfunc}[1]{f_{#1}} 
\newcommand{\temptempfunc}[1]{g_{#1}} 
\newcommand{\epsAT}{\varepsilon_\text{AT}}

\newcommand{\epsATb}{\varepsilon_\text{AT-b}}

\newcommand{\epsATa}{\varepsilon_\text{AT-a}}

\newcommand{\epsATc}{\varepsilon_\text{AT-c}}
\newcommand{\epsdecoy}{\varepsilon_\text{AT-d}}
\newcommand{\epsATsingle}{\varepsilon_\text{AT-s}}

\newcommand{\epsPA}{\varepsilon_{\text{PA}}}
\newcommand{\epsEV}{\varepsilon_{\text{EV}}}

\newcommand{\epsSec}{\varepsilon_{\text{secure}}}

\newcommand{\smoothmin}[1]{H^{#1}_{\text{min}}}
\newcommand{\smoothmax}[1]{H^{#1}_{\text{max}}}

\newcommand{\pzA}{p^{(A)}_{(Z)}}
\newcommand{\pxA}{p^{(A)}_{(X)}}
\newcommand{\pzB}{p^{(B)}_{(Z)}}
\newcommand{\pxB}{p^{(B)}_{(X)}}

\newcommand{\X}{\bm{X}}

\newcommand{\Y}{\bm{Y}}

\newcommand{\nO}{n_O}
\newcommand{\nOrv}{\bm{\nO}}
\newcommand{\nOmu}[1]{n_{O,\mu_{#1}}}
\newcommand{\nOmurv}[1]{\bm{\nOmu{#1}}}

\newcommand{\nOph}[1]{n_{O,{#1}}}
\newcommand{\nOphrv}[1]{\bm{\nOph{#1}}}

\newcommand{\nX}{n_{X}}
\newcommand{\nXrv}{\bm{\nX}}

\newcommand{\nXmu}[1]{n_{X,\mu_{#1}}}
\newcommand{\nXmurv}[1]{\bm{\nXmu{#1}}}
\newcommand{\nXph}[1]{n_{X,{#1}}}
\newcommand{\nXphrv}[1]{\bm{\nXph{#1}}}

\newcommand{\nXneqmu}[1]{n_{X_{\neq},\mu_{#1}}}
\newcommand{\nXneqmurv}[1]{\bm{\nXneqmu{#1}}}

\newcommand{\nXperf}{\tilde{n}_{X}}
\newcommand{\nXperfrv}{\bm{\nXperf}}
\newcommand{\eX}{e^{\text{obs}}_{X}  }
\newcommand{\eXrv}{\bm{\eX}}

\newcommand{\eXmu}[1]{e^{\text{obs}}_{X,\mu_{#1}}}
\newcommand{\eXmurv}[1]{\bm{\eXmu{#1}}}
\newcommand{\eXph}[1]{e^{\text{obs}}_{X,#1}}
\newcommand{\eXphrv}[1]{\bm{\eXph{#1}}}

\newcommand{\eXperf}{\tilde{e}^{\text{obs}}_{XX}  }
\newcommand{\eXperfrv}{\bm{\eXperf}}

\newcommand{\eZ}{e^{\text{obs}}_Z }

\newcommand{\eph}{e^{\text{key}}_X } 
\newcommand{\ephrv}{\bm{\eph}}
\newcommand{\ephph}[1]{e^\text{key}_{X,{#1}}}
\newcommand{\ephphrv}[1]{\bm{\ephph{#1}}}

\newcommand{\ephperf}{\tilde{e}^{\text{key}}_{ZX} } 
\newcommand{\ephperfrv}{\bm{\ephperf}}

\newcommand{\ephwierd}{\tilde{e}^\text{key}_{XX}}
\newcommand{\ephwierdrv}{\bm{\ephwierd}}

\newcommand{\len}{l}
\newcommand{\lenrv}{\bm{\len}}

\newcommand{\deltaone}{\delta_1}
\newcommand{\deltatwo}{\delta_2}
\newcommand{\cone}{c_1}
\newcommand{\ctwo}{c_2}

\newcommand{\nK}{n_{K}}
\newcommand{\nKrv}{\bm{\nK}}
\newcommand{\nKperf}{\tilde{n}_{K}}
\newcommand{\nKperfrv}{\bm{\nKperf}}

\newcommand{\nKmu}[1]{n_{K,\mu_{#1}}}
\newcommand{\nKmurv}[1]{\bm{\nKmu{#1}}}
\newcommand{\nKph}[1]{n_{K,{#1}}}
\newcommand{\nKphrv}[1]{\bm{\nKph{#1}}}

\newcommand{\pt}{p_\text{Z,T}}

\newcommand{\fullevent}{\event{\params}}

\newcommand{\event}[1]{\Omega_{(#1)} }
\newcommand{\EVevent}{\Omega_\text{EV}}
\newcommand{\params}{\nX,\nK,\eX,\eZ}
\newcommand{\decoyparams}{\nXmu{\allk},\nKmu{\allk},\eXmu{\allk},\eZ}

\newcommand{\leak}{\lambda_\text{EC}}
\newcommand{\ECost}{ \log(2/\epsEV) }

\newcommand{\Cfull}{C^nC_EC_P}

\newcommand{\BobPOVM}[2]{\Gamma^{(B)}_{(#1,#2)}}
\newcommand{\BobPOVMswap}[2]{\Gamma^{(B),(\text{swap})}_{(#1,#2)}}
\newcommand{\AlicePOVM}[2]{\Gamma^{(A)}_{(#1,#2)}}
\newcommand{\AliceBobPOVM}[2]{\Gamma_{(#1),(#2)}}

\newcommand{\AliceBobPOVMFilter}[2]{\tilde{F}_{(#1),(#2)}}
\newcommand{\AliceBobPOVMsecond}[2]{G^{\text{con}}_{(#1),(#2)}}

\newcommand{\AliceBobPOVMsecondsandwich}[2]{G^{\text{con,F}}_{(#1),(#2)}}
\newcommand{\AliceBobPOVMprimeFilter}[2]{{F}_{(#1),(#2)}}

\newcommand{\etaMax}{\eta_{\text{max}}} 
\newcommand{\etaMin}{\eta_{\text{min}}} 
\newcommand{\etaRenorm}{r_{\eta}} 
\newcommand{\etaAvg}{\eta_{\text{avg}}} 
\newcommand{\etaMultAvg}{\eta_{\text{mult}}} 

\newcommand{\intPOVM}{P} 
\newcommand{\dMin}{d_{\text{min}}} 
\newcommand{\dMax}{d_{\text{max}}} 
\newcommand{\dMultAvg}{d_{\text{mult}}} 

\newcommand{\dX}{d_{X}} 
\newcommand{\etaX}{\eta_X} 
\newcommand{\dZ}{d_{Z}} 
\newcommand{\etaZ}{\eta_Z} 

\newcommand{\Ap}{{A^\prime}}

\theoremstyle{definition} 
\newtheorem{theorem}{Theorem}

\newtheorem{lemma}{Lemma}
\newtheorem{corollary}{Corollary}
\newtheorem{remark}{Remark}

\begin{document}
	\title{Phase error rate estimation in QKD with imperfect detectors}  
	\author{Devashish Tupkary}
	\email{djtupkary@uwaterloo.ca}
	\affiliation{Institute for Quantum Computing and Department of Physics and Astronomy, University of Waterloo, Waterloo, Ontario, Canada, N2L 3G1}	
	
	\author{Shlok Nahar}
	\affiliation{Institute for Quantum Computing and Department of Physics and Astronomy, University of Waterloo, Waterloo, Ontario, Canada, N2L 3G1}	
	
	\author{Pulkit Sinha}
	\affiliation{Institute for Quantum Computing and School of Computer Science, University of Waterloo, Waterloo, Ontario, Canada, N2L 3G1}	
	
	\author{Norbert L\"utkenhaus}
	\email{nlutkenhaus.office@uwaterloo.ca}
	\affiliation{Institute for Quantum Computing and Department of Physics and Astronomy, University of Waterloo, Waterloo, Ontario, Canada, N2L 3G1}

	\begin{abstract}
		We present a finite-size security proof of the decoy-state BB84 QKD protocol against coherent attacks, using entropic uncertainty relations, for imperfect detectors. We apply this result to the case of detectors with imperfectly characterized basis-efficiency mismatch. Our proof works by obtaining a suitable bound on the phase error rate, without requiring any new modifications to the protocol steps or hardware. It is applicable to imperfectly characterized detectors, and only requires the maximum relative difference in detection efficiencies and dark count rates of the detectors to be characterized. Moreover, our proof allows Eve to choose detector efficiencies and dark count rates in their allowed ranges in each round, thereby addressing an important problem of detector side channels. We prove security in the variable-length framework, where users are allowed to adaptively determine the length of key to be produced, and number of bits to be used for error-correction, based on observations made during the protocol. We quantitatively demonstrate the effect of basis-efficiency mismatch by applying our results to the decoy-state BB84 protocol.
	\end{abstract}

	\maketitle
	
	\section{Introduction} \label{sec:introduction}
	Security proofs of QKD based on the entropic uncertainty relations (EUR) \cite{tomamichelUncertainty2011,tomamichelLargely2017,tomamichelTight2012,tupkary2025qkdsecurityproofsdecoystate}, and the phase error correction approach \cite{koashiSimple2005,koashiSimple2009a,tupkary2025qkdsecurityproofsdecoystate}, yield some of the highest key rates against coherent attacks in the finite-size regime. While source imperfections \cite{pereiraModified2023,tamakiLosstolerant2014,curras-lorenzoSecurity2024,zapateroImplementation2023} have been extensively studied within the phase error correction framework, detector imperfections have not yet been addressed in any meaningful sense in either security proof framework. 
	In particular, these proof techniques require the probability of detection in Bob's detection setup to be independent of basis choice. This assumption is referred to as ``basis-independent loss" in the literature, while the violation of this assumption is referred to as ``basis-efficiency mismatch" or ``detection-efficiency mismatch".
	Satisfying this assumption requires the efficiency and dark count rates of Bob's detectors to be \textit{exactly} identical. Therefore, justifying this assumption in practice requires \textit{exact} characterization of Bob's identical detectors. Either of these tasks are impossible in practice. Therefore, these proof techniques are not applicable to practical prepare-and-measure (and entangled-based) QKD scenarios involving realistic detectors. Note that MDI-QKD  \cite{loMeasurementdeviceindependent2012} is able to address \textit{all} detector imperfections and detector  side-channels, since it assumes the detectors to be completely in Eve's control. Morever, source imperfections can also be handled to a large degree. However, MDI-QKD is significantly more complex to implement as compared to prepare-and-measure QKD, and the latter remain dominant in real-world implementations. Thus, addressing detector imperfections within prepare-and-measure QKD is of paramout importance. To date, there is no security proof method that addresses this problem \textit{without} requiring significant protocol modifications. While some theoretical analyses of  basis-efficiency mismatch in the asymptotic setting exist for standard QKD \cite{lydersenSecurity2010,fungSecurity2009a,zhangSecurity2021,bochkovSecurity2019,trushechkinSecurity2022}, there is no work that handles basis-efficiency mismatch for coherent attacks in the finite-size regime.
	Meanwhile, there have been several experimental demonstrations \cite{gerhardtFullfield2011,makarovEffects2006,lydersenHacking2010,sajeedSecurity2015,qiTimeshift2007}, exploiting  basis-efficiency mismatch for attacks on QKD systems.
	
	In this work, we prove the security of the decoy-state BB84 protocol with an active detection setup \textit{without} assuming basis-independent loss. We do so by showing that the phase error rate can be suitably bounded even without the assumption. We explicitly define metrics $\deltaone$, $\deltatwo$ that quantify the deviation from the ideal case, and bound the phase error rate in terms of these deviations. Our framework is general, and can be applied to any (IID) detector model of one's choice, as long as the relevant metrics $\deltaone$, $\deltatwo$ can be suitable bounded. In general, this will require some model of the detectors and characterization of its properties. An important, non-trivial open question remains on how best to experimentally bound these quantities. We leave a full experimental analysis for future work, and restrict our attention in this work to the case where we utilise common models for detectors \footnote{We stress that characterisation will \textit{always} proceed by assuming some physically motivated model of the devices in terms of a small number of paramters. The characterisation experiment will then estimate these model parameters. The question that must then be answered is about which model best describes the physical implementation, and how to characterise the parameters of the chosen model.}, and bound $\deltaone, \deltatwo$ in terms of commonly measured model parameters \cite{etsi}.

	 In this work we explicitly compute these metrics for the case of detectors with basis-efficiency mismatch and unequal dark count rates. To do so, we assume the the canonical model of detectors described in \cref{subsec:detectors}. The block-diagonal structure of the detector POVMs significantly aids the computation of these metrics. Moreover, we compute these metrics directly from the experimental characterization of the detection efficiencies and dark count rates of the detectors. Our results extend the security of QKD to the following practical scenarios:
	
	\begin{enumerate}
		\item \label{caseone} Bob's detectors are \textit{not} identical, but the values of efficiency ($\eta_{b_i}$ for basis $b$ and outcome $i$) and dark count rates ($d_{b_i}$) are known. Note that while this is a useful toy model, such scenarios are impractical since they require $\eta_{b_i}, d_{b_i}$ to be known exactly.  We treat the dark count rate as a part of the POVM element, as described in \cref{subsec:detectors}.
		\item \label{casetwo} Bob's detectors are \textit{not} identical, and the values of efficiency and dark count rates are only known to be in some range $\eta_{b_i} \in   [\etadet (1 -\etachar),\etadet(1+\etachar)], d_{b_i} \in   [ \dcprob( 1-\dcchar), \dcprob (1+\dcchar)]$.  While this is again a useful toy model, a detectors response ($\eta_{b_i}, d_{b_i}$) to incoming photons typically depends on the spatio-temporal modes of incoming photons, which are in Eve's control.
		\item \label{casethree} Bob's detectors are \textit{not} identical, and the values of efficiency and dark count rates are only known to be in some range. Moreover, these values depend on the spatio-temporal modes (labelled by $\mode$) of the incoming photons, and can therefore be chosen by Eve \cite{makarovEffects2006,zhangSecurity2021,sajeedSecurity2015,qiTimeshift2007}. This is expressed mathematically as $\eta_{b_i}(\mode) \in [  \etadet(\mode) (1 -\etachar), \etadet(\mode) (1+\etachar)], d_{b_i} (\mode) \in [ \dcprob (1-\dcchar), \dcprob (1+\dcchar) ]$. Note that in this model, the range of allowed values of the loss can depend on the spatio-temporal mode, whereas the dark count rates for all the modes lie in the same range. 

	\end{enumerate}
	
	Our metrics $\deltaone,\deltatwo$ involve an optimization over all possible values of $\eta_{b_i}, d_{b_i}$ in their respective ranges. Moreover, for our model of  multi-mode detectors, we find that our methods  yield the same values for Case \ref{casetwo} and Case \ref{casethree}. Thus, our methods address one practically important detector side-channel as a by-product. We discuss this in greater detail in \cref{sec:detectorside}.
	
	We use entropic uncertainty relations \cite{tomamichelUncertainty2011} for our security proofs in this work. Since  both the entropic uncertainty relations approach and the phase error correction \cite{koashiSimple2005,koashiSimple2009a} approach involve bounding the phase error rate, we expect our techniques to also work when using the phase error correction approach for the security proof. In fact, one may also use the equivalence by Tsurumuru \cite{tsurumaruEquivalence2022,tsurumaruLeftover2020} to relate the two security proof approaches. 
	For example,  \cite[Section III]{tsurumaruLeftover2020} constructs an explicit phase error correction circuit whose failure probability can be bound by the smooth min entropy via \cite[Theorem 1 and Corollary 2]{tsurumaruLeftover2020}. Therefore one may indirectly prove security in the phase error correction framework by applying the above equivalence on the results of this work.
	
	Moreover, we prove security in the variable-length framework \cite{portmannSecurity2022,Benor2004}  which allows Alice and Bob to adaptively determine the number of bits to be used for error-correction, and the length of the output key, based on the observations during runtime. Such protocols are critical for practical implementations, where the honest behavior of the channel connecting Alice and Bob is difficult to determine in advance.  A security proof for variable-length QKD has been recently obtained for generic protocols in Ref.~\cite{tupkarySecurity2023} which utilizes an IID security proof followed by the postselection technique lift \cite{christandlPostselection2009,naharPostselection2024}  to coherent attacks.  The use of the postselection technique in Ref.~\cite{tupkarySecurity2023} leads to pessimistic key rates, which is avoided by this work. Our proof of variable-length security uses the same essential tricks as prior work on variable-length security in Refs.~\cite{hayashiConcise2012} and \cite[Chapter 3]{kawakamiSecurity} for the phase error correction framework, and is nearly identical to \cite[Supplementary Note A]{curras-lorenzoTight2021} for the EUR framework.

	This paper is organized as follows. We discuss similarities and differences with other related work \cite{lydersenSecurity2010,maroy_2010,trushechkinSecurity2022,bochkovSecurity2019,zhangSecurity2021,Marcomini2024,fungSecurity2009a} towards the end of this section. In \cref{sec:protocol}, we describe the QKD protocol that we consider in this work. For simplicity, we first consider the BB84 protocol where Alice prepares perfect single-photon signal states. We do this since there are many existing techniques for dealing with imperfect state preparation by Alice, and the goal of this work is to focus on detector imperfections.  We show that that the variable-length security of the QKD protocol follows if one is able to obtain suitable bounds on the phase error rate (\cref{eq:req}). In \cref{sec:perfectdetectors} we show how such bounds can be obtained in the case where the basis-independent loss assumption is satisfied. In \cref{sec:imperfectdetectors}, we show how such bounds can be obtained in scenarios where the basis-independent loss assumption is \textit{not} satisfied. In \cref{sec:decoy} we extend our analysis to prove the variable-length security of the decoy-state BB84 protocol with imperfect detectors. 
	In \cref{sec:results} we apply our results and compute key rates for the decoy-state BB84 protocol, and study the impact of basis-efficiency mismatch on key rates. We base our analysis of decoy-state BB84 on Lim et al \cite{limConcise2014}, and also point out and fix a few technical issues in that work. In \cref{sec:correlated} we outline an approach towards addressing correlated detectors. In \cref{sec:detectorside} we explain how our results can be applied to detector side channels (Case.~\ref{casethree} above).	
 In \cref{sec:conclusion} we summarize and present concluding remarks. Many details are relegated to the Appendices.

	Thus, in this work we 
	\begin{enumerate}
		\item Provide a method for phase error rate estimation in the presence of (bounded) detector imperfections, in the finite-size setting against coherent attacks.
		\item Address some detector side-channel vulnerabilities by allowing Eve to control (bounded) detector imperfections.
		\item Rigorously prove the variable-length security of  the decoy-state BB84 protocol in the framework of entropic uncertainty relations.
	\end{enumerate}
	
	We have attempted to write this paper in a largely modular fashion beyond \cref{sec:protocol}. For variable-length security, one can directly read \cref{app:varlengthproof,subsec:decoyvariable}. Similarly, for phase error estimation for the BB84 protocol, one can directly read \cref{sec:perfectdetectors} (for perfect detectors) and \cref{sec:imperfectdetectors} (for imperfect detectors). For phase error estimation in decoy-state BB84, one can read \cref{subsec:decoyphase,subsec:decoyanalysis}. For a recipe for applying our results to compute key rates in the presence of detector imperfections, one can refer to \cref{subsec:recipe}. For the application of our results to detectors with efficiency mismatch, one can refer to \cref{subsec:detectors,subsec:deltabounds,subsec:plots}. \cref{sec:correlated,sec:detectorside}  can be read independently, but require \cref{sec:imperfectdetectors} to be read first.

    \subsection{ Relation to other work on basis-efficiency mismatch} \label{subsec:relationtootherwork}

	\begin{table}[H]
		\centering
		
		\renewcommand\tabularxcolumn[1]{m{#1}}
		
		\newcolumntype{C}{>{\centering\arraybackslash}X}
		
		\begin{tabularx}{\textwidth}{|C|C|C|C|C|C|}
			\hline
			\textbf{Paper} & \textbf{Coherent Attacks} & \textbf{Finite-size} & \textbf{Eve has (bounded) control over detectors } & \textbf{Hardware Modifications} & \textbf{Notes} \\
			\hline
			Fung et al. \cite{fungSecurity2009a} & Yes & No & Yes & Detector Decoy \cite{moroder2009detector} (To remove Qubit assumption) & -  \\
			\hline
			Lydersen et al. \cite{lydersenSecurity2010} & Yes & No & Yes & None & Handles wide range of multi-mode models   \\
			\hline
			{\O ystein} {Mar{\o}y} et al. \cite{maroy_2010} & Yes & No & Yes & None & Handles wide range of multi-mode models  \\
			\hline
			Winick et al. \cite{winickReliable2018} & No & No & No & None & - \\
			\hline
			Zhang et al. \cite{zhangSecurity2021} & No & No & Yes* & None & * Only two modes \\
				\hline
	Bochkov et al.	\cite{bochkovSecurity2019} & No & No & No & None & Qubit assumption on Bob \\
				\hline
		Trushechkin et al.	\cite{trushechkinSecurity2022} & No & No & No & None & - \\
				\hline
		Grasselli et al.	\cite{Grasselli2024} & No & No & No & Detector decoy \cite{moroder2009detector}, requires tunable beam splitter  & Does not require detector characterization \\
			\hline
			Marcomini et al. 	\cite{Marcomini2024} & Yes  & No & No & No & Qubit assumption Bob, can handle some source imperfections.\\
				\hline
			This work & Yes & Yes & Yes & None &  \\
				\hline
		\end{tabularx}
		\caption{ Comparison of prior work on phase error rate estimation in the presence of basis efficiency mismatch. Note that Ref.~\cite{zhangSecurity2021} relies on numerical evidence for a part of the proof (bounding weight outside the subspace of low photon numbers). }
		\label{tab:qkd_security}
	\end{table}

		We will now discuss several prior works on addressing basis-efficiency mismatch in the literature (see \cref{tab:qkd_security}).	In a broad sense, the technique used in this work of reformulating the detector setup as first implementing a filtering step followed by further measurements is an important ingredient in many of these works (although the precise details may differ). However, all of them perform an asymptotic analysis, where there is no need for finite-size sampling arguments we use (such as \cref{lemma:sampling,lemma:pulkitgeneric} of this work). Instead one can directly associate the various error rates with POVM measurements on a single round state, and the analysis is greatly simplified.

		Ref.~\cite{fungSecurity2009a} proposed the first security proof of QKD in the presence of (bounded) Eve controlled basis-efficiency mismatch, in the asymptotic regime. It required the assumption that a qubit is received on Bob's side, for which it required the use of detector-decoy methods. Furthermore it also argued that the basis-efficiency mismatch can be removed entirely (for qubits received by Bob) if one randomly swaps the $0$ and the $1$ detector. Note that this trick, as argued in 	
		Ref.~\cite{fungSecurity2009a},   \textit{only works for the qubit subspace} and does not hold for higher photon numbers, which can be seen from our analysis in  \cref{subsec:deltabounds}.  Refs.~\cite{lydersenSecurity2010,maroy_2010} extended this work without having to assume qubit detections, but still performs an asymptotic analysis.
		In Ref.~\cite{winickReliable2018}, the numerical framework for key rate computations was proposed and used to compute IID asymptotic key rates for perfectly characterized and fixed basis-efficiency mismatch. In Ref.~\cite{zhangSecurity2021} the numerical framework \cite{winickReliable2018} was used to compute IID asymptotic key rates for a toy model where Eve was allowed to induce basis efficiency mismatch via the use of two spatio-temporal modes.  Recently, Ref.~\cite{bochkovSecurity2019} improved upon \cite{fungSecurity2009a} by obtaining tighter estimates  on the phase error rate in the presence of basis efficiency mismatch, but again assumed IID collective attacks in the asymptotic regime, and single qubits received on Bob's side. The assumption of single qubits received by Bob was later removed in the follow up work \cite{trushechkinSecurity2022}. In both \cite{bochkovSecurity2019,trushechkinSecurity2022}, Eve is not allowed to control the efficiency mismatch. A recent work \cite{Grasselli2024} again considers a scenario with IID collective attacks in the asymptotic regime, but has the advantage of not requiring prior characterization of the detector parameters. Another recent work \cite{Marcomini2024} combines qubit flaws in the source with efficiency mismatch in the detectors for coherent attacks, but is valid only in the asymptotic regime, and that Bob always receives a qubit.
			 Finally, MDI-QKD \cite{loMeasurementdeviceindependent2012} addresses \textit{all} detector side-channels in the finite-size regime against coherent attacks, but requires a drastically different protocol implementation compared to standard QKD. 
		
		In comparison (as will fully see in the coming sections), this work:
		\begin{enumerate}
			\item \textbf{Does not assume IID collective attacks.}
			\item \textbf{Is valid for finite-size settings. }
			\item \textbf{Does not assume that Bob receives a qubit. }
			\item \textbf{Requires no modifications or hardware changes to the protocol.}  
			\item \textbf{Also deals with dark counts.}
			\item \textbf{Allows Eve to control the efficiency mismatch via spatio-temporal modes.}
			\item \textbf{Can handle independent detectors (does not require IID detectors).}
		\end{enumerate}

	\section{Protocol Description} \label{sec:protocol}
	In this section we describe the steps of the QKD protocol we analyze.  
	
	\begin{enumerate}
		\item \textbf{State Preparation: } Alice decides to send states in the basis $Z$  ($X$) with probability $\pzA$($\pxA$). If she chooses the $Z$ basis, she sends states $\{\ket{0}_\Ap,\ket{1}_\Ap\}$ with equal probability. If she chooses the $X$ basis, she sends states $\{\ket{+}_\Ap,\ket{-}_\Ap\}$ with equal probability. She repeats this procedure $n$ times. Notice that this ensures 
		\begin{equation} \label{eq:sourcecondition}
			\rho_{\Ap | X} = \frac{\ketbra{+} + \ketbra{-}}{2}  = \frac{\ketbra{0} + \ketbra{1}}{2} = \rho_{\Ap | Z} = \frac{\id_{\Ap}}{2}
		\end{equation}
		where $\rho_{\Ap| b}$ denotes the the state sent out from Alice's lab given that she chooses a basis $b$. Essentially, \cref{eq:sourcecondition} says that the Alice's signal states leak no information about the basis chosen by Alice. This can be shown rigorously as follows.
		
		Using the source-replacement scheme \cite{curtyEntanglement2004,ferencziSymmetries2011}, Alice's signal preparation is equivalent to her first preparing the state $\ket{\Psi_+} = \frac{\ket{00}_{A\Ap} + \ket{11}_{A \Ap}}{\sqrt{2}}$ followed by measurements on the $A$ system.  Eve is allowed to attack the $A^{\prime n}$ system in any arbitrary (non-IID) manner, and forwards the system to $B$ to Bob. 	Furthermore, without loss of generality, this process can be viewed as Alice \textit{first} sending the system $A^\prime$ to Bob and \textit{then} measuring her system.

		Now, if Alice prepares the states from \cref{eq:sourcecondition}, her POVM elements corresponding to the basis $b$ signal states sum to $p^{(A)}_{(b)} \id_A$. Because of this fact, one can view Alice's measurement process,
		\textit{after} using the source-replacement scheme, as equivalent to choosing the basis $Z(X)$ with probability $\pzA(\pxA)$, followed by measuring using the POVM 
		$\{ \AlicePOVM{b}{0},\AlicePOVM{b}{1}\}$, for a given basis $b$. This reflects the fact that Eve has no knowledge of the basis used.
		
		The POVM elements are given by
		\begin{equation} \label{eq:alicePOVMs}
			\begin{aligned}
				\AlicePOVM{Z}{0} = \ketbra{0}, \quad  & \AlicePOVM{Z}{1} = \ketbra{1} \\
				\AlicePOVM{X}{0} = \ketbra{+}, \quad  & \AlicePOVM{X}{1} = \ketbra{-}.
			\end{aligned}
		\end{equation}
		
		Therefore, we now have a setup where the state $\rho_{A^nB^n}$ is shared between Alice and Bob, followed by basis choice and measurements by Alice. 
		
		\begin{remark} \label{remark:sourcereplacement}
			Without  loss of generality, one can always use the source-replacement scheme, and delay Alice's measurements until after Eve's attack has been completed, for any set of signal states. However, this process might result in POVM elements for Alice whose sum (for a specific basis) is not proportional to identity. In this case, Alice's measurements are \textit{incompatible with active basis choice} after the source-replacement scheme. We utilize the fact that Alice implements active basis choice when using the EUR statement (\cref{app:varlengthproof}), and in bounding the phase error rate (\cref{sec:perfectdetectors,sec:imperfectdetectors}). It is precisely for this reason that \cref{eq:sourcecondition} is needed. For methods to address imperfect state preparation, we refer the reader to \cite{curras-lorenzoSecurity2024,tamakiLosstolerant2014,pereiraModified2023,zapateroImplementation2023} (however we note that the analysis there is within the phase error framework).

		\end{remark}

		\item \textbf{Measurement: } Bob chooses to measure in the $Z$($X$) basis with probability $\pzB$($\pxB$). For each basis choice, Bob has two threshhold detectors, each of which can click or not-click. Bob maps double clicks to $0/1$ randomly (this is essential, see \cref{remark:doublecountremapping}), and thus has 3 POVM elements in each basis $b$, which we denote using  $\{ \BobPOVM{b}{\bot},\BobPOVM{b}{0},\BobPOVM{b}{1}  \}$ which correspond to the inconclusive-outcome, $0$-outcome, and the $1$-outcome. In this work, we will use the following notation to write joint POVM elements, 
		\begin{equation} \label{eq:alicebobpovms} \begin{aligned} 
				\AliceBobPOVM{b_A, b_B}{i,j} &\coloneq \AlicePOVM{b_A}{i} \otimes \BobPOVM{b_B}{j}, \quad \\ 
				\AliceBobPOVM{b_A,b_B}{\neq} &\coloneq \AlicePOVM{b_A}{0} \otimes \BobPOVM{b_B}{1} +  \AlicePOVM{b_A}{1} \otimes \BobPOVM{b_B}{0} ,\\
				\AliceBobPOVM{b_A,b_B}{=} & \coloneq \AlicePOVM{b_A}{0} \otimes \BobPOVM{b_B}{0} +  \AlicePOVM{b_A}{1} \otimes \BobPOVM{b_B}{1}, \\
				\AliceBobPOVM{b_A,b_B}{\bot} & \coloneq \id_A \otimes \BobPOVM{b_B}{\bot},
			\end{aligned}
		\end{equation}
		where Alice's POVMs are defined in \cref{eq:alicePOVMs}, and Bob's in \cref{subsec:detectors}.		\begin{remark} \label{remark:doublecountremapping}
			As we will see in \cref{sec:perfectdetectors}, the mathematical assumption on Bob's detector setup needed for phase error estimation is actually given by
			\begin{equation} \label{eq:bobcondition}
				\BobPOVM{X}{\bot}=\BobPOVM{Z}{\bot}.
			\end{equation} 
			This means that the probability of a round being inconclusive (i.e discarded) is independent of the basis for all input states. Notice that \cref{eq:bobcondition} depends on the choice of classical post-processing on Bob's side. In particular, it can be trivially satisfied by mapping no-click and double-click events to $0$ and $1$ randomly (so that $\BobPOVM{X}{\bot}=\BobPOVM{Z}{\bot}$ is zero). However, such a protocol cannot produce a key when loss is greater than $50\%$, and is therefore impractical. In general, if one assumes the canonical model of detectors (see \cref{subsec:detectors}), and maps double-clicks to 0/1 randomly, then \cref{eq:bobcondition} requires the loss and dark count rates in each detector-arm to be equal.  This is why this condition is referred to as \inquotes{basis-independent loss}, and its violation is referred to as \inquotes{detection-efficiency mismatch} in the literature. Note that even for identical detectors, one is forced remap double-click events to satisfy \cref{eq:bobcondition}.
		\end{remark}
		\item \textbf{Classical Announcements and Sifting: } For all rounds, Alice and Bob announce the basis they used. Furthermore, Bob announces whether he got a conclusive outcome ($\{ \BobPOVM{b}{0},\BobPOVM{b}{1} \}$), or inconclusive ($\{  \BobPOVM{b}{\bot} \}$). A round is said to be ``conclusive'' if Alice and Bob used the same basis, and Bob obtained a conclusive outcome. 
		
		On all the $X$ basis conclusive rounds, Alice and Bob announce their measurement outcomes. These rounds are used to estimate the phase error rate. We let $\nX$ be the number of $X$ basis conclusive rounds, and let $\eX$ be the observed error rate in these rounds. 
		
		On all $Z$ basis conclusive round, Alice and Bob announce their measurement outcomes with some small probability $\pt$. We let $\eZ$ denote the observed error rate in these rounds,  which is used to determine the amount of error-correction that needs to be performed. Note that this estimation need not be accurate for the purposes of proving security of the protocol. The remaining $\nK$ rounds are used for key generation. 
		
		All these classical announcements are stored in the register $C^n$.	The state of the protocol at this stage is given by  $\rho_{Z_A^{\nK} Z_B^{\nK} C^nE^n | \event{\nX,\nK,\eX,\eZ}}$, where $Z_A$ and $Z_B$ denote Alice and Bob's raw key register, and 	$\event{\nX,\nK,\eX,\eZ}$ denotes the event that $\nX,\nK,\eX,\eZ$ values are observed in the protocol.

		\begin{remark}
			In this work, we use bold letters, such as $\bm{x}$ to denote a classical random variable, and $x$ to denote a particular value it takes. Furthermore, we will use $\event{x}$ to denote the event that $\bm{x} =x$. 
			Thus our protocol involves random  variables $\nXrv,\nKrv,\eXrv,\ephrv$, which take  values $\nX,\nK,\eX,\eph$ in any given run.
		\end{remark}
		\item \textbf{Variable-Length Decision: } When event $\fullevent$ occurs, Alice and Bob compute $\leak(\params)$ (the number of bits to be used for one-way  error-correction) and $l(\params)$ (the length of the final key to be produced). Aborting is modeled as producing a key of length zero. 
		\begin{remark} \label{remark:errorcorrection}
			Note that current security proofs do not allow users to \textit{first} implement error-correction and \textit{then} take the number of bits actually used as $\leak{}$ in the security analysis.
			This is because the length of the error-correction string actually used in the protocol run leaks information about Alice and Bob's  raw key data. This is because Eve can simulate the same error-correction protocol on all possible classical strings to determine which strings are compatible with the length she observes. This leakage is difficult to incorporate in security proofs.
			
			The variable-length protocol we consider allows users to determine the length of the error-correction information as a function of observations on the announced data. This data is anyway known to Eve, and therefore this procedure does not leak information via the above mechanism. Thus in this work, one must fix $\leak(\params)$ to be a suitable function that determines the exact number of bits to be used for one-way error-correction, as a function of announcements. For more discussion, see footnote.~\footnote{ In general, the number of bits leaked during error-correction is equal to the length of the classical bit string needed to encode all possible transcripts of the error-correction protocol (which can be one-way or two-way). Thus, if one requires $\leak(...)$ to be an upper bound on the number of bits used, then we can proceed by noting that the number of bit strings of length up to some value $\leak$ is $2^{\leak+1}-1$, so a $(\leak+1)$-bit register suffices to encode
				all such bit strings. With this, it suffices to replace the
				$\leak(...)$ values in our subsequent key length formulas with $\leak(...)+1$. A similar analysis can be done for other error-correction protocols as well.}.
			
		\end{remark}
		\item \textbf{Error-correction and error-verification: } Alice and Bob implement error-correction using a one-way error-correction protocol with $\leak(\params)$ bits of information. They implement error-verification by implementing a common two-universal hash function to $\ECost$ bits, and comparing the hash values. We let $C_E$ be the classical register storing this communication, and note that it involves $\leak(\params) + \log(2/\epsEV)$ bits of communication (see footnote.~\footnote{Technically, one also has to include the choice of the hash function and one bit for the result of the hash comparison.	However, the choice of the hash function is independent of the QKD protocol, and thus gives no info to Eve. Moreover, the protocol aborts when hash comparison fails, and thus this extra bit takes a deterministic values and does not affect any entropies. }). We let $\EVevent$ denote the event that error-verification passes.
		
		\item \textbf{Privacy Amplification: } Alice and Bob implement a common two-universal hash function (communicated using the register $C_P$), and produce a key of length $l(\params)$. The state of the protocol at this stage is given by $\rho_{K_A K_B C^n C_E C_P E^n | \Omega(\params) \wedge \EVevent}$. 
	\end{enumerate}
	
	Notice here that we implement a variable-length protocol. Such protocols are advantageous to fixed-length protocols, since they do not require Alice and Bob to properly characterize their channel before runtime, and carefully choose the acceptance critera. Instead, they can adjust the length of the key produced to the appropriate length, depending on the observed values during runtime. For the above protocol, the variable-length decision is a function of $\params$.

	\subsection{Requirements on phase error estimation procedure and variable-length security}
We now turn our attention to the estimation of the phase error rate. Note that in a QKD protocol, one starts with a fixed but unknown state $\rho_{A^nB^nE^n}$ that represents Eve's attack. This state then gives rise to random variables $\nXrv,\nKrv,\eXrv,\ephrv$. Here $\ephrv$ denotes the random variable corresponding to the \inquotes{phase error rate} in the key-generation rounds, when Alice and Bob measure those rounds (virtually) in the $X$ basis. (The phase error rate is explained in greater detail in \cref{sec:perfectdetectors,sec:imperfectdetectors}). To obtain variable-length security, one must obtain a high probability upper bound on the phase error rate $\ephrv$.  We assume that one has a way to obtain the following statement (which we prove in \cref{sec:perfectdetectors,sec:imperfectdetectors}): 
	\begin{equation} \label{eq:req}
		\Pr(\ephrv \geq \Boundbasicdelta(\eXrv,\nXrv,\nKrv) ) \leq \epsAT^2.
	\end{equation}
	This states that the phase error rate is upper bounded (with high probability) by a suitable function $\Boundbasicdelta$ of the observed error rate in the $X$ basis rounds, and the number of test and key generation rounds. 	We will obtain a suitable $\Boundbasicdelta$ satisfying \cref{eq:req} in  \cref{sec:perfectdetectors,sec:imperfectdetectors}, with and without the basis-independent loss assumption. The function $\Boundbasicdelta$ depends on the metrics $\deltaone,\deltatwo$ that quantify the deviation from ideal behavior for a given protocol description. We compute explicit bounds for $\deltaone,\deltatwo$ for detectors with efficiency mismatch in \cref{subsec:deltabounds} using the recipe outlined in \cref{subsec:recipe}.

	\begin{remark} \label{remark:randomvariables}
		When working with random variables that are obtained via measurements on quantum states, the joint distributions of random variables can only be specified when those random variables \textit{can exist at the same time}, via some physical measurements on the state. For example, one cannot speak of the joint distribution of $X$ and $Z$ measurement outcomes on the \textit{same} state, since such a joint distribution does not exist. 
		In the entirety of this work, all the random variables whose joint distribution is used in our arguments can indeed exist at the same time.
	\end{remark}
	
	Given an upper bound on the phase error rate (\cref{eq:req}), we have  the following theorem regarding the variable-length security of the QKD protocol described above. The proof is essentially identical to \cite[Supplementary Note A]{curras-lorenzoTight2021}, and uses the same techniques as those in Refs.~\cite{hayashiConcise2012,kawakamiSecurity} and is included in \cref{app:varlengthproof}.

	\begin{restatable}{theorem}{variablelengthproof}[Variable-length security of BB84 with qubit source] \label{thm:variablesecurity}
		Suppose \cref{eq:req} is satisfied and let $\leak(\params)$ be a function that determines the number of bits used for error-correction. Define 	\begin{equation} \label{eq:lvalue}
			\begin{aligned}
				l(\params) &\coloneq  \max\Big(0, n_K\left(1- h \left( \Boundbasicdelta(\eX,\nX,\nK ) \right) \right) - \leak(\params)  \\
				&- 2\log(1/2\epsPA) - \ECost \Big),
			\end{aligned}
		\end{equation} 
		where $h(x)$ is the binary entropy function for $x\leq 1/2$, and $h(x) =  1$ otherwise.
		Then the variable-length QKD protocol that produces a key of length $l(\params)$ using $\leak(\params)$ bits for error-correction, 	
		upon the event $\fullevent \wedge \EVevent$ is $(2\epsAT+\epsPA+\epsEV)$-secure \footnote{ For pedagogical reasons, we ignore the issues arising from non-integer values of hash-lengths. Such issues can be easily fixed by suitable use of floor and ceiling functions.}. 
	\end{restatable}

	\section{Phase error estimation with basis-independent loss assumption} \label{sec:perfectdetectors}

	In this section, we will prove \cref{eq:req} for an implementation that satisfies the basis-independent loss assumption. It is useful to refer to \cref{fig:virtualprotperfect} for this section. To prove \cref{eq:req}, we will need to modify the actual protocol to an equivalent protocol (in the sense of being the same quantum to classical channel). To do so we will use \cref{lemma:twostep} below to reformulate Alice and Bob's measurements to consist of two steps. The first step will implement a basis-\textit{independent} filtering operation that discards the inconclusive outcomes, while the second step will complete the measurement procedure. Then the required claim will follow from random sampling arguments on the second step measurements. We start by explaining the two-step protocol measurements.

	\subsection{Protocol Measurements} \label{subsec:protmeasureperfect}
	
	We will first use the following lemma to divide Alice and Bob's measurement procedure into two steps. For the proof, we refer the reader to \cref{appendix:technical}. We will use $S_\bullet(A)$ and $S_\circ(A)$ to denote the set of sub-normalized and normalized states on the register $A$ respectively.

	\begin{restatable}{lemma}{twosteplemma}[Filtering POVMs] \label{lemma:twostep}
				Let $\{\Gamma_{k}  | k \in \mathcal{A}\}$ be a POVM on a register $Q$, and let $\{ \mathcal{A}_i\}_{i \in \mathcal{P}_\mathcal{A}}$ be a partition of $\mathcal{A}$, and let  $\rho \in S_\bullet(Q)$ be a state. The classical register storing the measurement outcomes when $\rho$ is measured using $\{\Gamma_k\}_{k \in \mathcal{A}}$ is given by
		\begin{equation}
			\rho_\text{final} \coloneq	\sum_{k \in \mathcal{A}} \Tr(\Gamma_{k} \rho)  \ketbra{k}.
		\end{equation}
		This measurement procedure is equivalent (in the sense of being the same quantum to classical channel) to the following two-step measurement procedure: First doing a coarse-grained ``filtering'' measurement of $i$, using POVM $\{ \tilde{F}_i \}_{ i \in \mathcal{P}_\mathcal{A}}$, where
		\begin{equation}
			\begin{aligned}
				\tilde{F}_i  &\coloneq \sum_{j \in \mathcal{A}_i} \Gamma_{j}, \quad \quad \text{leading to the post-measurement state} \\
				\rho^\prime_\text{intermediate} &= \sum_{i \in \mathcal{P}_{\mathcal{A}}}  \sqrt{\tilde{F}_i} \rho \sqrt{\tilde{F}_i}^\dagger  \otimes \ketbra{i}.
			\end{aligned}
		\end{equation}
		Upon obtaining outcome $i$ in the first step, measuring using  POVM $\{ G_{k} \}_{ k \in \mathcal{A}_i}$ where
		\begin{equation} \label{eq:twostepsecondstate}
			\begin{aligned}
				G_{k} &\coloneq \sqrt{\tilde{F}}^+_i \Gamma_{k} \sqrt{\tilde{F}}^+_i + P_{k} \quad \quad \text{leading to the post-measurement classical state} \\
				\rho^\prime_\text{final} &= \sum_{i \in \mathcal{P}_{\mathcal{A}}} \sum_{k \in \mathcal{A}_i}  \Tr( G_{k }\sqrt{\tilde{F}_i} \rho \sqrt{\tilde{F}_i})  \ketbra{k},
			\end{aligned}
		\end{equation}
		where $F^+$ denotes the pseudo-inverse of $F$, and $P_{k}$ are any positive operators satisfying $\sum_{k \in \mathcal{A}_i} P_k = \id - \proj_{\tilde{F}_i}$, where $\proj_{\tilde{F}_i}$ denotes the projector onto the support of $\tilde{F}_i$.

	\end{restatable}

	\begin{figure}
		\centering				\scalebox{0.6}{\begin{tikzpicture}

\begin{scope}[shift = {(-8,0)}]
    \pic {detector={det_full, $\{\Gamma_k\}_{k\in\mathcal{A}}$, black}};
    \coordinate (left_det_full) at ($(det_full) - (3,0)$);
    
    \draw[line width = 0.7mm] (left_det_full) -- ([xshift=0cm] det_full.west);
\end{scope}
        
\begin{scope}[shift={(8,0)}]
    \pic {detector={det_second_step, $\{G_k\}_{k\in\mathcal{A}_i}$,black}};
    \node[channel, left of = det_second_step, xshift = -4cm, font = {\fontsize{18}{23.76}\selectfont}](channel_first_step){$\{F_i\}_{i \in \mathcal{P}_\mathcal{A}}$\\ Measurement \\Channel};
    \coordinate (left_channel) at ($(channel_first_step) - (4.2,0)$);
    
    \draw[line width = 0.7mm] (left_channel) -- (channel_first_step.west);
    \draw[line width = 0.7mm] ([yshift = 0.2cm] channel_first_step.east) -- ([xshift=0.15cm,yshift = 0.2cm] det_second_step.west);
    \draw[dashed, line width = 0.7mm] ([yshift=-0.2cm] channel_first_step.east) -- ([yshift=-0.2cm, xshift=0.15cm] det_second_step.west) node[midway, below, yshift = -0.1cm, black, font = {\fontsize{18}{23.76}\selectfont}]{$i$};
\end{scope}

\node at (-2.5,0) {\scalebox{1}{\resizebox{2.5cm}{0.6cm}{$\Leftrightarrow$}}};

\end{tikzpicture}}
		\caption{Schematic for the two-step measurement procedure from \cref{lemma:twostep}. Note that the second step measurement $\{G_k\}_{k \in \mathcal{A}_i}$ depends on the outcome of the first step measurement.}
	\end{figure}
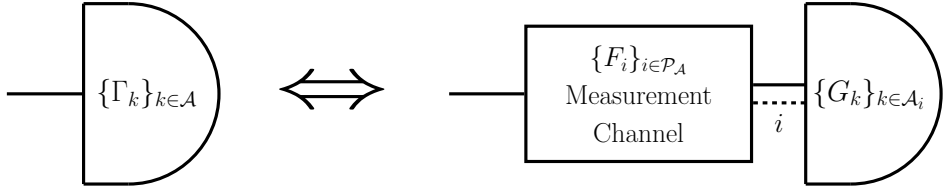
	
	Consider the POVMs $\{	\AliceBobPOVM{b_A,b_B}{\neq}, \AliceBobPOVM{b_A,b_B}{=}, \AliceBobPOVM{b_A,b_B}{\bot} \}$ defined in \cref{eq:alicebobpovms}, which correspond to Bob obtaining a conclusive outcome and Alice and Bob obtaining an error, Bob obtaining a conclusive outcome and Alice and Bob not obtaining an error, and Bob obtaining an inconclusive outcome respectively, for basis choices $b_A,b_B$. Without loss of generality, we can use \cref{lemma:twostep} to equivalently describe
	Alice and Bob's measurement procedure as consisting of two steps. 
	\begin{enumerate}
		\item First, they measure using POVM $\{\AliceBobPOVMFilter{b_A,b_B}{\con } ,  \AliceBobPOVMFilter{b_A,b_B}{\bot}\}$ which determines whether they obtain a conclusive and inconclusive measurement outcome. 
		\item Then, if they obtain a conclusive outcome, they measure using a second POVM \\ $\{ \AliceBobPOVMsecond{b_A,b_B}{=},\AliceBobPOVMsecond{b_A,b_B}{\neq} \}$.
	\end{enumerate}
	We use the convention that whenever an explicit basis $(X/Z)$ is written in the subscript of these POVMs, it refers to the basis used by both Alice and Bob.	
	We refer to the first-step measurements as \inquotes{filtering} measurements, since they determine whether Bob gets a conclusive outcome (which may be kept or discarded depending on basis choice), or an inconclusive outcome (which is always discarded).
	Furthermore, due to the construction of the POVM from \cref{lemma:twostep}, we have
	\begin{equation}
		\AliceBobPOVMFilter{b_A,b_B}{\bot} =  \id_A \otimes \BobPOVM{b_B}{\bot} .
	\end{equation}

	\subsection{Constructing an equivalent protocol} \label{subsec:equivprotperfect}
	
		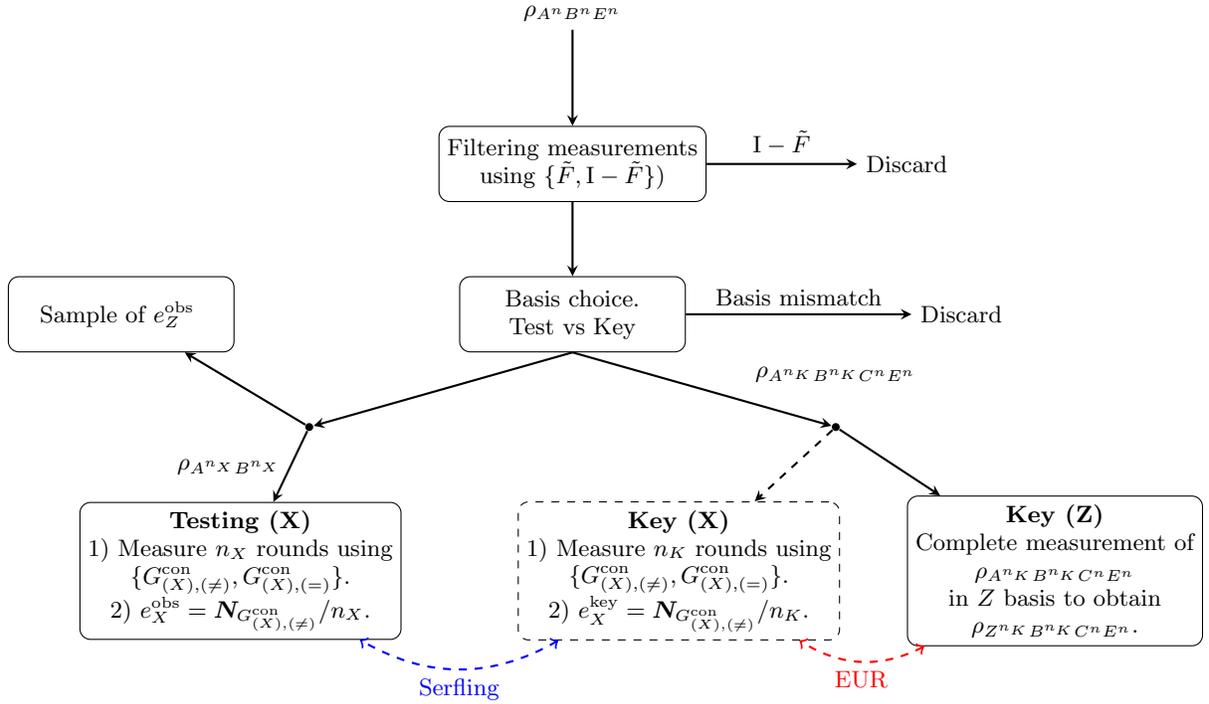
\begin{figure} 
        \hspace*{-0.8cm}
		\begin{tikzpicture}[node distance=2cm] 
			\node (start) {$\rho_{A^nB^nE^n}$ };
			\node (filtering) [process, below of=start] {Filtering measurements \\ using $\{ \tilde{F} , \id - \tilde{F}\}$)};
			
			\node (center) [process, below of=filtering]{Basis choice. \\ Test vs Key };

			\node (eZsample) [process, left of=center, xshift = -4cm] 
			{ Sample of $\eZ$ };

			\node (test) [process, below left of=center, xshift=-3cm, yshift=-2cm] 
			{ \textbf{Testing (X)} \\
				1) Measure $\nX$ rounds using \\
				$\{  \AliceBobPOVMsecond{X}{\neq}, \AliceBobPOVMsecond{X}{=}\}$. \\ 
				2) $\eX = \bm{N}_{ \AliceBobPOVMsecond{X}{\neq}} / \nX$. };
			
			\node(keystate) [point, right of=center,xshift = 1.5cm,yshift = -1.5cm] {};
			
			\node(teststate) [point, left of=center,xshift = -1.5cm,yshift = -1.5cm] {};
			
			\node(keystateabove) [emptyprocess, above of=keystate,yshift = -1.3cm] {$\rho_{A^{\nK}B^{\nK} C^n E^n}$};
			
			\node (keyX) [virtualprocess, below right of=center, xshift=0cm, yshift=-2cm] 
			{ \textbf{Key (X)} \\
				1) Measure $\nK$ rounds using \\
				$\{  \AliceBobPOVMsecond{X}{\neq}, \AliceBobPOVMsecond{X}{=}\}$. \\ 
				2) $\eph = \bm{N}_{ \AliceBobPOVMsecond{X}{\neq}} / \nK$. };

			\node (key) [process, below right of=center, xshift=5cm, yshift=-2cm] 
			{ \textbf{Key (Z)} \\
				Complete measurement of \\
				$\rho_{A^{\nK}B^{\nK}C^nE^n}$ \\
				in $Z$ basis to obtain \\
				$\rho_{Z^{\nK}B^{\nK}C^nE^n}$.};
			
			\draw [arrow] (start) --  ++(filtering); 
			\draw [arrow] (filtering) -- ++(center);

			\draw [arrow] (center.south) -- ++ (teststate) ;
			\draw [arrow] (center.south) -- ++ (keystate) ;
			
			\draw [arrow] (filtering.east) -- ++(2,0) node[midway, above] {$\id-\tilde{F}$} node[right] {Discard};

			\draw [arrow] (center.east) -- ++(3,0) node[midway, above] {Basis mismatch} node[right] {Discard};
			
			\draw [arrow] (teststate) -- ++(test) node[midway,left] {$\rho_{A^{\nX} B^{\nX}}$};;
			\draw [arrow] (teststate) -- ++(eZsample);
			
			\draw [arrow] (keystate) -- ++(key);
			\draw [dashedarrow] (keystate) -- ++(keyX);
			
			\draw (key) edge [dashed, <->, thick, bend left, color = red]  node[midway, below] {EUR} (keyX);
			
			\draw (keyX) edge [dashed, <->, thick, bend left, color = blue]  node[,yshift = 0cm, midway, below] {Serfling} (test);
		\end{tikzpicture}
		\caption{Protocol flowchart for the equivalent protocol from \cref{sec:perfectdetectors}, where basis-independent loss assumption (\cref{eq:bobcondition}) is satisfied. The dotted arrows and boxes represent virtual measurements that do not actually happen in the real protocol. Connections between different boxes are highlighted using curved arrows. We use the Serfling bound (\cref{lemma:sampling}) to obtain a bound on the phase error rate from observations. The phase error rate is then used to bound the smooth min entropy using the EUR statement. We use $\bm{N}_P$ to denote the number of $P$ measurement outcomes, where $P$ denotes a POVM element. For clarity, we have omitted the conditioning on events in the figure (but not in our proof). The basis used for measurements is indicated in each box, and refers to the basis used by \textit{both} Alice and Bob.}
		\label{fig:virtualprotperfect}
	\end{figure}
	We will now construct an equivalent protocol that is described in \cref{fig:virtualprotperfect}. 
	\begin{enumerate}
		\item If one has $\BobPOVM{X}{\bot} = \BobPOVM{Z}{\bot}$, then we find that the filtering measurements $\AliceBobPOVMFilter{b_A,b_B}{\con}$ is independent of the basis choices $(b_A,b_B)$. Let this basis-independent  POVM element be $\tilde{F}$. If the filtering measurement does not depend on the basis choice, then implementing the basis choice followed by filtering measurement is the same as implementing the filtering measurement followed by basis choice. Thus, we can delay basis choice until after the filtering measurements have been performed. This can also be formally argued using \cref{lemma:twostep}. This allows us to obtain the first node of \cref{fig:virtualprotperfect}, where we measure using $\{\tilde{F} , \id -\tilde{F}\}$.
		\item This is then followed by random basis choices and assignment to test vs key by Alice and Bob.  All $X$ basis rounds are used for testing, while most $Z$ basis rounds are used for key generation (and a small fraction is used to estimate $\eZ$). Thus, we get the second node of \cref{fig:virtualprotperfect}.  Note that the estimate $\eZ$ of the error rate in the key bits is only used to determine the amount of error-correction required and does not affect the secrecy of the protocol. However, $\eZ$ and the choice of error-correction protocol is important to ensure that error-verification succeeds with high probability.
		
		\item The $\nX$ testing rounds are measured using $\{ \AliceBobPOVMsecond{X}{=},\AliceBobPOVMsecond{X}{\neq} \}$, and the error rate in these rounds is denoted by $\eX$. This is the error rate we observe. This is the \textbf{Testing (X)} node of \cref{fig:virtualprotperfect}.
		
		\item The $\nK$ key generation rounds can be measured (virtually) using the same POVM \\ $\{ \AliceBobPOVMsecond{X}{=},\AliceBobPOVMsecond{X}{\neq} \}$. The error rate in these rounds is denoted by $\eph$ and is the phase error rate we wish to estimate. This is the \textbf{Key (X)} node of \cref{fig:virtualprotperfect}.
		
		\item The actual $\nK$ key generation rounds are measured in the $Z$ basis to obtain the raw key. This is the \textbf{Key (Z)} node of \cref{fig:virtualprotperfect}.
	\end{enumerate}
	
	\subsection{Sampling} \label{subsec:samplingperfect}
	We will now turn our attention to the sampling part of the argument, and obtain an estimate $\Boundbasiczero$ on the phase error rate that satisfies \cref{eq:req}. To do so, we will make use of the following Lemma, which uses the Serfling bound \cite{serflingProbability1974}. For the proof, we refer the reader to \cref{appsubsec:randomsampling}.

	\begin{restatable}{lemma}{serflinglemma}[Serfling with IID sampling] \label{lemma:sampling}
		Let $\X_1 \dots \X_{n}$ be bit-valued random variables. Suppose each position $i$ is mapped to the ``test set'' ($i \in \bm{J}_t$) with probability $p_t$, and the ``key set'' ($i \in \bm{J}_k$) with probability  $p_k$.  
		Let $\event{\nX,\nK}$ be the event that exactly $\nX$ positions are mapped to test, and exactly $\nK$ positions are mapped to key. Then, conditioned on the event $\event{\nX,\nK}$, the following statement is true:
		\begin{equation}
			\begin{aligned}
				\Pr\left(\sum_{i \in \bm{J}_k} \frac{\X_i}{\nK} \geq \sum_{i \in \bm{J}_t} \frac{\X_i}{\nX} + \gamma_\text{serf} \right)_{ | \event{\nX,\nK}} &\leq e^{-2 \gamma_\text{serf}^2 \fserf(\nX,\nK)} ,\\
				\fserf(\nX,\nK )&\coloneq \frac{\nK \nX^2}{(\nK+\nX)(\nX+1)}.
			\end{aligned}
		\end{equation}
	\end{restatable}
	
	To use the lemma, we will identify $X_i=1$ with error, and $X_i=0$ with the no-error  outcome, when the conclusive rounds are measured in the $X$ basis. The test data will correspond to $\eX$, whereas the key data will correspond to $\eph$. 
	
	\begin{remark}
		There are two important aspects to the sampling argument. First, the Serfling bound applies in the situation where one chooses a random subset of \textit{fixed-length} for testing. However, the above procedure (and many QKD protocols) randomly assigns each round to testing vs key generation. Thus, Serfling must be applied with some care, and that is what is done here (see footnote.~\footnote{It is also worthwhile to note that if one is interested in estimating the QBER \textit{independent} of basis, then the standard serfling argument is directly applicable (for instance in \cite{tomamichelLargely2017}).}). This observation has been missing in many prior works. 
		Second, since we are interested in a variable-length protocol, we require slightly different statements than standard fixed-length security proofs (\cref{eq:req}). However, these can also be obtained by simple (almost trivial) modifications to existing arguments and yield the same results as before. Both these issues are addressed in the proof of \cref{lemma:sampling} in \cref{appendix:sampling}.
	\end{remark}

		Let us consider the second node in the equivalent protocol constructed in \cref{fig:virtualprotperfect}, where rounds are now randomly assigned for testing ($X$ basis) or key generation ($Z$ basis and key generation). (The remaining rounds are used for estimating the $Z$ basis error rate or discarded and are unimportant for this discussion).  Consider the state $\rho_{ | \event{\nX,\nK}}$, where the number of rounds to be used to testing and key generation is fixed. Using \cref{lemma:sampling} on this state, we obtain
		\begin{equation} \label{eq:samplinguseperfect}
			\begin{aligned}
				\Pr(\ephrv \geq \eXrv + \gamma_\text{serf})_{ | \event{\nX,\nK}} &\leq  e^{-2 \gamma_\text{serf}^2 \fserf(\nX,\nK)} ,
			\end{aligned}
		\end{equation}
		Furthermore we can choose
		\begin{equation} \label{eq:gammaserfdefined}
			\gamma^{\epsAT}_\text{serf}(\nX,\nK) \coloneq \sqrt{\frac{\ln(1/\epsAT^2) }{ 2 \fserf(\nX,\nK)} } \implies e^{-2 \left(		\gamma^{\epsAT}_\text{serf}(\nX,\nK) \right)^2 \fserf(\nX,\nK)} = \epsAT^2.
		\end{equation}
		Thus we can choose
		\begin{equation} \label{eq:estimateperfect}
			\Boundbasiczero(\eX,\nX,\nK) = \eX + \gamma^{\epsAT}_\text{serf} (\nX,\nK)
		\end{equation}
		to be our bound for the phase error rate,  where the $(0,0)$ subscript indicates that the bound is only valid when there is no deviation from basis-independent loss. Finally, since the bound is valid for any event $\Omega(\nX,\nK)$, we can get rid of this conditioning in  \cref{eq:samplinguseperfect}, to obtain \cref{eq:req} via
		\begin{equation} 
			\begin{aligned}
				\Pr(\ephrv \geq 	\Boundbasiczero(\eXrv,\nXrv,\nKrv)  )& = \sum_{\nX,\nK} \Pr(\event{\nX,\nK}) \Pr(\ephrv \geq 	\Boundbasiczero(\eXrv,\nX,\nK)  )_{ | \event{\nX,\nK}}  \\
				&\leq \sum_{\nX,\nK} \Pr(\event{\nX,\nK})  \epsAT^2 \\
				& = \epsAT^2
			\end{aligned}
		\end{equation}
		(In this work,  we will use the convention that $\sum_{x}$ denotes the sum over all possible values $x$ can take). 
		Thus, for the above choice of $\Boundbasiczero(\eX,\nX,\nK)$, the variable-length security of the protocol follows from \cref{thm:variablesecurity}.

		\section{Phase error estimation without basis-independent loss assumption}
		\label{sec:imperfectdetectors}

		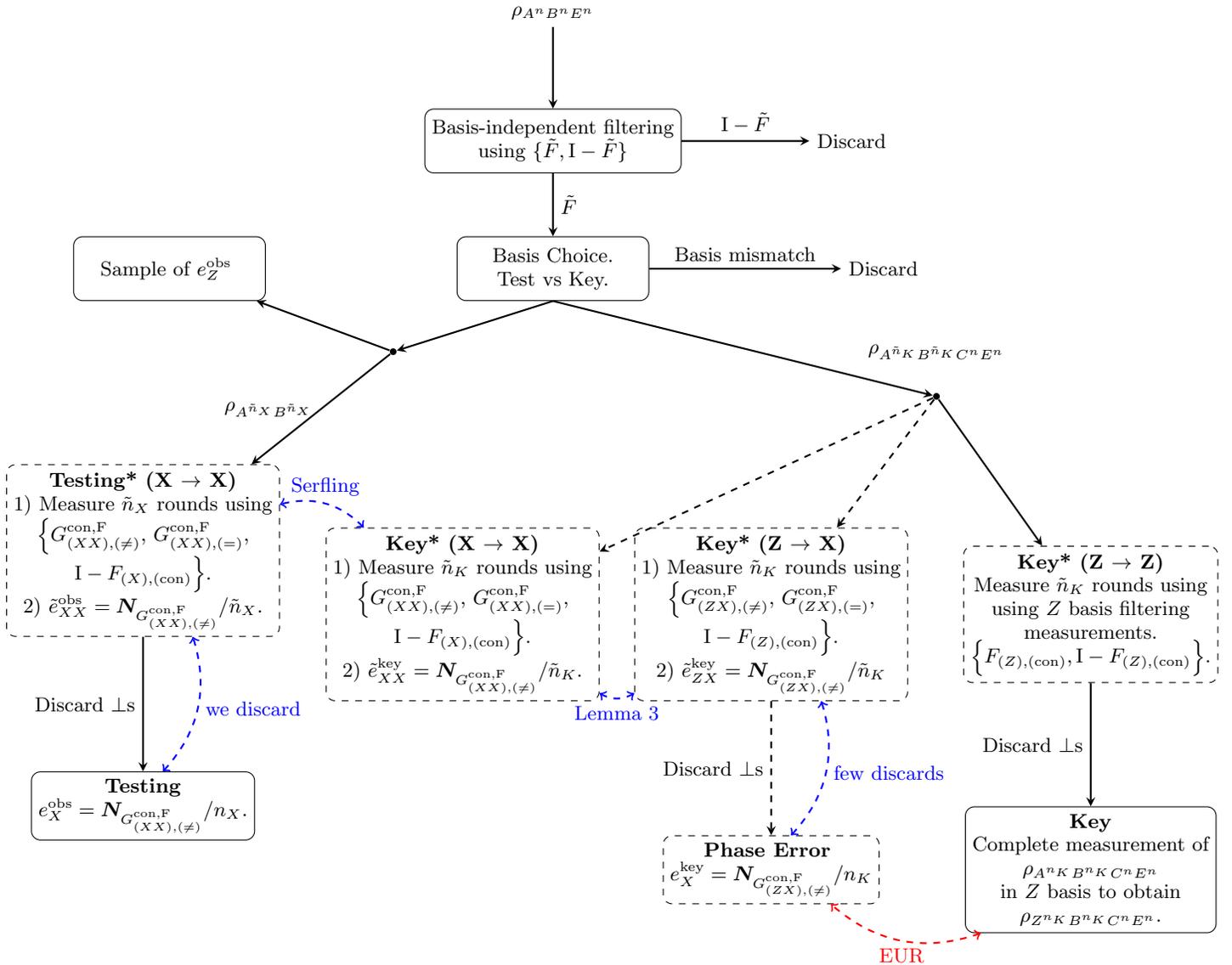
\begin{figure} 
         \hspace*{-2cm} 
			\begin{tikzpicture}[node distance=2cm] 
				\node (start) {$ \rho_{A^nB^nE^n}$ };
				\node (filtering) [process, below of=start] {Basis-independent filtering \\
					using $\{ \tilde{F} , \id - \tilde{F}\}$};
				
				\node (center) [process, below of=filtering] {Basis Choice. \\ Test vs Key. };

				\node (eZsample) [process, left of=center, xshift = -4cm] 
				{ Sample of $\eZ$ };

				\node(keystate) [point, right of=center,xshift = 4cm,yshift = -2cm] {};
				\node(teststate) [point, right of=center,xshift = -4.5cm,yshift = -1.3cm] {};
				
				\node(keystateabove) [emptyprocess, above of=keystate,yshift = -1.3cm] {$\rho_{A^{\nKperf}B^{\nKperf} C^n E^n}$};

				\node (testvirt) [virtualprocess, below left of=center, xshift=-5cm, yshift=-3cm] 
				{ \textbf{Testing* (X $\rightarrow$ X)} \\
					1) Measure $\nXperf$ rounds using \\
					$\Big\{ \AliceBobPOVMsecondsandwich{XX}{\neq},$ 
					$  \AliceBobPOVMsecondsandwich{XX}{=},$ \\$
					\id - \AliceBobPOVMprimeFilter{X}{\con}  \Big\}$. \\
					2) 	$\eXperf = \bm{N}_{ \AliceBobPOVMsecondsandwich{XX}{\neq}} / \nXperf$. };

				\node (test) [process, below of=testvirt,  yshift= -2cm] 	{ \textbf{Testing} \\
					$\eX = \bm{N}_{ \AliceBobPOVMsecondsandwich{XX}{\neq}} / \nX$. };

				\node (keyvirtXX) [virtualprocess, below left of=keystate, xshift=-6cm, yshift=-2cm] 
				{ \textbf{Key* (X $\rightarrow$ X)} \\
					1) Measure $\nKperf$ rounds using \\
					$\Big\{ \AliceBobPOVMsecondsandwich{XX}{\neq},$ 
					$  \AliceBobPOVMsecondsandwich{XX}{=},$ \\ $ \id - \AliceBobPOVMprimeFilter{X}{\con}  \Big\}$. \\
					2) $\ephwierd = \bm{N}_{ \AliceBobPOVMsecondsandwich{XX}{\neq}} / \nKperf$. };

				
				\node (keyvirtZX) [virtualprocess, below right of=keystate, xshift=-4cm, yshift=-2cm] 
				{ \textbf{Key* (Z $\rightarrow$ X)} \\
					1) Measure $\nKperf$ rounds using \\
					$\Big\{ \AliceBobPOVMsecondsandwich{ZX}{\neq},$ 
					$  \AliceBobPOVMsecondsandwich{ZX}{=},$ \\ $ \id - \AliceBobPOVMprimeFilter{Z}{\con} \Big\}$. \\ 
					2) $\ephperf = \bm{N}_{\AliceBobPOVMsecondsandwich{ZX}{\neq}} / \nKperf$};
				\node (keyZX) [virtualprocess, below of=keyvirtZX, yshift= -2cm] 
				{ \textbf{Phase Error } \\
					$\eph = \bm{N}_{\AliceBobPOVMsecondsandwich{ZX}{\neq}} / \nK$ };

				\node (keyvirt) [virtualprocess, below right of=keystate, xshift=1cm, yshift=-2cm] 
				{ \textbf{Key* (Z $\rightarrow$ Z)} \\
					Measure $\nKperf$ rounds using \\
					using  $Z$ basis filtering \\
					measurements.  \\
					$\Big\{\AliceBobPOVMprimeFilter{Z}{\con}, \id- \AliceBobPOVMprimeFilter{Z}{\con} \Big\}$. };

				\node (key) [process, below  of=keyvirt,yshift = -2cm] 
				{ \textbf{Key} \\
					Complete measurement of \\
					$\rho_{A^{\nK} B^{\nK} C^n E^n}$  \\
					in  $Z$ basis to obtain \\
					$\rho_{Z^{\nK} B^{\nK} C^n E^n}$.};

				\draw [arrow] (start) -- ++ (filtering); 
				\draw [arrow] (filtering) -- ++ (center) node[midway,right] {$\tilde{F}$};
				\draw [arrow] (teststate) -- (testvirt) node[midway,left] {$\rho_{A^{\nXperf} B^{\nXperf}}$};
				\draw [arrow] (teststate) -- (eZsample);
				\draw [arrow] (center.south) -- (keystate) ;
				\draw [dashedarrow] (keystate) -- (keyvirtXX);
				\draw [dashedarrow] (keystate) -- (keyvirtZX);
				\draw [arrow] (keystate) -- (keyvirt);
				\draw [arrow](center.south) -- (teststate);
				
				\draw [arrow] (testvirt) -- (test) node [midway, left] {Discard  $\bot$s};
				
				\draw [dashedarrow] (keyvirtZX) -- (keyZX) node [midway, left] {Discard  $\bot$s}  ;
				\draw [arrow] (keyvirt) -- (key) node [midway, left] {Discard  $\bot$s};

				\draw [arrow] (filtering.east) -- ++(2,0) node[midway, above] {$\id-\tilde{F}$} node[right] {Discard};
				
				\draw [arrow] (center.east) -- ++(3,0) node[midway, above] {Basis mismatch} node[right] {Discard};


				\draw (key) edge [dashed, <->, thick, bend left, color = red]  node[midway, below] {EUR} (keyZX);
				
				\draw (keyZX) edge [dashed, <->, thick, bend right, color = blue]  node[midway, right] {few discards} (keyvirtZX);

				\draw (keyvirtZX) edge [dashed, <->, thick, bend left, color = blue]  node[midway, below] {\cref{lemma:pulkitgeneric}} (keyvirtXX);

				\draw (keyvirtXX) edge [dashed, <->, thick, bend right, color = blue]  node[midway, above] {Serfling} (testvirt);

				\draw (test) edge [dashed, <->, thick, bend right, color = blue]  node[midway, right] {we discard} (testvirt);
				
			\end{tikzpicture}
			\caption{Protocol flowchart for the equivalent protocol from \cref{sec:imperfectdetectors}, where basis-independent loss assumption (\cref{eq:bobcondition}) is not satisfied. The dotted arrows and boxes represent virtual measurements that do not actually happen in the real protocol. Connections between the error rates in different boxes are highlighted using curved arrows. We use $\bm{N}_P$ to denote the number of $P$ measurement outcomes, where $P$ denotes a POVM element. For the POVMs, the reader may refer to \cref{table:povms} or \cref{subsec:protmeasure}. For clarity, we have omitted the conditioning on events in the figure (but not in our proof). Compared to \cref{fig:virtualprot}, the testing and key generation rounds go through an additional second step filtering measurement that depends on the basis used, which typically results in a few rounds being discarded, before undergoing the final measurement. The basis used in these measurements in indicated in each box, and indicates the basis used by \textit{both} Alice and Bob.}
			\label{fig:virtualprot}
		\end{figure}
		In this section, we will prove \cref{eq:req} for an implementation that \textit{does not satisfy} the basis-independent loss assumption. The argument is similar to the one presented in \cref{sec:perfectdetectors}, with important additions. It is helpful to refer to \cref{fig:virtualprot} for this section. We will first explain the idea behind the proof, before stating the proof itself.

		\subsubsection*{Proof Idea}
		We will use \cref{lemma:twostep}  in \cref{subsec:protmeasure} to construct an equivalent measurement procedure (in the sense that it is the same quantum to classical channel) for the protocol, which consists of three steps. The first step measurement is done using the POVM $\{ \tilde{F}, \id - \tilde{F}\}$ and implements basis-\textit{independent} filtering (discarding) operations. ($\tilde{F}$ here plays the same role as in \cref{sec:perfectdetectors}, but is defined differently). In particular it is the largest common filtering operation over both basis choices.
		
		Due to basis-efficiency mismatch, we will have a second step measurement that implements filtering operations that depends on the basis choice. (This will typically result in a small number of discards for a small amount of basis-dependent loss in the detectors). Once both filtering steps are done, the measurements on the remaining rounds can be completed using the third step  measurements which determines the exact measurement outcomes on the detected rounds.

		Turning our attention to \cref{fig:virtualprot}, the state first undergoes the basis-independent filtering measurement in the first node. This is then followed by random basis choice and assignment to testing and key generation at the second node. The testing rounds are further measured using second step $X$ basis filtering POVM and third step $X$ basis POVM at the \textbf{Testing* ($X \rightarrow X$)} node. Similarly, the key generation rounds are measured using second step $Z$ basis filtering POVM  and third step $Z$ basis POVM. Note that we use $(b_\text{2nd} \rightarrow b_\text{3rd})$ to denote the basis choice $b_\text{2nd}$ for the second step filtering measurement, and basis choice $b_\text{3rd}$ for the third step measurement, for both Alice and Bob.

		We will consider virtual measurements on the key generation rounds corresponding to  $X\rightarrow X$ and $Z\rightarrow X$. These are represented using dotted boxes and lines in the figure. These measurements are not performed in the protocol, but are only required in our proof. 
		We will then associate an error rate with all these choices of measurements, which corresponds to the number of rounds that resulted in an error divided by the total number of rounds on which the measurements were done.
		
		We see a variety of error rates in \cref{fig:virtualprot}. These errors are classified based on three criteria: 1) The basis used by Alice \textit{and} Bob in the second and third step measurements (written in the subscript), 2) Whether the $\bot$s due to the second step measurements have been discarded from the total number of rounds or not ($e$ vs $\tilde{e}$), 3) Whether they were done on testing rounds ($\text{obs}$ in superscript), or key generation rounds ($\text{key}$ in superscript). The proof will follow by building a connection from our observed error rate ($\eX$), to the phase error rate  ($\eph$). These connections are highlighted using curved blue arrows in the figure. Note that we only observe the error rate $\eX$ in the protocol. 
		
		In particular, we will relate the error rates before and after discarding for the testing rounds ($\eXperf \leftrightarrow \eX$) by simply noting that we discard rounds in the second-step measurements. On the other hand, we will relate $\ephperf
		\leftrightarrow \eph$ by bounding the number of discards that can happen in the second step filtering measurements. This relation will depend on $\delta_2$, which will be the metric that quantifies the ``smallness" of the POVM element corresponding to the discard outcome.  $\eXperf$ and $\ephwierd$ correspond to error rates corresponding to exactly the same measurement, and assigned to test vs key randomly. Thus, they can be related using Serfling (\cref{lemma:sampling}), exactly as in \cref{sec:perfectdetectors}.  $\ephwierd$ and $\ephperf$ correspond to error rates on the same state, but with slightly different POVMs, and thus are expected to be similar. This can be rigorously argued using \cref{lemma:pulkitgeneric}, where we use $\deltaone$ to quantify the ``closeness" of these POVMs. Combining all these relations, we will ultimately obtain \cref{eq:estimateimperfect}.

		We will now convert the above sketch into a rigorous proof. We start by explaining the three step protocol measurements.

		\subsection{Protocol Measurements} \label{subsec:protmeasure}
		Fix the basis $b_A,b_B$ used by Alice and Bob.
		As in \cref{subsec:protmeasureperfect}, consider the POVM \\ $\{\AliceBobPOVM{b_A,b_B}{\neq},
		\AliceBobPOVM{b_A,b_B}{=}, \AliceBobPOVM{b_A,b_B}{\bot}\}$ defined in \cref{eq:alicebobpovms}, which correspond to Bob obtaining a conclusive outcome (and Alice and Bob obtaining an error), Bob obtaining a conclusive outcome (and Alice and Bob not obtaining an error), and Bob obtaining an inconclusive outcome respectively.  Since $\AliceBobPOVM{b_A,b_B}{\bot} $ now depends on the basis choices, we cannot proceed in the same way as before. This reflects the fact that the discarding is basis dependent. Thus we will reformulate the measurement process in a different way.
		
		To do so, consider a $\tilde{F}$ such that 
		\begin{equation}\label{eq:Ftildedefined}
			\tilde{F} \geq	\AliceBobPOVM{b_A,b_B}{=} + \AliceBobPOVM{b_A,b_B}{\neq}  \quad \forall (b_A,b_B)
		\end{equation}
		This $\tilde{F}$ will play the role of a common  \inquotes{basis-independent filtering measurement}.	While any choice satisfying the above requirement will suffice, for the best results, $\tilde{F}$ must fulfil \cref{eq:Ftildedefined} as tightly as possible.
		\begin{remark}
			Since basis-mismatch rounds are discarded anyway, it is possible to argue that we only need $\tilde{F}$ to satisfy $	\AliceBobPOVM{b_A,b_B}{=}  +\AliceBobPOVM{b_A,b_B}{\neq} \leq \tilde{F}$ for $b_A = b_B$. This involves constructing a slightly different equivalent protocol where the first node decides basis match vs mismatch. The basis match events then undergo the usual filtering followed by basis choice, while the mismatch events are discarded without any filtering. If this modified requirement results in a value of $\tilde{F}$ that is ``smaller" then the original choice, then this will lead to tighter key rates. Intuitively, this is due to the fact that a smaller value of $\tilde{F}$ means that more loss is attributed to the basis-independent filtering.
		\end{remark}

		To reformulate the measurement procedure, start by considering the four-outcome POVM given by $\{\id - \tilde{F},  \tilde{F} - \AliceBobPOVM{b_A,b_B}{=}  - \AliceBobPOVM{b_A,b_B}{\neq} , \AliceBobPOVM{b_A,b_B}{=} , \AliceBobPOVM{b_A,b_B}{\neq}   \}$,  where the first two outcomes correspond to discard, the third correspond to a conclusive no-error outcome, and the fourth corresponds to a conclusive error. This  four-outcome measurement followed by classical grouping of the first two outcomes is then equivalent to the original three-outcome measurement in the protocol.
		
		Now, we can use \cref{lemma:twostep} to reformulate the four-outcome measurement as occurring in two steps. In the first step, Alice and Bob measure using POVM $\{ \tilde{F} , \id - \tilde{F} \}$ and discard the latter outcomes. If they obtain the $\tilde{F}$ outcome, they then complete the measurement using POVM
		$\{  \AliceBobPOVMFilter{b_A,b_B}{\bot}, \AliceBobPOVMFilter{b_A,b_B}{=}, \AliceBobPOVMFilter{b_A,b_B}{\neq}\}$, corresponding to discard, conclusive no-error and conclusive error outcomes respectively. 
		
		We then use \cref{lemma:twostep} again to reformulate this three-outcome measurement to consist of two steps. First, they measure using $\{\AliceBobPOVMprimeFilter{b_A,b_B}{\con} ,  \AliceBobPOVMprimeFilter{b_A,b_B}{\bot}\}$ which determines whether they obtain a conclusive or inconclusive measurement outcome. Then, if they obtain a conclusive outcome, they measure using the POVM $\{ \AliceBobPOVMsecond{b_A,b_B}{=},\AliceBobPOVMsecond{b_A,b_B}{\neq} \}$. Thus we now have a three-step measurement procedure, described in \cref{table:povms}.

		Since basis-mismatch signals are anyway discarded in the protocol, from this point onwards, we will only be concerned with POVMs that correspond to Alice and Bob choosing the same basis. As before, we will use the convention that whenever a basis is explicitly written as $X /Z$ (or denoted using $b_1  b_2$), it represents \textit{both} Alice and Bob's basis choices.
		
		It will be convenient to recombine the second and third step measurement into a single measurement step with three outcomes. For brevity we introduce the following notation to write this POVM $\{\AliceBobPOVMsecondsandwich{b_1 b_2}{\neq},\AliceBobPOVMsecondsandwich{b_1 b_2}{=} ,\id - \AliceBobPOVMprimeFilter{b_1}{\con}  \}$ where
		\begin{equation} \label{eq:notationGsandwich}
			\begin{aligned}
				\AliceBobPOVMsecondsandwich{b_1b_2}{\neq} &= \sqrt{\AliceBobPOVMprimeFilter{b_1}{\con}}\AliceBobPOVMsecond{b_2}{\neq} \sqrt{\AliceBobPOVMprimeFilter{b_1}{\con}}, \\
				\AliceBobPOVMsecondsandwich{b_1b_2}{=} &= \sqrt{ \AliceBobPOVMprimeFilter{b_1}{\con}}\AliceBobPOVMsecond{b_2}{=} \sqrt{\AliceBobPOVMprimeFilter{b_1}{\con}}. 
			\end{aligned}
		\end{equation}
		where the subscript $b_1 b_2$ determines the basis for the second step and third step measurements by \textit{both} Alice and Bob, and the superscript $F$ indicates the merging of the two measurement steps. (Note that if $b_1 = b_2 = b$, then this simply reverses the earlier action of \cref{lemma:twostep} that split $\{  \AliceBobPOVMFilter{b,b}{\bot}, \AliceBobPOVMFilter{b,b}{=}, \AliceBobPOVMFilter{b,b}{\neq}\}$ to generate the second and third-step measurements. However, we will consider fictitious measurements where $b_1 \neq b_2$ in our proof. To describe such measurements, it is indeed necessary to split $\{  \AliceBobPOVMFilter{b_A,b_B}{\bot}, \AliceBobPOVMFilter{b_A,b_B}{=}, \AliceBobPOVMFilter{b_A,b_B}{\neq}\}$  into two separate steps.)

        \begin{table}[H]
    \centering
    \renewcommand{\arraystretch}{1.3} 
    \begin{tabularx}{\textwidth}{>{\raggedright\arraybackslash}p{4cm} X}
        \toprule
        \textbf{Symbol} & \textbf{Meaning} \\
        \midrule
				$\{ \tilde{F}, \id -\tilde{F}\}$ & First step measurement. Implements basis-independent filter. \\
			
				$\{\AliceBobPOVMprimeFilter{b_A,b_B}{\con}, \id- \AliceBobPOVMprimeFilter{b_A,b_B}{\con} \}$.  & Second step measurement. Implements filtering that is basis dependent.  \\
			
				$\{ \AliceBobPOVMsecond{b_A b_B}{=},\AliceBobPOVMsecond{b_A b_B}{\neq} \}$  & Third step measurement corresponding to no-error and error. \\
				
				$\{ \AliceBobPOVMsecondsandwich{b_1 b_2}{\neq} $ ,
				$ \AliceBobPOVMsecondsandwich{b_1 b_2}{=} ,\id - \AliceBobPOVMprimeFilter{b_1}{\con}  \}$ & Combined second and third step measurement, corresponding to no-error, error and discard. \\ 
				
				$\nXperf$ & Number of testing rounds after basis-independent filter only \\
				
				$\nKperf$ & Number of key generation rounds after basis-independent filter only \\
				
				$\nX$ & Actual number of testing rounds \\
				
				$\nK$ & Actual number of key generation rounds \\
				\bottomrule
		   \end{tabularx}
			\caption{Different symbols used in our proof. Note that $b_A,b_B$ refer to basis choice of Alice and Bob. However, $b_1, b_2$ refer to the basis used by \textit{both} Alice and Bob, for the second and third step measurements. Whenever a basis is explicitly written as $X /Z$ (or $b_1,b_2$ ) it represents \textit{both} Alice and Bob's basis choices.
			}\label{table:povms}
		\end{table}

		\subsection{Constructing an equivalent protocol} \label{subsec:equivprot}
		We will now construct the equivalent protocol from \cref{fig:virtualprot}. The construction is similar to the one from \cref{subsec:equivprotperfect}, albeit with some important modifications. 
		\begin{enumerate}
			\item As in \cref{subsec:equivprotperfect}, we observe that the first step measurement is conducted using $\{\tilde{F},\id -\tilde{F}\}$ and is independent of basis. Therefore, we can delay basis choice until after this measurement has been completed, and the $\id -\tilde{F}$ outcomes are discarded. That is the first node of \cref{fig:virtualprot}.
			\item The remaining rounds undergo random basis choice. Basis mismatch rounds are discarded, all $X$ basis rounds are used for testing, while $Z$ basis rounds are probabilistically chosen for testing and key generation. This allows us to obtain the second node of \cref{fig:virtualprot}.  Again, as in \cref{subsec:equivprotperfect}, the estimate we obtain on $\eZ$ does not affect the secrecy claim of the protocol, since $\eZ$ is only used to determine the amount of error-correction to be performed. 
		\end{enumerate}
		Note that unlike \cref{subsec:equivprotperfect}, we have to perform \textit{two} measurements on the testing and key generation rounds after the second node, and these rounds are \textit{not} guaranteed to result in a conclusive outcome. We describe these measurements in detail below.
		\subsubsection{Testing Rounds after basis-independent Filter}
		We will now complete the measurement steps on the test rounds (which take place in the \textbf{Testing* ($X \rightarrow X$)} box in \cref{fig:virtualprot}).
		Let us consider the $X$ basis rounds used for testing at this stage. Let $\nXperf$ be the number of such rounds. Note that some of these rounds will be discarded during the remainder of the protocol, and therefore we do not know the value of $\nXperf$ in the actual protocol. However, we will see that we do not need to.
		
		These rounds must undergo the second step filtering measurement using $\{\AliceBobPOVMprimeFilter{X}{\con}, \id- \AliceBobPOVMprimeFilter{X}{\con} \}$, where the rounds which yield the latter outcome are discarded. Now, the remaining rounds are measured using the third step $\{ \AliceBobPOVMsecond{X}{=},\AliceBobPOVMsecond{X}{\neq} \}$ that determines whether Alice and Bob observe an error or no error. Recall that we use the convention that whenever a basis is explicitly written as $X/Z$, it refers to \textit{both} Alice and Bob measuring in the same basis.

		Combining the second and third measurement step, we see that measuring $\nXperf$ rounds using the above two-step procedure is equivalent to measuring directly using 
		$\Big\{ \AliceBobPOVMsecondsandwich{XX}{\neq} $ ,
		$ \AliceBobPOVMsecondsandwich{XX}{=} ,\id - \AliceBobPOVMprimeFilter{X}{\con}  \Big\}$ (see \cref{eq:notationGsandwich}), with the outcomes corresponding conclusive and error, conclusive and no-error and inconclusive respectively.  
		We write   $\eXperf$ be the error rate in these rounds, which is the fraction of rounds that resulted in the $\AliceBobPOVMsecondsandwich{XX}{\neq}$-outcome. The subscript $XX$ reflects the fact that this is the error rate when the second step and third step measurements are in $X$ basis. Note that we do not actually observe this error rate in the protocol. We write $\eX$ as the error rate in these rounds after discarding the $\bot$ outcomes. This is the error rate we actually observe in the protocol.
		
		\subsubsection{Key Generation Rounds after basis-independent Filter}
		We will now complete the virtual measurement steps on the key generation rounds, that lead to the phase error rate (which take place in the \textbf{Key* ($Z \rightarrow X$)} box in \cref{fig:virtualprot}). Let us consider the $Z$ basis rounds selected for key generation at this stage. Let $\nKperf$ be the number of such rounds. Note that some of these rounds will be discarded during the remainder of the protocol, and therefore we do not actually know the value of $\nKperf$ in the protocol. However, as in the case of $\nXperf$, we do not need to.
		
		These rounds must undergo the second step filtering measurement using $\{\AliceBobPOVMprimeFilter{Z}{\con}, \id- \AliceBobPOVMprimeFilter{Z}{\con} \}$, where the rounds which yield the latter outcome are discarded. Now, we wish to obtain the phase error rate when the remaining rounds are measured using the third step $\{ \AliceBobPOVMsecond{X}{\neq},\AliceBobPOVMsecond{X}{=} \}$ that determines whether Alice and Bob observe an error or no error. 
		
		Again, the above two-step measurement procedure is equivalent to measuring directly using
		$\Big\{ \AliceBobPOVMsecondsandwich{ZX}{\neq},$ 
		$ \AliceBobPOVMsecondsandwich{ZX}{=},$ $\id - \AliceBobPOVMprimeFilter{Z}{\con}  \Big\}$ (see \cref{eq:notationGsandwich}), with the outcomes corresponding conclusive and error, conclusive and no-error and inconclusive respectively. 
		We let  $\ephperf$ be the error rate in these rounds, which is the fraction of rounds that resulted in the $\ \AliceBobPOVMsecondsandwich{ZX}{\neq}$-outcome. Again, the subscripts denote the fact that this is the error rate when the second step measurement is in the $Z$ basis and the third step measurement is in the $X$ basis. 
		The phase error rate $\eph$ is the error rate in these rounds after discarding the $\bot$ outcomes.
		\begin{remark}
			When basis-efficiency mismatch is present, one must figure out the phase error rate in the key generation rounds, which are filtered using the $Z$ basis. However the rounds for testing are filtered using the $X$ basis. These filtering steps are not identical. Therefore it becomes very difficult to prove rigorous bounds on the phase error rate based on the observed data. One of the main contributions of this work is a rigorous derivation of such bounds, without relying on asymptotic behavior or IID assumptions.
		\end{remark}

		Since the measurements in the key generation rounds leading to $\ephperf$ are not identical to the one in the testing rounds which leads to $\eXperf$, one cannot directly use Serfling (\cref{lemma:sampling}) to relate the two, as we did in \cref{subsec:samplingperfect}.
		Therefore, we introduce another set of virtual measurements (which take place in the  \textbf{Key* ($X \rightarrow X$)} box in \cref{fig:virtualprot}), corresponding to $X$ basis second and third step measurements. Thus we obtain
		another error rate $\ephwierd$. This is the error rate corresponding to the case where these $\nKperf$ rounds are measured using 	$\Big\{ \AliceBobPOVMsecondsandwich{XX}{\neq},$ 
		$ \AliceBobPOVMsecondsandwich{XX}{=} , \id- \AliceBobPOVMprimeFilter{X}{\con}\Big\}$ (the same measurement that testing rounds are subject to).

		\subsection{Cost of removing the basis-independent loss assumption} \label{subsec:assumptions}
		In removing the basis-independent loss assumption from phase error estimation, we will need to define metrics $\deltaone,\deltatwo$, which will quantify the deviation from ideal behavior. We will now explain how these metrics are defined. 
		
		Consider the POVM elements $\AliceBobPOVMsecondsandwich{ZX}{\neq}$ and $ \AliceBobPOVMsecondsandwich{XX}{\neq}$ defined via \cref{eq:notationGsandwich}, which combine the second and third step measurements. In \cref{sec:perfectdetectors} they were exactly equal. 	We define $\deltaone$ to quantify the closeness of these POVM elements as
		\begin{equation} \label{eq:deltaonedefone}
			\delta_1 \coloneq 2 \norm{\AliceBobPOVMsecondsandwich{ZX}{\neq} - \AliceBobPOVMsecondsandwich{XX}{\neq} }_\infty,
		\end{equation}
		and use it in \cref{lemma:pulkitgeneric} (to be discussed later) in our proof. 
		
		Consider the second step measurements, where outcomes corresponding to POVM element $ \id - \AliceBobPOVMprimeFilter{Z}{\con}$ are discarded. In \cref{sec:perfectdetectors}, there was no need of the second step filtering measurement, which is equivalent to having $\AliceBobPOVMprimeFilter{Z}{\con} = \id$. We define $\deltatwo$ to quantify the amount of deviation from this case as
		\begin{equation} \label{eq:deltatwodef}
			\deltatwo \coloneq	\norm{\id - \AliceBobPOVMprimeFilter{Z}{\con} }_\infty .
		\end{equation}
		Thus $\deltatwo$ controls the likelihood of discards in the second step filtering measurements.
		
		Having defined $\deltaone,\deltatwo$ as metrics of the deviation from the basis-independent loss assumption, we now move on to consider the relations between the error rates in the next subsection.

		\subsection{Sampling} \label{subsec:sampling}
		Let us recall the error-rates we have defined so far:
		\begin{enumerate}
			\item $\eX $ is the fraction of the $\nX$ testing rounds that resulted in the $\AliceBobPOVMsecondsandwich{XX}{\neq}$ outcome. We have access to $\eX$ in the protocol, since it is something we actually observe. 
			\item $\eXperf $ is the fraction of the $\nXperf$ testing rounds (after basis-independent filter only) that result in $\AliceBobPOVMsecondsandwich{XX}{\neq}$-outcome. $\eX$ is obtained from $\eXperf$ after some rounds are discarded in the second step measurements.
			\item $\ephwierd$ is the fraction of the $\nKperf$ key generation rounds (after basis-independent filter only) that result in $\AliceBobPOVMsecondsandwich{XX}{\neq}$-outcome. 
			\item $\ephperf$ is the fraction of the $\nKperf$ key generation rounds (after basis-independent filter only) that result in $\AliceBobPOVMsecondsandwich{ZX}{\neq}$-outcome. 
			\item $\eph$ is the fraction of the $\nK$ key generation rounds that result in $\AliceBobPOVMsecondsandwich{ZX}{\neq}$-outcome. This is the quantity we wish to estimate. $\eph$ is obtained from $\ephperf$ after some rounds are discarded in  the second step measurements.
		\end{enumerate}
		
		We wish to prove \cref{eq:req} that relate $\eXrv$ to $\ephrv$. We do this by relating the various error-rates together as  $\eXrv \leftrightarrow \eXperfrv \leftrightarrow \ephwierdrv \leftrightarrow\ephperfrv\leftrightarrow\ephrv$. We will consider the event $\event{\nXperf,\nKperf}$, even though we do not actually observe it in the protocol. In the end, all random variables and events not directly observed in the protocol will disappear from our final expressions.
		
		\begin{itemize}
			\item $\eXrv \leftrightarrow \eXperfrv$: Recall from the \textbf{Testing*($X \rightarrow X$)} node in \cref{fig:virtualprot}, that $\eXrv = \num_{\AliceBobPOVMsecondsandwich{XX}{\neq}} / \nXrv$ and $\eXperfrv = \num_{\AliceBobPOVMsecondsandwich{XX}{\neq}} / \nXperfrv$. The required relation follows from the fact that we discard rounds to go from $\eXperfrv$ to $\eXrv$, i.e we have $\Pr(\nXrv \leq \nXperfrv)_{ | \event{\nXperf,\nKperf}} = 1$.  Therefore, we obtain 
			\begin{equation} \label{eq:boundone}
				\Pr( \eXperfrv \geq \eXrv )_{ | \event{\nXperf,\nKperf}} = 0.
			\end{equation}

			\item $\eXperfrv \leftrightarrow \ephwierdrv$ : These error rates correspond to measurement outcomes using the \textit{same} POVM, but with random assignment to testing vs key generation. Thus we can apply \cref{lemma:sampling} (Serfling) in  exactly the same manner as in \cref{subsec:samplingperfect}, conditioned on the event $\event{\nXperf,\nKperf}$. In doing so, we obtain
			\begin{equation} \label{eq:samplinguseimperfect}
				\begin{aligned}
					\Pr(\ephwierdrv \geq \eXperfrv + \gamma_\text{serf})_{ | \event{\nXperf,\nKperf}} &\leq  e^{-2\gamma_\text{serf}^2 \fserf(\nXperf,\nKperf)}.
				\end{aligned}
			\end{equation}
			Using the definition from \cref{eq:gammaserfdefined}, we have 
			\begin{equation} \label{eq:gammavalueimperfect}
				\gamma^{\epsATa}_\text{serf}(\nXperf,\nKperf) = \sqrt{\frac{\ln(1/\epsATa^2) }{ 2  \fserf(\nXperf,\nKperf)} } \implies e^{ - \left( \gamma^{\epsATa }_\text{serf}(\nXperf,\nKperf) \right)^2 2 \fserf(\nXperf,\nKperf)} = \epsATa^2.
			\end{equation}
			Therefore, we obtain
			\begin{equation} \label{eq:boundtwo}
				\begin{aligned}
					\Pr(\ephwierdrv \geq \eXperfrv + \gamma^{\epsATa}_\text{serf}   (\nXperf,\nKperf)   )_{ | \event{\nXperf,\nKperf}} &\leq  \epsATa^2
				\end{aligned}
			\end{equation}

			\item $ \ephwierdrv \leftrightarrow \ephperfrv$: We utilize the definition of $\deltaone$ stated in \cref{subsec:assumptions}. Since the POVM elements  generating $\ephperfrv$ ($\AliceBobPOVMsecondsandwich{ZX}{\neq} $) and $\ephwierdrv$ ($\AliceBobPOVMsecondsandwich{XX}{\neq} $) are close, we expect the bounds obtained on $\ephperfrv$ and $\ephwierdrv$ to also be close. This is made precise in the following lemma proved in \cref{subsec:pulkitlemmas}.

			\begin{restatable}{lemma}{pulkitlemmageneric}[Similar measurements lead to similar observed frequencies]  \label{lemma:pulkitgeneric} Let $\rho_{Q^n} \in S_\circ(Q^{\otimes n})$ be an arbitrary state. Let $\{P,\id - P\}$ and $\{P^\prime,\id-P^\prime\}$ be two sets of POVM elements, such that $\norm{P^\prime-P}_\infty \leq \delta$. 
				Then,
				\begin{equation} \label{eq:pulkitequationgeneric}
					\Pr(\frac{\num_{P^\prime}}{n} \geq e +  2 \delta+c) \leq \Pr(\frac{\num_P}{n} \geq e) + \binfunction{n}{2\delta}{c},
				\end{equation}
				for $e \in [0,1]$, where $\num_P$ is the number of $P$-outcomes when each subsystem of $\rho_{Q^n}$ is measured using POVM $\{P,\id - P\}$, and
				\begin{equation}
					\binfunction{n}{\delta}{c} \coloneq  \sum_{i = n (\delta+c)}^{n} {n \choose i} \delta^ i (1-\delta)^{n-i}. 
				\end{equation}
			\end{restatable}

			Thus, using  \cref{lemma:pulkitgeneric} and $\deltaone$ defined in \cref{eq:deltaonedefone}, we obtain
			\begin{equation} 
				\begin{aligned}
					\Pr(\ephperfrv \geq e + \deltaone + \cone)_{ | \event{\nXperf,\nKperf} } &\leq \Pr(\ephwierdrv \geq  e)_{| \event{\nXperf,\nKperf}}  + \binfunction{\nKperf}{ \deltaone}{ \cone}.
				\end{aligned}
			\end{equation}
			
			We would like $\binfunction{\nKperf}{ \deltaone}{ \cone}$ to be equal to a constant $\epsATb^2$ on the right hand side of the above expression. To do so, we note that $\binfunction{\nKperf}{ \deltaone}{ \cone}$ is a monotonic (and therefore invertible) function of $\cone$. Thus, we can choose  $c_1$ to be a function $	 \gamma^{\epsATb}_\text{bin}(\nKperf,\deltaone)$ such that 
			\begin{equation} \label{eq:gammabindefined}
				\binfunction{\nKperf}{ \deltaone}{ \gamma^{\epsATb}_\text{bin}(\nKperf,\deltaone})= \epsATb^2.
			\end{equation} 
			Using this as a definition $\gamma^{\epsATb}_\text{bin}(\nKperf,\deltaone)$, we obtain
			\begin{equation}  \label{eq:boundthree}
				\Pr(\ephperfrv \geq e + \deltaone + \gamma^{\epsATb}_{\text{bin}}(\nKperf,\deltaone) )_{ | \event{\nXperf,\nKperf}}\leq \Pr(\ephwierdrv \geq e)_{| \event{\nXperf,\nKperf}}  +\epsATb^2.
			\end{equation}
			Note that $\gamma_{\text{bin}}$ can be easily computed numerically by relating $\binfunction{n}{\delta}{c}$ (and its inverse) to the cumulative binomial distribution and using root finding algorithms.

			\item   $\ephperfrv \leftrightarrow \ephrv$:  We will use the fact that the filtering measurements result in a very small number of discards.
			
			First, note that $\ephperfrv =  \bm{N}_{\AliceBobPOVMsecondsandwich{ZX}{\neq}} / \nKperfrv$, and $\ephrv = \bm{N}_{\AliceBobPOVMsecondsandwich{ZX}{\neq}} / \nKrv$. Thus, we have $\ephperfrv	/ \ephrv = \nKrv /\nKperfrv$.

			Recall that $\nKrv$ is obtained by discarding rounds from $\nKperfrv$ based on $\{ \AliceBobPOVMprimeFilter{Z}{\con}, \id - \AliceBobPOVMprimeFilter{Z}{\con}\}$ measurements. We will essentially show that very few rounds are discarded in this step, using \cref{eq:deltatwodef}. To do so, we prove the following Lemma in \cref{appendix:sampling}.
			
			\begin{restatable}{lemma}{smallPOVM}[Small POVM measurement]
				\label{lemma:smallPOVM}
				Let $\rho_{Q^n} \in S_\circ(Q^{\otimes n})$ be an arbitrary state. Let $\{P,I - P\}$ be a POVM such that $\norm{P}_\infty \leq \delta$. Then
				\begin{equation}
					\Pr(\frac{\num_P}{n} \geq \delta+c) \le \binfunction{n}{\delta}{c} \coloneq  \sum_{i = n (\delta+c)}^{n} {n \choose i} \delta^ i (1-\delta)^{n-i} ,
				\end{equation}
				where $\num_P$ is the number of $P$-outcomes when each subsystem of $\rho_{Q^n}$ is measured using POVM $\{P,\id - P\}$.
			\end{restatable} 
			Then, using \cref{lemma:smallPOVM} with $P=\id - \AliceBobPOVMprimeFilter{Z}{\con}$ and $\deltatwo$ defined in \cref{eq:deltatwodef}, we obtain
			\begin{equation} 
				\begin{aligned}
					\Pr( \ephperfrv \leq \ephrv (1-\deltatwo-\ctwo))_{\event{\nXperf,\nKperf}} &=	\Pr(\frac{\nKperf - \nKrv}{\nKperf} \geq \deltatwo+\ctwo)_{ | \event{\nXperf,\nKperf}} \\
					&=	\Pr(\frac{\num_{\id - \AliceBobPOVMprimeFilter{Z}{\con} }}{\nKperf} \geq \deltatwo+\ctwo)_{ | \event{\nXperf,\nKperf}} \\
					&\leq \binfunction{\nKperf}{\deltatwo}{\ctwo}. \\
				\end{aligned}
			\end{equation}

			Again, we would like $\binfunction{\nKperf}{\deltatwo}{\ctwo}$ to be a constant value $\epsATc^2$. Thus, we replace $c_2$ with $\gamma^{\epsATc}_{\text{bin}}(\nKperf,\deltatwo)$
			and obtain
			
			\begin{equation} \label{eq:boundfour} 
				\begin{aligned}
					\Pr( \ephperfrv \leq \ephrv (1-\deltatwo- \gamma^{\epsATc}_{\text{bin}}(\nKperf,\deltatwo) ))_{ | \event{\nKperf,\nKperf}} &\leq \epsATc^2
				\end{aligned}
			\end{equation} 
			
		\end{itemize}
		
		Thus we have relationships \cref{eq:boundone,eq:boundtwo,eq:boundthree,eq:boundfour} between all the error rates, whose complements hold with high probability. These can all be combined using straightforward but cumbersome algebra (see \cref{appendix:combining}), to obtain
		\begin{equation} \label{eq:finalbound}
			\Pr (   \ephrv \geq \frac{\eXrv+  \gamma^{\epsATa}_\text{serf}   (\nXperf,\nKperf) + \deltaone + \gamma^{\epsATb}_{\text{bin}}(\nKperf,\deltaone)}{ (1-\deltatwo- \gamma^{\epsATc}_{\text{bin}}(\nKperf,\deltatwo) )}    )_{ | \event{\nXperf,\nKperf}} \leq \epsATb^2 + \epsATa^2 + \epsATc^2.
		\end{equation}
		Using the above expression requires us to know the values of $\nKperf$ and $\nXperf$ which we do  not. This problem is easily resolved by noting all the $\gamma$s are decreasing functions of $\nKperf$ and $\nXperf$, and that $\nKperf (\nXperf)$ cannot be smaller than $\nK (\nX)$ (since we discard rounds to from the former to the latter) . Thus, we can replace $\nKperf$ with $\nKrv$ and $\nXperf$ with $\nXrv$ and obtain
		
		\begin{equation} \label{eq:finalboundusable}
			\Pr (   \ephrv \geq \frac{\eXrv+  \gamma^{\epsATa}_\text{serf}   (\nXrv,\nKrv) + \deltaone + \gamma^{\epsATb}_{\text{bin}}(\nKrv,\deltaone)}{ (1-\deltatwo- \gamma^{\epsATc}_{\text{bin}}(\nKrv,\deltatwo) )}    )_{ | \event{\nXperf,\nKperf}} \leq \epsATb^2 + \epsATa^2 + \epsATc^2
		\end{equation}
		We set $\epsATa^2+\epsATb^2+\epsATc^2 = \epsAT^2$, and obtain the choice of $\Boundbasicdelta$:
		\begin{equation} \label{eq:estimateimperfect}
			\Boundbasicdelta(\eX,\nX,\nK) \coloneq  \frac{\eX+  \gamma^{\epsATa}_\text{serf}   (\nX,\nK) + \deltaone + \gamma^{\epsATb}_{\text{bin}}(\nK,\deltaone)}{ (1-\deltatwo- \gamma^{\epsATc}_{\text{bin}}(\nK,\deltatwo) )},
		\end{equation}
		where functions $\gamma_\text{bin},\gamma_\text{serf}$ are defined in \cref{eq:gammabindefined} and \cref{eq:gammaserfdefined} respectively.
		Since \cref{eq:finalboundusable} is valid for all events $\event{\nXperf,\nKperf}$, the above choice satisfies \cref{eq:req} via
		\begin{equation} 
			\begin{aligned}
				\Pr(\ephrv \geq 	\Boundbasicdelta(\eXrv,\nXrv,\nKrv)  ) &\leq \sum_{\nXperf,\nKperf} \Pr(\event{\nXperf,\nKperf}) 	\Pr(\ephrv \geq 	\Boundbasicdelta(\eXrv,\nXrv,\nKrv)  )_{ | \event{\nXperf,\nKperf}} \\
				&\leq  \sum_{\nXperf,\nKperf} \Pr(\event{\nXperf,\nKperf})  \epsAT^2  = \epsAT^2.
			\end{aligned}
		\end{equation}

		\begin{remark} Let us investigate the behavior of \cref{eq:estimateimperfect} in the limit  $\deltaone,\deltatwo \rightarrow 0$.
			Recall that $\gamma_\text{bin}^{\epsAT} (n,\delta)$ was defined as the value of $c$ such that $\binfunction{n}{\delta}{c} = \sum_{i=n(\delta+c)}^{n} {n \choose k} \delta^i (1-\delta)^{n-i} \leq \epsAT^2$.
			However, notice that $\delta \rightarrow  0 \implies \binfunction{n}{\delta}{c} \rightarrow 0$ for any value of $c$. Therefore, $\delta \rightarrow 0 \implies \gamma_\text{bin}(n,\delta) \rightarrow 0$. Setting these limits in \cref{eq:estimateimperfect}, we recover the result \cref{eq:estimateperfect} for the case where the basis-independent loss assumption is satisfied.
		\end{remark}

		Thus we now have a phase error estimation bound that is valid even in the presence of basis-efficiency mismatch. We summarize the results of this section in the following theorem. Note that the results of \cref{sec:perfectdetectors} are a special case ($\deltaone,\deltatwo=0$) of the following theorem.
		\begin{theorem}[Sampling with different filtering measurements] \label{theorem:samplingimperfect}
			Let $\rho_{A^nB^n} \in S_\circ(A^nB^n)$ be an arbitrary state representing $n$ rounds of the QKD protocol. Suppose each round is assigned to test with some probability and key with some probability (and discarded with some probability). 
			
			The test rounds undergo the following measurement procedure: 
			\begin{enumerate}
				\item Measurement using $\{\tilde{F},\id - \tilde{F}\}$ and discarding the latter outcomes.
				\item Measurement using $\{F^\text{test}_\text{con},\id - F^\text{test}_\text{con}\}$ and discarding the latter outcomes. We let $\bm{n_T}$ be the number of remaining rounds at this stage. 
				\item Measurement using $\{G^\text{test}_{\neq},G^\text{test}_=\}$. We let $\bm{e}^\text{obs} = \bm{N}_{G^\text{test}_{\neq}} / \bm{n_T}$ be the error rate in these rounds. 
			\end{enumerate}
			
			The key generation rounds undergo the following measurement procedure:
			\begin{enumerate}
				\item Measurement using $\{\tilde{F},\id - \tilde{F}\}$ and discarding the latter outcomes.
				\item Measurement using $\{F^\text{key}_\text{con},\id - F^\text{key}_\text{con}\}$ and discarding the latter outcomes. We let $\bm{n_K}$ be the number of remaining rounds at this stage. 
				\item Measurement using $\{G^\text{key}_{\neq},G^\text{key}_=\}$. We let $\bm{e}^\text{key} = \bm{N}_{G^\text{key}_{\neq}} / \nKrv$ be the error rate in these rounds. 
			\end{enumerate}
			
			Then, the following equation holds
			\begin{equation}
				\Pr(\bm{e}^\text{key} \geq 	\Boundbasicdelta(\bm{e}^\text{obs},\bm{n_T}, \bm{n_K})  ) \leq  \epsATa^2+\epsATb^2+\epsATc^2,
			\end{equation}
			where 
			\begin{equation}
				\begin{aligned}
					\Boundbasicdelta(e^\text{obs},n_T,n_K) &=  \frac{e^\text{obs}+  \gamma^{\epsATa}_\text{serf}   (n_T,n_K) + \deltaone + \gamma^{\epsATb}_{\text{bin}}(n_K,\deltaone)}{ (1-\deltatwo- \gamma^{\epsATc}_{\text{bin}}(n_K,\deltatwo) )} , \\
					\deltaone & =  2 \norm{  \sqrt{F^\text{key}_\text{con}} G^\text{key}_{\neq} \sqrt{F^\text{key}_\text{con}} - \sqrt{F^\text{test}_\text{con}}G^\text{test}_{\neq} \sqrt{F^\text{test}_\text{con}}     }_\infty, \\
					\deltatwo & = \norm{\id - F^\text{key}_\text{con} }_\infty,
				\end{aligned}
			\end{equation}
			and $\gamma_\text{bin},\gamma_\text{serf}$ are defined in \cref{eq:gammabindefined} and \cref{eq:gammaserfdefined} respectively. 
		\end{theorem}
\textit{Proof sketch.} 
		The rigorous proof follows from the analysis already seen in \cref{sec:imperfectdetectors}.  Essentially, we consider several error rates corresponding  to various measurement choices, as described in \cref{fig:virtualprot}, and use \cref{lemma:sampling,lemma:pulkitgeneric,lemma:smallPOVM} to relate the various error rates together.
	
		\begin{remark}
			Note that in our analysis in this section, we actually had $G^\text{key}_{\neq} = G^\text{test}_{\neq}$, i.e the third step measurements were identical in the key and test rounds. However, our result holds even if these measurements are different, by going through the same steps in the proof. Hence, we state \cref{theorem:samplingimperfect} in full generality.	
		\end{remark}

		\section{Application to Decoy-state BB84} \label{sec:decoy}
		So far in this work, we have focused our attention on the BB84 protocol implemented using perfect single-photon sources for pedagogical reasons. In this section, we will extend our techniques and obtain a variable-length security proof for decoy-state BB84 \cite{hwangQuantum2003,loDecoy2005,maPractical2005,hayashiSecurity2014,curtyFinitekey2014} with imperfect detectors. We base our security proof approach on that of Lim et al \cite{limConcise2014}, with the following differences.
        
		First, we rigorously prove the security for the variable-length decoy-state BB84 protocol in the entropic uncertainty relations framework (note that Ref.~\cite{limConcise2014} actually implicitly implements a variable-length protocol). In fact \cite{limConcise2014} considers a protocol with iterative sifting \footnote{This has been formulated in a variety of ways in the literature. In general, we use this phrase for protocols that have a loop phase, where some actions are taken repeatedly until certain conditions are met.}  where the total number of rounds in the protocol is not fixed a priori, and depends on observations made during the protocol, which are announced in a round-by-round manner. However a justification for this step is not provided. This is important because certain kinds of iterative sifting can lead to subtle issues in the security analysis (see Ref.~\cite{pfisterSifting2016} for some issues and Ref.~\cite{tamakiSecurity2018} for solutions). We do not fix this problem directly in this work. Instead, we consider a protocol with a fixed total number of signals sent, which avoids this problem. Second, we make rigorous certain technical steps in \cite{limConcise2014} (regarding entropic calculations on states conditioned on events), which we point out where applicable (see \cref{remark:zerophotonkey}). Third, our phase error estimates do not require the assumption of basis-independent loss unlike that of \cite{limConcise2014}. We also avoid a particular Taylor series approximation used by \cite{limConcise2014} (\cite[Eq. 22]{fungPractical2010}), and therefore our phase error estimation procedure yields a true bound without any approximations. Finally we also clarify certain aspects of the decoy analysis undertaken in \cite{limConcise2014}. In particular, we careful differentiate between random variables and an observed value of the random variable, and also properly condition on relevant events in our presentation. We stress that while we clarify and make rigorous	 certain steps in \cite{limConcise2014} (see also \cite[Section 6.1]{tupkary2025qkdsecurityproofsdecoystate}), our main contribution in this work is the variable-length security proof of decoy-state BB84 in the presence of detector imperfections.

		\begin{remark} Recently, a more accessible version of the security proof in Ref.~\cite{limConcise2014} was written in Ref.~\cite{wiesemannConsolidated2024}. While Ref.~\cite{wiesemannConsolidated2024}[Version 2] addresses many of the above concerns for fixed-length protocols, it does not deal with detector imperfections or variable-length protocols. 
		\end{remark}

		We start by first specifying the decoy-state BB84 protocol we study in \cref{subsec:decoyprotocol}. We will then explain the required bounds on the phase error rate in \cref{subsec:decoyphase}. We explain decoy analysis in \cref{subsec:decoyanalysis}, and state the security of our variable-length protocol in \cref{subsec:decoyvariable}. Some proofs are delegated to \cref{app:decoyanalysis}.

		\subsection{Protocol specification} \label{subsec:decoyprotocol}
		The decoy-state BB84 protocol modifies the following steps of the protocol described in \cref{sec:protocol}.
		\begin{enumerate}
			\item \textbf{State Preparation:} Alice decides to send states in the $Z(X)$ basis with probability $\pzA$ ($\pxA$). She additionally chooses a signal intensity $\mu_k \in \{\mu_1,\mu_2,\mu_{3}\}$ with some predetermined probability $p_{\mu_k}$ \footnote{This probability can depend on the basis used without affecting the results of this work. To incorporate this, one simply has to track the correct probability distribution through all the calculations.}. She prepares a phase-randomized weak laser pulse based on the chosen values, and sends the state to Bob.  We assume $\mu_1 > \mu_2 + \mu_3$ and $\mu_2 > \mu_3 \geq 0$. This requirement on the intensity values, as well as the total number of intensities, is not fundamental. It is used in deriving the analytical bounds in the decoy-state analysis.
			\item \textbf{Measurement:}  Bob chooses basis the basis $Z$($X$) with probability with $\pzB$($\pxB$) and measures the incoming state. This step of the protocol is identical to \cref{sec:protocol}.
			\item \textbf{Classical Announcements and Sifting:}  For all rounds, Alice and Bob announce the basis they used. Furthermore, Bob announces whether he got a conclusive outcome ($\{ \BobPOVM{b}{0},\BobPOVM{b}{1} \}$), or an inconclusive outcome ($\{  \BobPOVM{b}{\bot} \}$). A round is said to be ``conclusive'' if Alice and Bob used the same basis, and Bob obtained a conclusive outcome. 
			
			On all the $X$ basis conclusive rounds, Alice and Bob announce their measurement outcomes and intensity choices. We let $\nXmu{k}$ be the number of $X$ basis conclusive rounds where Alice chose intensity $\mu_k$, and let $\eXmu{k}$ be the observed error rate in these rounds. For brevity, we use the notation $\nXmu{\allk} = ( \nXmu{1} \dots \nXmu{3})$ to denote observations from all intensities. (We use similar notation for $\eXmu{\allk}$, $\nKmu{\allk}$ etc).
			
			On all $Z$ basis conclusive round, Alice and Bob announce their measurement outcomes with some small probability $\pt$. We let $\eZ$ denote the observed error rate in these rounds (intensity is ignored),  which is used to determine the amount of error-correction that needs to be performed. For the remaining $\nK$ rounds, Alice announces her intensity choices, and these rounds
			are used for key generation. 
			
			All announcements are stored in the register $C^n$. We use $\event{\decoyparams}$ to denote the event that $\decoyparams$ values are observed in the protocol.
			
		\end{enumerate}
		The remaining steps of the protocol are the same as in \cref{sec:protocol}. In particular, based on the observations $\decoyparams$, Alice and Bob implement one-way error-correction using  $\leak(\decoyparams)$ bits of communication, followed by error-verification, and privacy amplification to produce a key of $l(\decoyparams)$ bits. This requirement of one-way communication is not fundamental, and more complicated error-correction protocols can be accommodated in a straightforward manner (see   \cref{remark:errorcorrection}). Additionally note that our protocol generates key from \textit{all} intensities, instead of having a single ``signal'' intensity for key generation.

		\subsection{Required and actual phase error estimation bound} \label{subsec:decoyphase}
		In order to prove security for our decoy-state QKD protocol, we will need to bound two quantities. First, we must obtain a lower bound on the number of single-photon events that lead to key generation $\nKphrv{1}$. Second, we must obtain an upper bound on the phase error rate within these single-photon key generation rounds, given by $\ephphrv{1}$. This can be represented mathematically as
		\begin{equation} \label{eq:decoyreq}
			\Pr(\ephphrv{1} \geq \Bound_e(\eXmurv{\allk}, \nXmurv{\allk}, \nKmurv{\allk} ) \quad \lor \quad  \nKphrv{1} \leq \Bound_1(\nKmurv{\allk} )  ) \leq \epsAT^2,
		\end{equation}
		where	$\lor$ denotes the logical OR operator, and $\Bound_e,\Bound_1$ are functions that provide these bounds as a function of the observed values.
		
		This statement will be used in the proof of \cref{thm:decoyvariablesecurity} to prove the variable-length security of our protocol. We will derive the required bounds ($\Bound_e,\Bound_1$) in \cref{eq:decoyreq} in two steps. First we will use decoy analysis to convert from observations corresponding to different intensities (which we have access to) to those corresponding to different photon numbers (which we do not have access to).   We will be concerned with three outcomes $ \{X_{\neq},X,K\}$, corresponding to $X$ basis conclusive error outcome, $X$ basis conclusive outcome, and $Z$ basis conclusive outcome used for key generation respectively.
		Thus, at the end of the first step we will obtain
		
		\begin{equation} \label{eq:decoyreqstepone}
			\begin{aligned}
				\Pr \Big(  & \eXphrv{1} \geq \frac{	\Bounddecoymax{1} (\nXneqmurv{\allk})  }{  	\Bounddecoymin{1} (\nXmurv{\allk})}
				\quad \lor \quad   \nXphrv{1} \leq    \Bounddecoymin{1}( \nXmurv{\allk})   \quad \lor \quad 
				\nKphrv{1} \leq   \Bounddecoymin{1}(\nKmurv{\allk}  ) \Big) \leq  9 \epsdecoy^2
			\end{aligned}
		\end{equation}
		where $\Bounddecoymin{m}$ and $\Bounddecoymax{m}$ are functions that compute bounds on the $m$-photon components of the input statistics. Note that we use $ \nXneqmurv{\allk} =  (\nXmurv{1} \times \eXmurv{1}, \dots, \nXmurv{\nint}  \times \eXmurv{\nint})$ to denote the number of rounds resulting both Alice and Bob using the $X$ basis and obtaining an error,  for each intensity (and we will assume implicit conversion between these two notations). The $9$ on the RHS comes from the fact that we implement decoy analysis on $3$ different events and we have $3$ intensities. We will prove \cref{eq:decoyreqstepone} in \cref{subsec:decoyanalysis}.
		
		\begin{remark}
			Note that the only parameters actually observed in the protocol are given by $\decoyparams$. Variables like $\eXphrv{1}$ are not actually directly observed, but instead are derived from observations.
		\end{remark}
		
		In this second step, we will use $\eXphrv{1},\nXphrv{1},\nKphrv{1}$ to bound the single photon phase error rate $\ephphrv{1}$. Notice this is exactly what we showed \cref{sec:perfectdetectors,sec:imperfectdetectors}.  In particular, with $	\Boundbasicdelta$ directly obtained from \cref{eq:estimateimperfect}, we have
		\begin{equation} \label{eq:decoyreqsteptwo}
			\Pr(\ephphrv{1} \geq \Boundbasicdelta(\eXphrv{1},\nXphrv{1},\nKphrv{1}) ) \leq  \epsATsingle^2.
		\end{equation}
		where $\epsATsingle^2$ denotes the failure probability of the \inquotes{single-photon} part of our estimation.

		However, note that we do not directly observe $\eXphrv{1},\nXphrv{1},\nKphrv{1}$ in the decoy-state protocol (unlike \cref{sec:imperfectdetectors}). Thus we would like to replace these values with the bounds computed from our decoy analysis (\cref{eq:decoyreqstepone}). This is straightforward to do, since $\Boundbasicdelta$ is an increasing function of $\eXphrv{1}$, and decreasing function of $\nXphrv{1},\nKphrv{1}$.  This can be done formally by a straightforward application of the union bound for probabilities ($\Pr(\Omega_1 \lor \Omega_2) \leq \Pr(\Omega_1)+\Pr(\Omega_2)$) applied to \cref{eq:decoyreqstepone,eq:decoyreqsteptwo}. Doing so allows us to conclude that the probability of \textit{any} of the bounds in \cref{eq:decoyreqstepone,eq:decoyreqsteptwo} failing is smaller than $9 \epsdecoy^2 + \epsATsingle^2$.  Then we use the fact that if \textit{none} of the bounds inside the probabilities in \cref{eq:decoyreqstepone,eq:decoyreqsteptwo} fail, then this implies that the bounds inside the probability in \cref{eq:decoyreqcombined} below must hold. Formally, we obtain
		\begin{equation} \label{eq:decoyreqcombined}
			\begin{aligned}
				\Pr \Bigg(  & \ephphrv{1} \geq  \Boundbasicdelta \Bigg( \frac{	\Bounddecoymax{1} (\nXneqmurv{\allk})  }{  	\Bounddecoymin{1} (\nXmurv{\allk})}   ,  \Bounddecoymin{1}( \nXmurv{\allk})   ,  \Bounddecoymin{1}(\nKmurv{\allk})     \Bigg)   \quad \lor \quad \\
				& \nKphrv{1} \leq  \Bounddecoymin{1}(\nKmurv{\allk}  )\Bigg)  \leq 9\epsdecoy^2 +\epsATsingle^2 \eqcolon \epsAT^2 
			\end{aligned}
		\end{equation}
		which is the required statement.  Thus, it is now enough to prove \cref{eq:decoyreqstepone} in order to prove \cref{eq:decoyreqcombined} (equivalently \cref{eq:decoyreq}), for which we turn to decoy analysis in the next section.

		\subsection{Decoy Analysis} \label{subsec:decoyanalysis}
		Let $O$ denote a specific outcome of a given round, and let $\nO$ denote the number of rounds that resulted in the outcome $O$. For instance, it could denote that both Alice and Bob measured in the $X$ basis and obtained a detection (in which case $\nO = \nX$). We will perform a general decoy analysis for any outcome $O$.  Let $\nOmu{k}$ denote the number of rounds that resulted in the outcome $O$ where Alice used intensity $\mu_k$. We have access to this information during the protocol. Let $\nOph{m}$ denote the number of rounds that resulted in the outcome $O$ where Alice prepared a state of $m$ photons. We wish to obtain bounds on $\nOph{m}$ using $\nOmu{k}$.
		
		In practice, Alice first chooses an intensity $\mu_k$ of the pulse, which then determines the photon number $m$ of the pulse, via the Poissonian distribution, independently for each round. Thus we have
		\begin{equation} \label{eq:poissonian}
			p_{m | \mu_k} = e^{-\mu_k}  \frac{\mu_k^m}{m!}.
		\end{equation}
		The probability of $m$-photons being emitted, can be obtained via
		\begin{equation} 
			\tau_{m} = \sum_{\mu_k} p_{\mu_k} p_{m | \mu_k}  = \sum_{\mu_k}  p_{\mu_k} e^{-\mu_k}  \frac{\mu_k^m}{m!}.
		\end{equation}
		Now, without loss of generality, we can view Alice as \textit{first} choosing the photon number $m$, and \textit{then} choosing a intensity setting $\mu_k$ with probability given by 
		\begin{equation}
			p_{\mu_k | m} = p_{\mu_k} p_{m | \mu_k} / \tau_m.
		\end{equation} 
		This is the fundamental idea used by \cite{limConcise2014,hayashiSecurity2014,curtyFinitekey2014}.
		In this case, due to the fact that each signal is mapped to an intensity independently of other signals, one can apply the Hoeffdings inequality to these independent events, and obtain
		\begin{equation} \label{eq:decoy1}
			\Pr( \abs{ \nOmurv{k} - \sum_{m=0}^{\infty} p_{\mu_k | m} \nOphrv{m} } \geq \sqrt{ \frac{\nOrv}{2} \ln( \frac{2}{\epsdecoy^2}) } ) \leq  \epsdecoy^2.
		\end{equation}
		
		\begin{remark}
			The application of Hoeffdings inequality here is subtle, and is made rigorous in \cref{lemma:hoeffdings,lemma:decoy} in \cref{app:decoyanalysis} (see also Ref.~\cite{curtyFinitekey2014}). Note that in general, the photon numbers of every pulse in the protocol are chosen independently, since Alice chooses intensity independently for each pulse. However, here we are interested in photon numbers corresponding to rounds that led to a specific outcome $O$. Since we postselect pulses based on the outcome, we can no longer claim 
	that the photon numbers of these pulses (pulses that led to outcome $O$) are sampled independently, or that intensities of these pulses are chosen independently. This is because they now depend on Eve's attack.
		Rather, \cref{lemma:hoeffdings,lemma:decoy} rely on exploiting the fact that conditioned on  \textit{any fixed sequence} of photon numbers of the pulses, the intensities are chosen independently of one another. One can therefore apply Hoeffdings inequality. Then, since the resulting statements holds for any fixed sequence of photon numbers, the conditioning on this event can be removed. 
		\end{remark}	
		We can now combine \cref{eq:decoy1} for all intensities $\mu_k$ using the union bounds for probabilities ($\Pr(\Omega_1 \land \Omega_2) \geq 1 - \Pr(\Omega_1^c) - \Pr(\Omega_2^c)$). Reformulating the expressions, we obtain
		\begin{equation} 
        \begin{aligned}
        \label{eq:decoy1allintensities}
			&\Pr(  \nOmurv{k} - \sqrt{ \frac{\nOrv}{2} \ln( \frac{2}{\epsdecoy^2}) } \leq  \sum_{m=0}^{\infty} p_{\mu_k | m} \nOphrv{m}    \leq   \nOmurv{k} +  \sqrt{ \frac{\nOrv}{2} \ln( \frac{2}{\epsdecoy^2}) }\quad \forall k \in \{1,2,3\} ) \\
            &\geq  1- 3  \epsdecoy^2.
            \end{aligned}
		\end{equation}
		To obtain \cref{eq:decoyreqstepone}, we will apply decoy analysis (\cref{eq:decoy1allintensities}) for three separate events: conclusive $Z$ basis rounds selected for key generation (denoted by $K$), conclusive $X$ basis rounds (denoted by $X$), and conclusive $X$ basis rounds leading to an error (denoted by $X_{\neq}$).	Then, \cref{eq:decoy1allintensities} can be applied these events (again using the union bound for probabilities) to obtain:
		\begin{equation} \label{eq:decoy3}
			\begin{aligned}
				&\Pr \Bigg(  \nOmurv{k} - \sqrt{ \frac{\nOrv}{2} \ln( \frac{2}{\epsdecoy^2}) }   \leq  \sum_{m=0}^{\infty} p_{\mu_k | m} \nOphrv{m}    \leq  \nOmurv{k} + \sqrt{ \frac{\nOrv}{2} \ln( \frac{2}{\epsdecoy^2}) }\quad \\
                & \forall k \in \{1,2,3\} , \quad \forall O \in \{X_{\neq},X,K\} \Bigg) \geq  1-9 \epsdecoy^2.
			\end{aligned}
		\end{equation} 
		Let $\constraints$ denote the set of inequalities inside the probability in the above expressions. Therefore we have $\Pr(\constraints) \geq 1 - 9 \epsdecoy^2$.

		\subsection{Bounds on zero and one photon statistics} 
		For any event $O \in \{X,X_{\neq} , K\}$, the relevant bounds on the zero-photon and single-photon components can be obtained by algebraic manipulation of the expressions in $\constraints$. In general, any method for bounding the relevant zero-photon and single-photon components using $\constraints$ suffices. In this work, we follow exactly the steps taken by Ref.~\cite[Appendix A]{limConcise2014} to obtain these bounds.  Thus, we only write the final expressions here. We define
		\begin{equation}\label{eq:decoydefexpressions} 
			\begin{aligned}
				\nOmurv{k}^{\pm} &\coloneq \frac{e^{\mu_k}}{p_{\mu_k}}  \left( \nOmurv{k} \pm \sqrt{ \frac{\nOrv}{2} \ln( \frac{2}{\epsdecoy^2}) }  \right)
			\end{aligned}
		\end{equation}

		The lower bound on the zero-photon component is given by \cite[Eq. 2]{limConcise2014}
		\begin{equation} \label{eq:decoyboundone}
			\constraints \implies \nOphrv{0} \geq \Bounddecoymin{0} (\nOmurv{\allk}) \coloneq \tau_0 \frac{ \mu_2 \nOmurv{3}^- - \mu_3 \nOmurv{2}^+} {\mu_2 - \mu_3}.
		\end{equation}
		
		The lower bound on the one-photon component is given by \cite[Eq. 3]{limConcise2014}
		
		\begin{equation} \label{eq:decoyboundtwo}
			\begin{aligned}
				\constraints \implies \nOphrv{1}  \geq \Bounddecoymin{1} (\nOmurv{\allk}) &\coloneq \left( \frac{\mu_1 \tau_1} {\mu_1 (\mu_2 - \mu_3) - \mu^2_2 + \mu_3^2}  \right) \times \\
				& \left(      \nOmurv{2}^- - \nOmurv{3}^+ - \frac{\mu_2^2 - \mu_3^2}{\mu_1^2} \left(  \nOmurv{1}^+ -   \Bounddecoymin{0} (\nOmurv{\allk}) / \tau_0   \right)            \right)  .
			\end{aligned}
		\end{equation}

		The upper bound on the one-photon component is given by \cite[Eq.  4]{limConcise2014}

		\begin{equation} \label{eq:decoyboundthree}
			\begin{aligned}
				\constraints \implies \nOphrv{1}  \leq \Bounddecoymax{1} (\nOmurv{\allk}) &\coloneq \tau_1 \frac{\nOmurv{2}^+ -\nOmurv{3}^-}{\mu_2 - \mu_3}.
			\end{aligned}
		\end{equation}

		Since $\Pr(\constraints) \geq 1 - 9 \epsdecoy^2$, and \cref{eq:decoyboundone,eq:decoyboundtwo,eq:decoyboundthree} follow from the expressions in $\constraints$, we obtain
		
		\begin{equation}
			\Pr \Big(   \eXphrv{1} \geq \frac{	\Bounddecoymax{1} (\nXneqmurv{\allk})  }{  	\Bounddecoymin{1} (\nXmurv{\allk})}
			\quad \lor \quad   \nXphrv{1} \leq    \Bounddecoymin{1}( \nXmurv{\allk})   \quad \lor \quad 
			\nKphrv{1} \leq   \Bounddecoymin{1}(\nKmurv{\allk}  ) \Big) \leq  9 \epsdecoy^2
		\end{equation}

		\subsection{Variable-length security statement for decoy-state}		 \label{subsec:decoyvariable}
		
		Having proved \cref{eq:decoyreq}, we now have the following theorem regarding variable-length security of the decoy-state BB84 protocol.

		\begin{restatable}{theorem}{variablelengthproofdecoy}[ Variable-length security of decoy-state BB84] \label{thm:decoyvariablesecurity}
			Suppose \cref{eq:decoyreq} is satisfied and let $\leak(\decoyparams)$ be a function that determines the number of bits used for error-correction in the QKD protocol. Define 	\begin{equation} \label{eq:lvaluedecoy}
				\begin{aligned}
					l(\decoyparams) &\coloneq  \max\Bigg(0, \Bound_1 \left(\nKmu{\allk} \right) \left(1- h \left( \Bound_e \left(\eXmu{\allk},\nXmu{\allk}, \nKmu{\allk}   \right) \right) \right)  \\
					&- \leak(\decoyparams)					- 2\log(1/2\epsPA) - \ECost \Bigg) 
				\end{aligned}
			\end{equation}
			where $h(x)$ is the binary entropy function for $x\leq 1/2$, and $h(x) =  1$ otherwise.
			Then the variable-length decoy-state QKD protocol that produces a key of length $l(\decoyparams)$ using $\leak(\decoyparams)$ bits for error-correction, 	
			upon the event $\event{\decoyparams} \wedge \EVevent$ is $(2 \epsAT+\epsPA+\epsEV)$-secure.
		\end{restatable}

		\begin{remark}
			The decoy bounds used in this work requires the use of three total intensities to provide usable bounds. Later, this was improved to only require two total intensities in Ref.~\cite{ruscaFinitekey2018} (see also \cite{kaminImproved2024} for recent improvements in decoy analysis). 
			In this work we did \textit{not} follow the two intensities analysis of ~\cite{ruscaFinitekey2018}. This is due to complications stemming from the fact that 	
			this improved analysis requires the knowledge of error rates in the key generation rounds for various intensities (which is not announced). Although this issue can be resolved with additional reasoning, addressing it would divert from the primary focus of this work (which is imperfect detectors).  
			
			For instance, one way to avoid this problem is to argue that Bob can compare his raw key before and after error-correction to calculate the number of errors in the key generation rounds (assuming error-correction succeeded). This can indeed be made rigorous by arguing that if error-correction fails, the protocol aborts with high probability anyway (due to error-verification). However an additional issue remains.
			For variable-length protocols, Bob must \textit{announce} either  the number of errors he observes, or the length of key he wishes to produce, to Alice. This additional announcement leaks information to Eve which must be accounted for. Assuming that Bob announces the final output key length, a naive analysis would reduce the key length by an additional $\log( n_\text{len})$ where $n_\text{len}$ denotes the number of allowed output key length. These observations are missing in Ref.~\cite{ruscaFinitekey2018}.
			
			Note that this problem is avoided by this work since the length of output key is a function only of the public announcements during  Step 3 of the protocol (\cref{subsec:decoyprotocol}). 
			
		\end{remark}

		\section{Results} \label{sec:results} 
		We will now apply our results to a decoy-state BB84 protocol with realistic detectors. To do so, we start by outlining a recipe for using this work to compute key rates in \cref{subsec:recipe}. We will then specify the canonical model for our detectors with efficiency mismatch in \cref{subsec:detectors}. We will apply the recipe to our model in \cref{subsec:deltabounds}. Finally, we will plot the key rate we obtain in \cref{subsec:plots}.

		\subsection{Recipe for computing key rates in the presence of basis-efficiency mismatch} \label{subsec:recipe} 
		In this subsection, we provide straightforward instructions for using the results of this work to compute key rates for decoy-state BB84 in the presence of basis-efficiency mismatch (see the end of this subsection for a pointer to the exact expressions). We will start by explaining the computation of (upper bounds on) $\deltaone, \deltatwo$ for a given model of the measurement POVMs in the protocol. To do so, one has to break up the measurement process implemented by Alice and Bob into multiple steps via multiple uses of \cref{lemma:twostep}. This is done as follows:

		\begin{enumerate}
			\item Start with POVM $\{ \AliceBobPOVM{b_A,b_B}{\neq}, \AliceBobPOVM{b_A,b_B}{=}, \AliceBobPOVM{b_A,b_B}{\bot} \}$  which describe Alice and Bob measuring in the $(b_A,b_B)$ basis, and obtaining a conclusive error, a conclusive no-error, and an inconclusive outcome respectively. (In this work, we apply this recipe on the POVMs defined in \cref{eq:alicebobpovmmodel}.)
			
			\item Pick a $\tilde{F} \geq \AliceBobPOVM{b_A,b_B}{\neq}+\AliceBobPOVM{b_A,b_B}{=}$ for all $(b_A,b_B)$. Consider the four-outcome POVM \\
			$\{ \id - \tilde{F} , \tilde{F} - \AliceBobPOVM{b_A,b_B}{=} -\AliceBobPOVM{b_A,b_B}{\neq}, \AliceBobPOVM{b_A,b_B}{=},\AliceBobPOVM{b_A,b_B}{\neq} \}$. Group the last three outcomes together, and use \cref{lemma:twostep} to divide this measurement into two steps. In the first step, $\{ \tilde{F}, \id - \tilde{F}  \}$ is measured and latter outcomes discarded. The remaining rounds are measured using $\{ \AliceBobPOVMFilter{b_A,b_B}{\bot}, \AliceBobPOVMFilter{b_A,b_B}{=}, \AliceBobPOVMFilter{b_A,b_B}{\neq}\}$ where 
			\begin{equation} \label{eq:filteredFilters}
				\begin{aligned}
					\AliceBobPOVMFilter{b_A,b_B}{\bot}&=  \sqrt{\tilde{F}}^+ (\tilde{F} -  \AliceBobPOVM{b_A,b_B}{\neq} -  \AliceBobPOVM{b_A,b_B}{\neq})   \sqrt{\tilde{F}}^+  + \id - \proj_{\tilde{F}} \\
					\AliceBobPOVMFilter{b_A,b_B}{\neq} &= \sqrt{\tilde{F}}^+  \AliceBobPOVM{b_A,b_B}{\neq} \sqrt{\tilde{F}}^+ \\
					\AliceBobPOVMFilter{b_A,b_B}{=} &= \sqrt{\tilde{F}}^+  \AliceBobPOVM{b_A,b_B}{=}  \sqrt{\tilde{F}}^+
				\end{aligned}
			\end{equation}
			where $\proj_{\tilde{F}}$ denotes the projector onto the support of $\tilde{F}$.
			
			\item Consider the new POVM $\{ \AliceBobPOVMFilter{b_A,b_B}{\bot}, \AliceBobPOVMFilter{b_A,b_B}{=}, \AliceBobPOVMFilter{b_A,b_B}{\neq}\}$ .  Using \cref{lemma:twostep} again, divide this POVM measurement into two steps. The first step is implemented using $\{	\AliceBobPOVMprimeFilter{b_A,b_B}{\con}, 	\AliceBobPOVMprimeFilter{b_A,b_B}{\bot}\}$ and decides whether the outcome is conclusive or inconclusive. The conclusive outcomes are further measured using  $\{	\AliceBobPOVMsecond{b}{\neq} ,	\AliceBobPOVMsecond{b}{=} \}$. These POVM elements are given by
			\begin{equation} \label{eq:recipeone}
				\begin{aligned}
					\AliceBobPOVMprimeFilter{b_A,b_B}{\con} &= \AliceBobPOVMFilter{b_A,b_B}{\neq} + \AliceBobPOVMFilter{b_A,b_B}{=}  \\
					\AliceBobPOVMprimeFilter{b_A,b_B}{\bot} &= \AliceBobPOVMFilter{b_A,b_B}{\bot}  \\
					\AliceBobPOVMsecond{b_A,b_B}{\neq} &= \sqrt{\AliceBobPOVMprimeFilter{b_A,b_B}{\con}}^+ \AliceBobPOVMFilter{b_A,b_B}{\neq}  \sqrt{\AliceBobPOVMprimeFilter{b_A,b_B}{\con}}^+ \\
					\AliceBobPOVMsecond{b_A,b_B}{=} &= \sqrt{\AliceBobPOVMprimeFilter{b_A,b_B}{\con}}^+ \AliceBobPOVMFilter{b_A,b_B}{=}  \sqrt{\AliceBobPOVMprimeFilter{b_A,b_B}{\con}}^+ + \id - \proj_{\AliceBobPOVMprimeFilter{b_A,b_B}{\con}} \quad (= \id - 	\AliceBobPOVMsecond{b_A,b_B}{\neq})
				\end{aligned}
			\end{equation}
			where $\proj_{\AliceBobPOVMprimeFilter{b_A,b_B}{\con}} $ is the projector onto the support of $\AliceBobPOVMprimeFilter{b_A,b_B}{\con}$. This projector plays a trivial rule in the measurement itself, and is only included to ensure that we obtain a valid POVM.
			
			\item Compute
			\begin{equation} \label{eq:deltaCostsRecipe}
				\begin{aligned}
					\deltaone =& 2 \norm{\sqrt{\AliceBobPOVMprimeFilter{Z}{\con}}\AliceBobPOVMsecond{X}{\neq} \sqrt{\AliceBobPOVMprimeFilter{Z}{\con}} -  \sqrt{\AliceBobPOVMprimeFilter{X}{\con}}\AliceBobPOVMsecond{X}{\neq} \sqrt{\AliceBobPOVMprimeFilter{X}{\con}} }_\infty \\
					\deltatwo =&	\norm{\id - \AliceBobPOVMprimeFilter{Z}{\con} }_\infty 
				\end{aligned}
			\end{equation}
			where we recall that whenever the basis is explicitly written as $X/Z$, it represents both Alice and Bob’s basis choices.
			\item For the analysis of practical scenarios, where $\eta_{b_i}, d_{b_i}$ are not known exactly but are instead known to be in some range, one must also additionally maximize \cref{eq:deltaCostsRecipe} over all possible choices of $\eta_{b_i}, d_{b_i}$. 
		\end{enumerate}

		\textbf{ Once $\deltaone,\deltatwo$ are computed via the procedure above, we can compute key rates as follows. The key rate expression for the decoy-state BB84 protocol is given by \cref{eq:lvaluedecoy}. To use this expression,  refer to  \cref{eq:decoyreqcombined,eq:decoyreq} (which are notationally equivalent). The bounds for the decoy analysis in \cref{eq:decoyreqcombined} are in turn found in \cref{eq:decoyboundone,eq:decoyboundtwo,eq:decoyboundthree}, whereas the bound for the phase error estimation is found in \cref{eq:estimateimperfect}.  For the BB84 protocol where Alice sends single photons, the key rate is given by \cref{eq:lvalue,eq:estimateimperfect}. }

        \subsection{Assumptions} \label{subsec:assumptions}
        Note that the recipe is derived for the BB84 protocol (qubit or decoy-state) under the assumption that Alice's source is perfect. That is, Alice sends either perfect qubit BB84 states, or perfectly phase-randomized weak coherent pulses for the decoy-state protocol. The recipe is valid for all active-choice detector models, and yields non-trivial results as long as one can suitably bound $\deltaone,\deltatwo$. For the explicit calculations in this work, we consider the canonical model of detectors in the next section, and then compute the values of $\deltaone,\deltatwo$ for this model in \cref{subsec:deltabounds}. Therefore, the bounds in \cref{eq:finaldeltas} are derived assuming that Bob's detector POVMs are given by \cref{eq:bobpovmmodel} and characterized upto \cref{eq:etadcmodel}.
		
\subsection{Detector Model} \label{subsec:detectors}

In this section, we specify the canonical model of Bob's detectors (for active BB84) we use in this work. Let $\eta_{b_i}, d_{b_i}$ denote the efficiency and dark count rate of Bob's POVM corresponding to basis $b$, and bit $i$. We first define Bob's double click POVM for basis $b\in\{Z, X\}$ to be $\BobPOVM{b}{\dc} = \sum_{N_0, N_1 =0}^\infty (1-(1-d_{b_0})(1-\eta_{b_0})^{N_0}) (1-(1-d_{b_1})(1-\eta_{b_1})^{N_1}) \ketbra{N_0,N_1}_b$ ,	where $\ketbra{N_0, N_1}_b$ is the state with $N_0$ photons in the mode 1, and $N_1$ photons in mode 2, where the modes are defined with respect to basis $b$. For example, for polarization-encoded BB84, $\ketbra{2,1}_{Z}$ would signify the state with $2$ horizontally-polarised photons and $1$ vertically polarised photon. Recall that double clicks are mapped to single clicks randomly in our protocol. 
		Thus, we can write Bob's POVM elements as
		\begin{equation} \label{eq:bobpovmmodel}
			\begin{aligned}
				\BobPOVM{b}{\bot} &= \sum_{N_0, N_1=0}^\infty (1-d_{b_0})(1-d_{b_1})(1-\eta_{b_0})^{N_0} (1-\eta_{b_1})^{N_1} \ketbra{N_0,N_1}_b\\
				\BobPOVM{b}{0} &= (1-d_{b_1})\sum_{N_0, N_1 =0}^\infty (1-(1-d_{b_0})(1-\eta_{b_0})^{N_0}) (1-\eta_{b_1})^{N_1} \ketbra{N_0,N_1}_b + \frac{1}{2}\BobPOVM{b}{\dc}\\
				\BobPOVM{b}{1} &= (1-d_{b_0})\sum_{N_0, N_1=0}^\infty (1-\eta_{b_0})^{N_0} (1-(1-d_{b_1})(1-\eta_{b_1})^{N_1}) \ketbra{N_0,N_1}_b+\frac{1}{2}\BobPOVM{b}{\dc}.
			\end{aligned}
		\end{equation}
		Decoy methods allow us to restrict out attention to rounds where Alice sent single photons. Thus her Hilbert space is qubit while Bob holds two optical modes. The joint Alice-Bob POVM elements for the basis $b$ can be constructed via \cref{eq:alicePOVMs,eq:alicebobpovms,eq:bobpovmmodel} and are given by
		\begin{equation} \label{eq:alicebobpovmmodel}
			\begin{aligned}
				\AliceBobPOVM{b,b}{\bot} =& \id_A\otimes \sum_{N_0, N_1=0}^\infty (1-d_{b_0})(1-d_{b_1})(1-\eta_{b_0})^{N_0} (1-\eta_{b_1})^{N_1} \ketbra{N_0,N_1}_b\\
				\AliceBobPOVM{b,b}{\neq} =& \ketbra{0}_b \otimes (1-d_{b_0})\sum_{N_0, N_1=0}^\infty (1-\eta_{b_0})^{N_0} (1-(1-d_{b_1})(1-\eta_{b_1})^{N_1}) \ketbra{N_0,N_1}_b\\
				&+ \ketbra{1}_b \otimes (1-d_{b_1})\sum_{N_0, N_1 =0}^\infty (1-(1-d_{b_0})(1-\eta_{b_0})^{N_0}) (1-\eta_{b_1})^{N_1} \ketbra{N_0,N_1}_b\\
				&+ \id_A \otimes \frac{1}{2}\BobPOVM{b}{\dc}\\
				\AliceBobPOVM{b,b}{=} =& \ketbra{0}_b \otimes (1-d_{b_1})\sum_{N_0, N_1 =0}^\infty (1-(1-d_{b_0})(1-\eta_{b_0})^{N_0}) (1-\eta_{b_1})^{N_1} \ketbra{N_0,N_1}_b\\
				&+ \ketbra{1}_b \otimes (1-d_{b_0})\sum_{N_0, N_1=0}^\infty (1-\eta_{b_0})^{N_0} (1-(1-d_{b_1})(1-\eta_{b_1})^{N_1}) \ketbra{N_0,N_1}_b\\
				&+ \id_A \otimes \frac{1}{2}\BobPOVM{b}{\dc},
			\end{aligned}
		\end{equation}
	where $\ketbra{0}_b$ on Alice's system is the $\ket{0}$ state encoded in basis $b$. Note that this is different from the vacuum state $\ketbra{0,0}_b$ on Bob's system, the state with 0 photons in all modes.
		
		In any practical protocol, the detection efficiencies $\eta_{b_i}$ and dark count rates $d_{b_i}$ cannot be characterized exactly. Therefore, instead of assuming exact knowledge of these parameters, we assume that they are characterized upto some tolerances $\etachar, \dcchar$ given by
		\begin{equation}\label{eq:etadcmodel}
			\begin{aligned}
				\eta_{b_i} &\in [\etadet(1-\etachar ), \etadet(1+\etachar )], \\
				d_{b_i} &\in [\dcprob(1-\dcchar ), \dcprob(1+\dcchar )].
			\end{aligned}
		\end{equation}

		\subsection{Computing bounds on $\deltaone,\deltatwo$} \label{subsec:deltabounds}
		In this subsection, we will compute upper bounds on $\deltaone, \deltatwo$ by following the recipe in \cref{subsec:recipe}. 
			
		\subsubsection{Active BB84 detection setup without any hardware modification}
	
	In this case the POVMs used by Alice and Bob are exactly given by \cref{eq:alicebobpovmmodel}. We construct the POVMs from \cref{eq:recipeone,eq:filteredFilters} in \cref{app:povmcalcs}.  To bound $\deltaone,\deltatwo$,  we use the fact that all POVMs are block-diagonal in the total photon number, and bound the $\infty$-norm of each block separately. Note that we can always treat the common value of loss in the detectors to be a part of the channel \cite[Section III C]{zhangEntanglement2017}.  This means that we pull out $(\max_{b,i} \{ \eta_{b_i}\})$, and treat it as a part of the channel. (This is equivalent to giving the $\{\tilde{F},\id-\tilde{F}\}$ measurement to Eve.)	 This computation of $\deltaone,\deltatwo$ using the above steps is quite cumbersome, and is explained in \cref{subsec:deltaonecalcs,subsec:deltatwocalcs}. Finally, we obtain
		\begin{equation} \label{eq:finaldeltas}
			\begin{aligned}
				\deltaone  &\leq  \max \left\{  \left(1-\frac{1-(1-\dMin)^2}{1-(1-\dMax)^2}\right)\frac{\dMax(2-\dMin)}{1-(1-\dMin)^2},  4  \abs{ 1- \sqrt{ 1-(1-\dMin)^2(1-\etaRenorm)}} \right\}, \\
				\deltatwo &\leq \max \left\{1-\frac{1-(1-\dMin)^2}{1-(1-\dMax)^2},(1-\dMin)^2 (1-\etaRenorm) \right\}, \\
			\end{aligned}
		\end{equation}
		where
		\begin{equation} \label{eq:boundsetadc}
			\begin{aligned}
				\etaRenorm &= \etaMin / \etaMax \\
				\dMax &= \max\{d_{X_0},d_{X_1},d_{Z_0},d_{Z_1}\} \leq \dcprob (1+\dcchar), \quad \text{ and } \dMin = \min \{d_{X_0},d_{X_1},d_{Z_0},d_{Z_1}\} \geq \dcprob(1-\dcchar), \\ 
				\etaMax &= \max\{\eta_{X_0},\eta_{X_1},\eta_{Z_0},\eta_{Z_1}\} \leq  \etadet(1+\etachar), \quad \text{ and } \etaMin = \min \{\eta_{X_0},\eta_{X_1},\eta_{Z_0},\eta_{Z_1}\} \geq \etadet(1-\etachar). 
			\end{aligned}
		\end{equation}
		Thus, upper bounds on $\deltaone,\deltatwo$ can be computed using \cref{eq:finaldeltas} and the bounds in \cref{eq:boundsetadc}. It is these bounds that we use to compute key rates.

		\subsubsection{Random Swapping of 0 and 1 Detectors}
		
		In \cite{fungSecurity2009a} it was argued that random swapping of the 0 and the 1 detector can be used to remove basis-efficiency mismatch for single-photon pulses entering Bob's detectors. Note that this trick \textit{only works for the single-photon subspace}. We will now adapt our analysis to the case where Bob randomly swaps the $0$ and the $1$ detector. 
		
		In the scenario where we randomly swap the $0$ and the $1$ detectors, we make certain physically motivated assumptions (\cref{eq:etaXIsEtaZ}) about the detector setup. In particular, we assume that the dark count rate is a property of the detector only. Furthermore, we assume that the basis choice setting does not change the detector parameters. This means that  the dark count rate and detection efficiency in both bases is the same (though these can be different for each detector).  Thus, we have
		\begin{equation} \label{eq:etaXIsEtaZ}
			\begin{aligned}
				\eta_{X_0} &= \eta_{Z_0}\eqqcolon \eta_0\\
				\eta_{X_1} &= \eta_{Z_1}\eqqcolon \eta_1\\
				d_{X_0} &= d_{Z_0}\eqqcolon d_0\\
				d_{X_1} &= d_{Z_1}\eqqcolon d_1.
			\end{aligned}
		\end{equation}

		We will see that this indeed allows us to obtain improved results, even though it does not completely remove efficiency mismatch. In particular, the leading order terms in $\deltaone, \deltatwo$ are improved in the new bounds obtained  in \cref{eq:deltacalcsnew,eq:deltacalcsnewbound}.
		 Note that our metrics $\deltaone, \deltatwo$ do not improve unless we make these assumptions. These assumptions are also implicit in the claims presented in Ref.~\cite{fungSecurity2009a}.
		
		If the random swapping is implemented with probability $p$, the Bob's POVM elements are given by
		\begin{equation} \label{eq:bobswappovmmodel}
			\begin{aligned}
				\BobPOVMswap{b}{\bot} &= \sum_{N_0, N_1=0}^\infty (1-d_{b_0})(1-d_{b_1})\left((1-p)(1-\eta_{b_0})^{N_0}(1-\eta_{b_1})^{N_1}+p(1-\eta_{b_1})^{N_0}(1-\eta_{b_0})^{N_1}\right) \ketbra{N_0,N_1}_b\\
				\BobPOVMswap{b}{0} &= \Big((1-p)(1-d_{b_1})\sum_{N_0, N_1 =0}^\infty (1-(1-d_{b_0})(1-\eta_{b_0})^{N_0}) (1-\eta_{b_1})^{N_1}\\
				&+p (1-d_{b_0})\sum_{N_0, N_1 =0}^\infty (1-(1-d_{b_1})(1-\eta_{b_1})^{N_0}) (1-\eta_{b_0})^{N_1}\Big)\ketbra{N_0,N_1}_b + \frac{1}{2}\BobPOVM{b}{\dc}\\
				\BobPOVMswap{b}{1} &= \Big((1-p)(1-d_{b_0})\sum_{N_0, N_1=0}^\infty (1-\eta_{b_0})^{N_0} (1-(1-d_{b_1})(1-\eta_{b_1})^{N_1})\\
				&+p(1-d_{b_1})\sum_{N_0, N_1=0}^\infty (1-\eta_{b_1})^{N_0} (1-(1-d_{b_0})(1-\eta_{b_0})^{N_1})\Big)\ketbra{N_0,N_1}_b+\frac{1}{2}\BobPOVM{b}{\dc},
			\end{aligned}
		\end{equation}
		analogously to \cref{eq:bobpovmmodel}.
		Alice and Bob's joint POVM elements can be constructed from \cref{eq:alicePOVMs,eq:alicebobpovms,eq:bobswappovmmodel} analogously to \cref{eq:alicebobpovmmodel}.
		Therefore we can repeat the calculations for $\deltaone,\deltatwo$ using the recipe from  \cref{subsec:recipe}. We explain these computations in \cref{app:deltarandomswaps} and obtain (for swap probability $p=1/2$)

		\begin{equation} \label{eq:deltacalcsnew}
			\begin{aligned}
			\deltaone &\leq 4 \left(1-\sqrt{1-(1-\dMultAvg)^2\frac{(1-\etaRenorm)^2}{2}}\right), \\
				\deltatwo &\leq (1-\dMultAvg)^2\frac{(1-\etaRenorm)^2}{2},
				\end{aligned}
		\end{equation}
		where
		\begin{equation} \label{eq:deltacalcsnewbound}
			\begin{aligned}
			 \dMultAvg &= 1-\sqrt{(1-d_0)(1-d_1)} \geq \dMin ,\\
			 \etaRenorm &= \frac{\etaMin}{ \etaMax} \geq \frac{1-\etachar}{1+\etachar}.
			\end{aligned}
		\end{equation}
			Thus, upper bounds on $\deltaone,\deltatwo$ in case of random swapping of detectors can be computed using \cref{eq:deltacalcsnew} and the bounds in \cref{eq:deltacalcsnewbound}. 
		We see that these bounds are better than the earlier bounds from \cref{eq:finaldeltas}. On inspecting our calculations from \cref{app:deltarandomswaps}, we find that the zero-photon component of $\deltaone,\deltatwo$ goes to zero due to  $d_X = d_Z$. Furthermore, random swapping in addition to the assumption of $\eta_X = \eta_Z$ leads to the single-photon contribution also being zero. Thus, we are left with the two-photon contribution.
	
		\subsection{Plots}   \label{subsec:plots}
				\begin{figure} 
			\includegraphics[width=\linewidth]{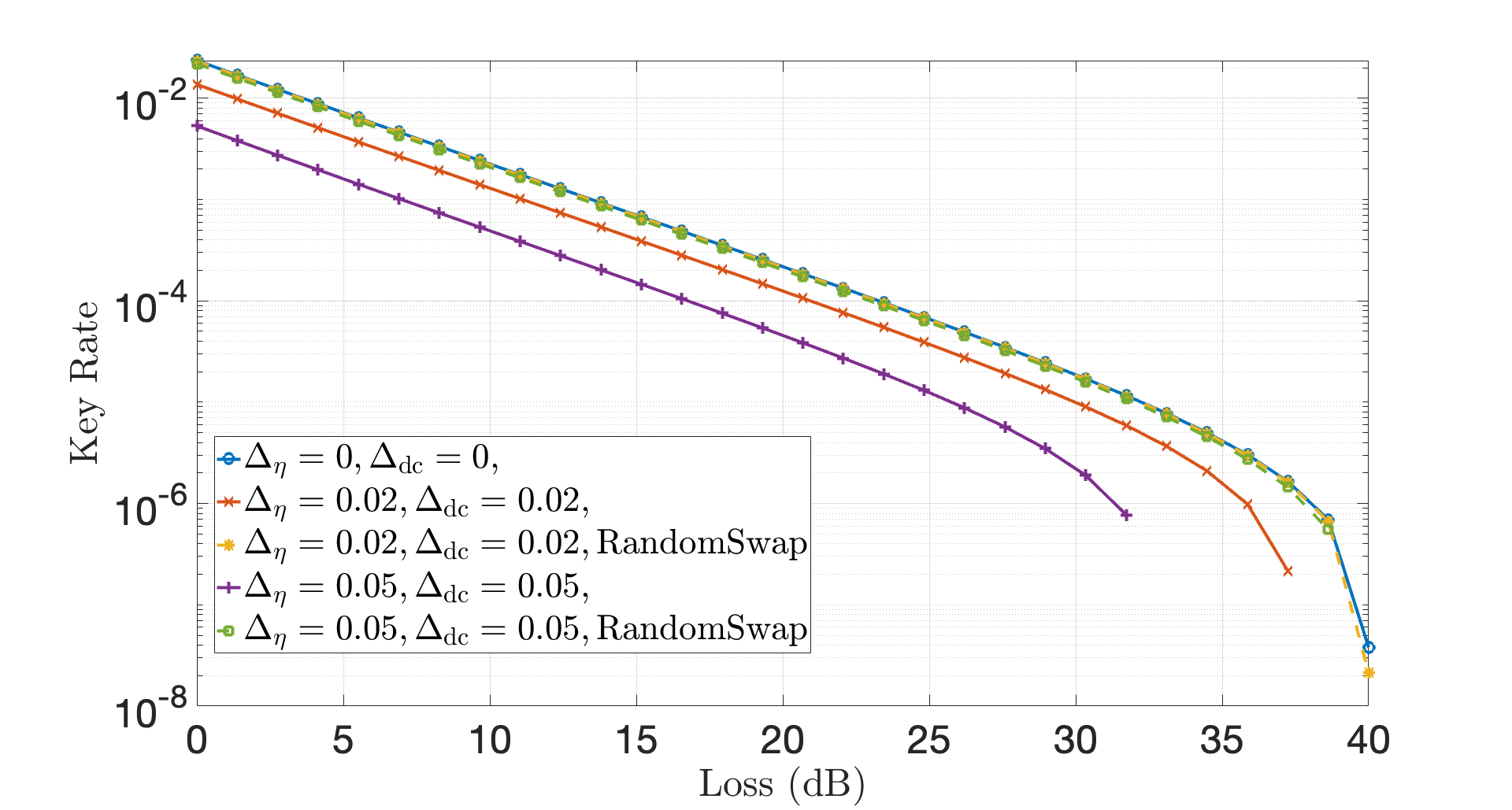}
			\caption{Finite-size key rates in the presence of basis-efficiency mismatch, for the decoy-state BB84 protocol, against loss. We plot key rates for $\ntot = 10^{12}$ number of total signals sent, for various values of $\etachar,\dcchar$. We find that random swapping of the $0$ and $1$ detectors drastically improves the key rates obtained.}  \label{fig:figOne}
		\end{figure}

		We plot finite size key rates for the decoy-state BB84 protocol described in \cref{subsec:decoyprotocol}. We choose typical protocol parameters and plot the key rate for the expected observations for a given channel model. For best results, one would optimize over the protocol parameter choices. For all plots, we set the basis choice probabilities to be $\pzA = \pzB = 0.5$ and $\pxA = \pxB = 0.5$, and $\pt = 0.05$ (probability of $Z$ basis rounds used for testing). 
		We set the detector parameters to be $\etadet = 0.7$ and $\dcprob = 10^{-6}$. 
		We set the misalignment angle $\theta$ to be $2^\circ$. We set the number of bits used for error-correction to be $ \leak(\decoyparams) = f_{\text{EC}} \nK h(\eZ)$, where $f_\text{EC} = 1.16$ is the error-correction efficiency.  The decoy intensities are chosen to be $\mu_1 = 0.9$, $\mu_2 = 0.1$, and $\mu_3=0$. Each intensity is chosen with equal probability.
		We set $\epsATa = \epsATb= \epsATc = \epsdecoy = \epsEV = \epsPA = 10^{-12}$. This leads to a value of $\epsAT = \sqrt{12} \times 10^{-12}$. The overall security parameter is then given by $(2 \sqrt{12} + 2) 10^{-12}$.  Due to machine precision issues arising from small values of $\epsAT^2$, we use Hoeffdings inequality to bound $\gamma_{\text{bin}}$ (\cref{eq:gammabindefined}) instead of using the cumulative binomial distribution (which is tighter). 
		
\begin{figure} 
			\includegraphics[width=\linewidth]{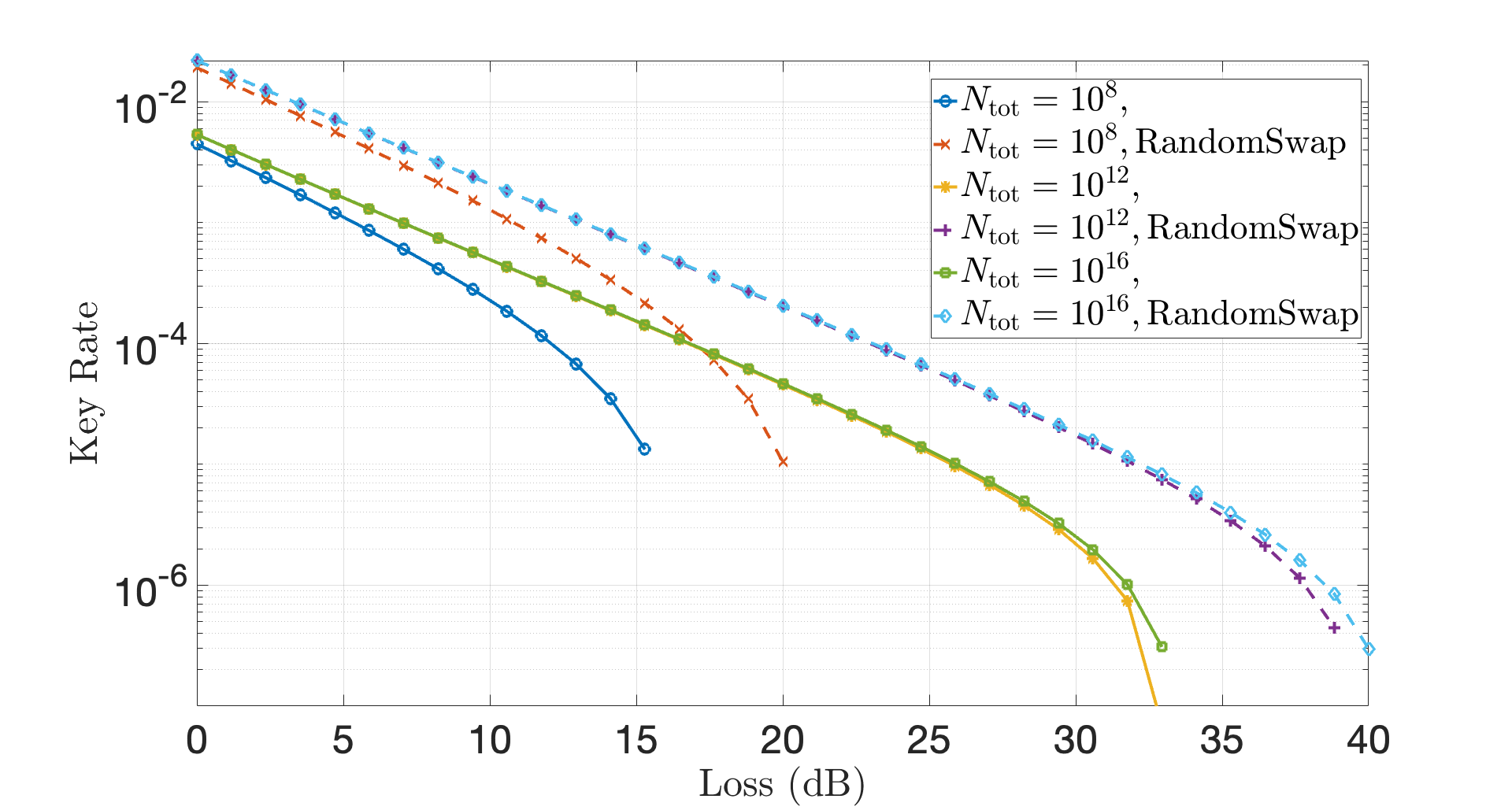}
			\caption{Finite-size key rates in the presence of basis-efficiency mismatch, for the decoy-state BB84 protocol against loss. We plot key rates for various values of  total number of signals sent ($\ntot$), for $\etachar=\dcchar=0.05$.}  \label{fig:figTwo}
		\end{figure}

\begin{figure} 
	\includegraphics[width=\linewidth]{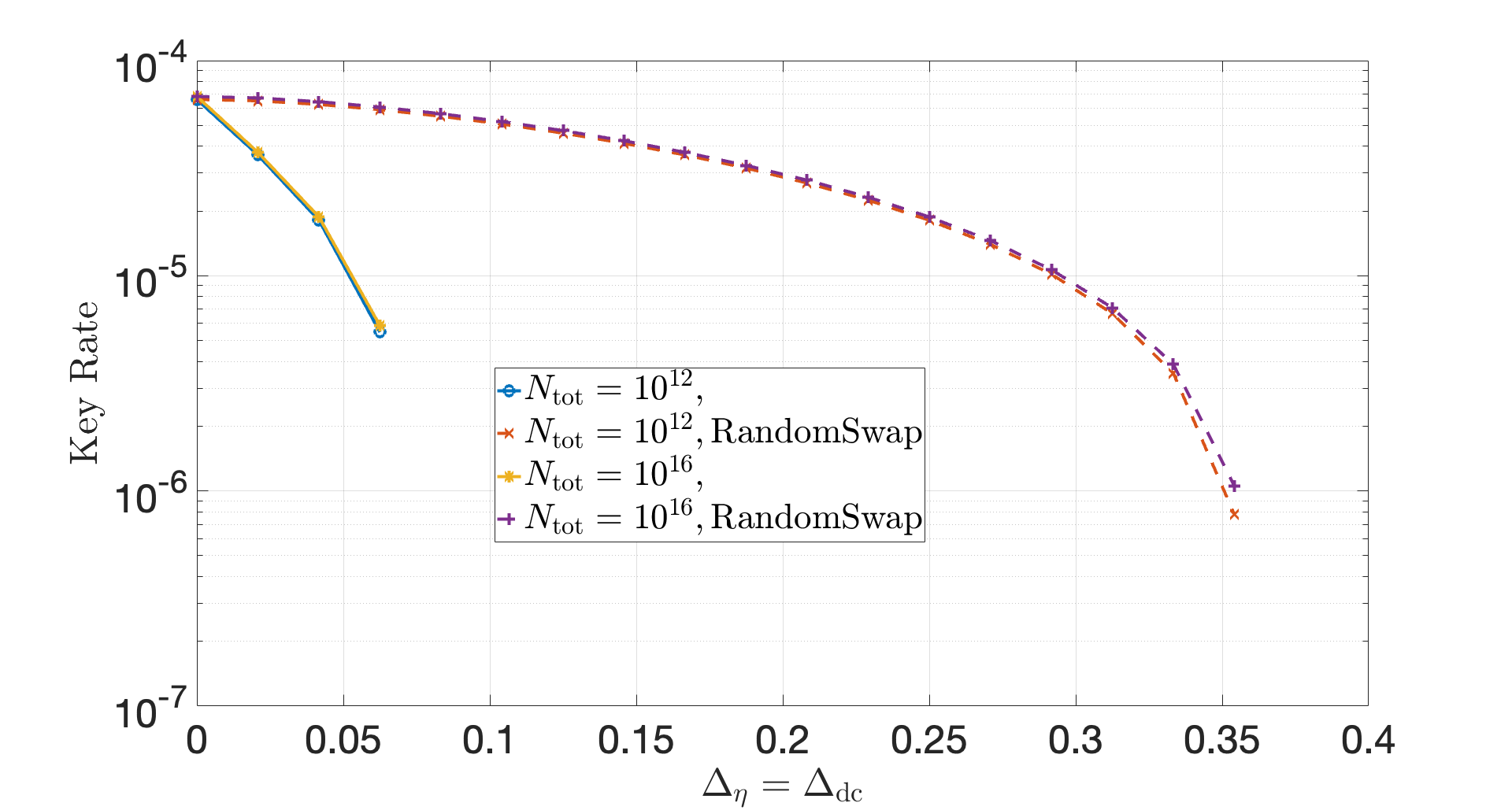}
	\caption{Finite-size key rates in the presence of basis-efficiency mismatch, for the decoy-state BB84 protocol against detector characterization parameters $\etachar,\dcchar$. We plot key rates for a channel with $25$dB loss.}  \label{fig:figThree}
\end{figure}

\begin{figure} 
    \includegraphics[width=\linewidth]{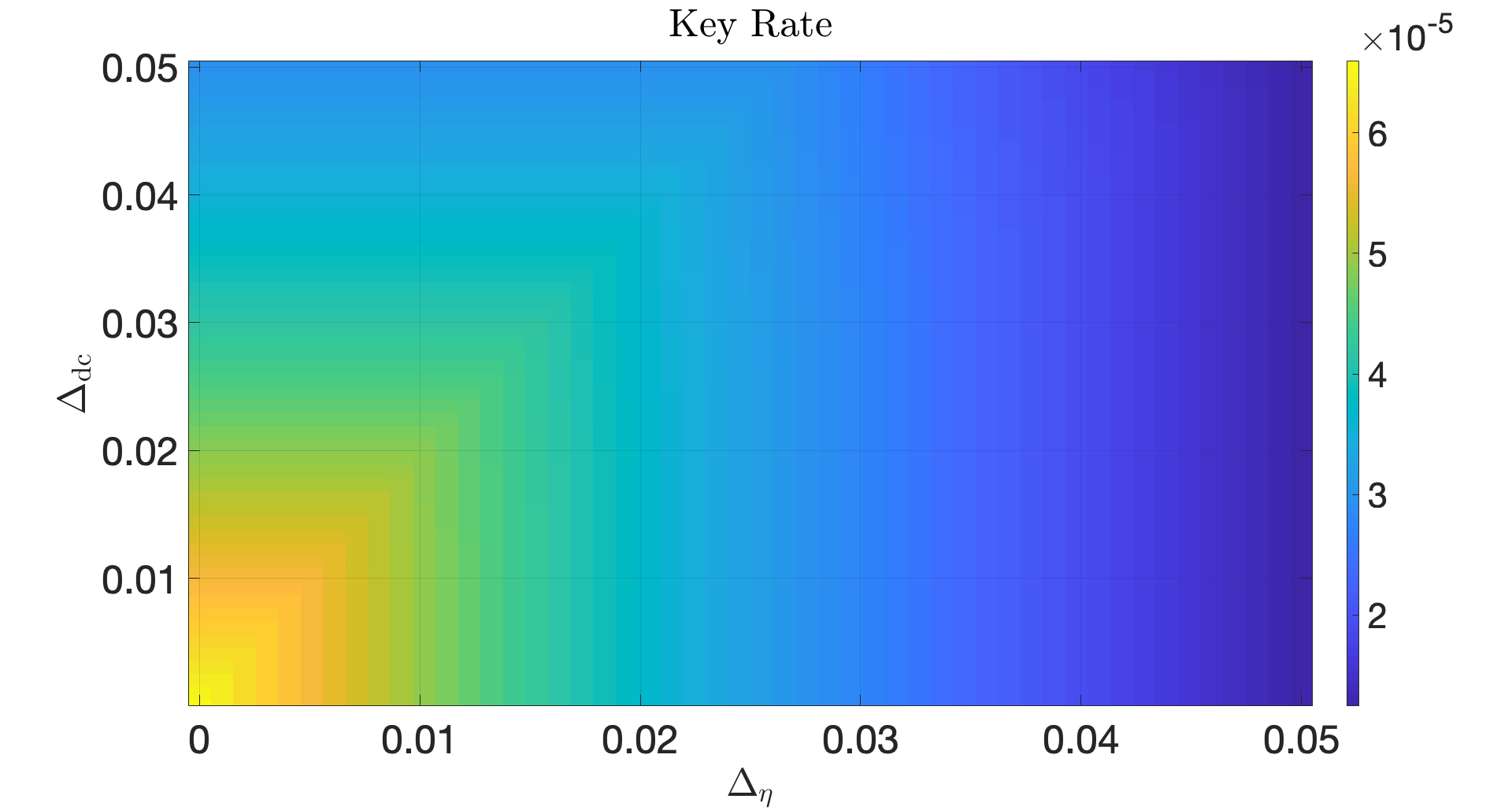}
    \caption{ Finite-size key rates for various values of $\etachar,\dcchar$ for the decoy-state BB84 protocol, for $\ntot=10^{12}$ and $25$dB loss. We find that both $\etachar,\dcchar$ have comparable impact on the keyrate, although $\dcchar$ penalizes the key rate less than $\etachar$. The above plot is interpolated from key rate calculations of $2500$ points, and the detectors are not swapped randomly.}  \label{fig:figFour}
\end{figure}

		\begin{enumerate}
			\item 	In \cref{fig:figOne}, we plot the finite size key rate against loss for various values of detector characterizations $\etachar, \dcchar$ for $n_\text{total} = 10^{12}$ number of total signals. 
			For $\etachar = \dcchar = 0$, we have $\deltaone = \deltatwo = 0$. Therefore the phase error rate bound from \cref{eq:estimateimperfect}	reduces to the scenario where the basis-independent loss assumption is satisfied (\cref{eq:estimateperfect}). For non-zero values of $\etachar, \dcchar$, the key rate is reduced. This is mostly due to the increase in the bound for the phase error rate from \cref{eq:estimateimperfect} from $\deltaone$. We find that random swapping leads to a dramatic improvement in performance.
			\item 
			In \cref{fig:figTwo}, we plot the finite size key rate against loss for various values of total signals sent. We set  $\etachar = \dcchar = 0.05$. 
			We find that we get close to asymptotic key rates already at $\ntot =10^{12}$ signals sent. 
			\item 
			In \cref{fig:figThree}, we plot the finite size key rate against detector characterization parameters $\etachar,\dcchar$. We find that our methods can tolerate a significant amount of error in detector characterization. In fact, with random swapping of detectors, we get positive key rate for $\ntot=10^{12}$ signals sent for $\etachar,\dcchar$  upto $0.35$.
			\item 	In \cref{fig:figFour}, we plot the finite size key rate against both detector characterization parameters $\etachar,\dcchar$ independently. We find that both the parameters $\etachar,\dcchar$ lead to comparable penalties in the key rate, although $\dcchar$ penalizes the key rate less than $\etachar$. Note that the values for $\deltaone,\deltatwo$ also depend on the dark count rate $\dcprob$.
				
		\end{enumerate}
	
	
We end this this section by considering a scenario where the detector behavior is independent (but not identical) in each round.
In this case each round has a well-defined POVM that is independent of those in other rounds, i.e the POVM is tensor product with other rounds. Here we simply note that all the statistical claims used in our phase error estimation proofs (from \cref{subsec:pulkitlemmas}) remain true even if the POVM measurements are independent (but not IID). We comment on this and restate some of our lemmas for independent (but not identical) measurements in \cref{app:pulkitindependent}.
Moreover the Serfling statement (\cref{lemma:sampling}) does not assume any IID property of the input string. Thus, our bounds on the phase error rate remain unchanged as long as  $\deltaone,\deltatwo$ can be bounded for the independent POVMs in each round.
Such a scenario is of practical importance, as detection setups are never perfectly IID \cite[Fig. 3(a)]{dixon2015high}.
Another important practical consideration is that of correlated detectors, which we now discuss in the next section.

\section{Application to Correlated effects} \label{sec:correlated}

We now turn our attention to detectors exhibiting correlated behavior across the rounds. We begin by formalizing our model for such detectors. Let the outcome of round~$i$ for Bob be denoted by $k_i$, and define $k_i^j$ as the sequence of outcomes from round $i$ to round $j$ ($i\leq j$). Recall that $\bot$ corresponds to the no-detect outcome. Correlated effects such as afterpulsing and detector dead times can be modeled by allowing the POVM used in later rounds to depend on the outcome --- in particular, detection events --- of previous rounds. This is the scenario we are interested in.
More formally, in the $i$th round, the POVM used for the measurement is given by $\{ \Gamma_{k_{i}}^{(k^{i-1}_{i-l_c}) } \}$, where $l_c$ denotes the correlation length. Note that $l_c$ here denotes the correlation length in units of the time slots (i.e, the unit is the time between successive measurements in the QKD protocol). We use the convention that when all $k_{i-l_c}^{i-1} = \bot s$, then the superscript can be omitted.

Our approach involves a protocol modification in which Alice and Bob retain only those rounds where previous $l_c$ rounds involved only no-click events. All other rounds are `rejected', i.e thrown away. In our terminology, a round can be rejected in this manner (due to the specific postprocessing described above), but may also be discarded later based on its detection outcome -— for instance, if it results in a no-detection event. The idea is that after this postprocessing, the remaining rounds can be thought of as all being measured using uncorrelated POVM $\{ \Gamma_{k_{i}}  \}$. 

The key idea is to introduce a generalized filtering step (analogous to $\tilde{F}$ in the round-by-round case) that performs the minimum amount of measurements necessary, and uses the outcomes to reject rounds based on the above postprocessing. Once the rounds to be rejected are fixed, the remaining rounds can be measured fully using uncorrelated POVMs, and the analysis on \textit{these} rounds can follow the same approach as the usual EUR analysis.

However, note that the postprocessing described above depends on the detection events, which in turn depend on the choice of basis, and thus can only be determined \textit{after} basis choice. Thus, in general, a basis choice needs to be made \textit{before} rounds can be rejected. This is the core difficulty in proving security under detector correlations within the EUR framework, as the EUR statement must be applied to the state \textit{before} choosing the basis.

We can now present a two-step proof  sketch to address the above issue (we elaborate on each step in later subsections):
\begin{enumerate}
    \item \label{step:noDCFiltering} Assuming the detectors have no loss or dark counts, the rejection step can be performed in a basis-independent manner (on the rounds which will not be rejected). Thus, the EUR statement can be applied to those rounds by completing the measurements later. We explicitly construct this procedure in \cref{subsubsec:perfectcorr}, and show that is rejects the correct rounds (see \cref{lem:corrFilter}).  Note that here we consider a setting in which the detectors themselves have no intrinsic loss or dark counts, i.e., they are perfect in the absence of correlated effects arising from clicks in previous rounds. However, the detectors suffer from afterpulsing and dead times due to correlated effects. 
    \item We lift the assumption (of no loss and no dark counts) made in the first step above using techniques from Ref.~\cite{nahar2025imperfectdetectorsadversarialtasks}. Intuitively, Ref.~\cite{nahar2025imperfectdetectorsadversarialtasks} allows one to incorporate the effects of basis-dependent dark counts and efficiencies at some cost (to be discussed in \cref{subsubsec:reductiontoperfect}). Thus, this step can be done first, reducing the problem to the analysis described in the earlier step.
\end{enumerate}

\subsection{Perfect correlated detectors} \label{subsubsec:perfectcorr}

The intuition behind the first step relies on the following two observations. First, detections in the rounds that will not be rejected can be inferred from the photon number of the incoming state, independent of basis choice. This is because the previous rounds have had no clicks, so the POVM used in the current is the uncorrelated, perfect one. Moreover, due to the block-diagonal nature of the POVMs, this photon number measurement does not affect the measurement statistics. Second, the EUR statement is not used on the rounds that are rejected. Thus, we are allowed to complete the measurement on these rounds.   This can be formalized in the following lemma.
\begin{lemma} \label{lem:corrFilter}
    Consider the state $\rho_{Q^n}$, and let the $i$th subsystem be measured using POVM $\{ \Gamma_{k_{i}}^{(k^{i-1}_{i-l_c}) } \}$, where $k_i$ denotes the outcome of the $i$th measurement, and $k^{i-1}_{i-l_c}$ denotes the string of outcomes in the previous $l_c$ rounds. Suppose that $\{ \Gamma_{k_{i}}\}$ (corresponding to $k^{i-1}_{i-l_c} = \bot s)$ is such that it can be described by a two-step measurement, where the first measurement $\{\tilde{F},\id-\tilde{F}\}$ determines the detect vs no-detect, followed by a second step measurement (see \cref{fig:correlated}).
    Consider the state obtained after rejecting all rounds $i$ for which a detection occurred in the previous $l_c$ rounds. Then this state can be obtained via  a procedure that only performs the $\{\tilde{F}, \id -\tilde{F} \}$ measurements on the rounds that are not rejected.
\end{lemma}
\begin{proof}
    For simplicity, we assume a correlation length $l_c$ of $1$. The extension to larger, finite correlation lengths is straightforward. We will prove the required claim by explicit construction of the procedure. We will prove it sequentially, from round $1$ to round $n$.
    
     First note that round $1$ will never be rejected.
  The $\{\tilde{F}, \id -\tilde{F} \}$ measurement in round $1$ directly tells us whether or not the round will result in a detection event. Thus, this tells us whether or not the next round ($i=2$) will be rejected. (Note that this measurement was performed without choosing a measurement basis for round $1$.) 

    For round $i \geq 3$, assume rounds $1$ to $i{-}1$ already have definite reject/no-reject tags, with only the $\{ \tilde{F}, \id - \tilde{F} \}$ measurement being performed on the rounds that are not rejected. Then, the round $i$ can be given the correct reject/no-reject tag as follows. 
    \begin{enumerate}
        \item To determine whether round~$i$ should be kept, check the reject/no-reject tag of round~$i-1$ (which is guaranteed to be correct).
        \item If round $i-1$ is to be rejected, complete the measurement on it. This determines whether or not there is a detection event in this round, which inturn determines whether or not round $i$ will be rejected.
        \item  If round~$i-1$ is not rejected, then round~$i-2$ must have had no detection (since otherwise the round $i-1$ would be rejected). This means that the POVM to be used in round $i-1$ is the uncorrelated one ($\{ \Gamma_{k_{i-1}}\}$). Thus, measuring $\{ \tilde{F}, \id-\tilde{F} \}$ directly tells us whether or not round $i-1$ will result in a detection event. This, in turn, determines whether or not round $i$ will be rejected.
    \end{enumerate}
    By construction, the above procedure only implements the $\{ \tilde{F}, \id-\tilde{F} \}$ on the rounds that are not rejected, and results in the correct rounds being rejected. This concludes our proof.

    \begin{figure} 
    \hspace*{-0.5cm}
        \centering
        \scalebox{0.6}{\begin{tikzpicture}[
  transform shape,
  keepbox/.style={draw=green!50!black, fill=green!10, thick, minimum width=2.8cm, minimum height=1.6cm, align=center},
  discardbox/.style={draw=red!60!black, fill=red!10, thick, minimum width=2.8cm, minimum height=1.6cm, align=center},
  continuation/.style={draw=gray!50!black, dashed, fill=gray!10, very thick, minimum width=3.2cm, minimum height=5.5cm, align=center},
  detectstyle/.style={draw=blue!60!black, fill=blue!10, rounded corners=2pt, minimum width=1.8cm, minimum height=0.8cm},
  nodetectstyle/.style={draw=red!60!black, fill=red!10, rounded corners=2pt, minimum width=1.8cm, minimum height=0.8cm},
  process/.style={->, thick, double, double distance=1pt, >=Stealth},
  decision/.style={->, thick, >=Latex},
  every node/.style={font=\small, text centered},
  measurementlabel/.style={align=center, text width=2.5cm}
]

\node[keepbox] (r1) at (0,0.5) {Keep};
\node[discardbox, right=3.5cm of r1] (r2) {Discard};
\node[keepbox, right=3.5cm of r2] (r3) {Keep};

\node[above=0.3cm of r1] {\textbf{Round $i{-}2$}};
\node[above=0.3cm of r2] {\textbf{Round $i{-}1$}};
\node[above=0.3cm of r3] {\textbf{Round $i$}};

\node[detectstyle, below=2.1cm of r1] (d1) {Detect};
\node[nodetectstyle, below=2.1cm of r2] (d2) {No-detect};
\node[nodetectstyle, below=2.1cm of r3] (d3) {No-detect};

\node[continuation] (prev) at ($(r1) + (-5.5, -0.8)$) {Earlier\\rounds};
\node[continuation] (next) at ($(r3) + (5.5, -0.8)$) {Later\\rounds};

\draw[decision] ([xshift=1.6cm, yshift=-2.5cm] prev.center) -- ++(0.8,0) |- (r1.west);
\draw[decision] (d3.east) -- ++(1.4,0) |- ([yshift=0.8cm]next.west);

\draw[process] (r1.south) -- node[measurementlabel, left=3mm] {Photon-number\\measurement} (d1);
\draw[process] (r2.south) -- node[measurementlabel, left=3mm] {Complete\\measurement} (d2);
\draw[process] (r3.south) -- node[measurementlabel, left=3mm] {Photon-number\\measurement} (d3);

\draw[decision] (d1.east) -- ++(1.4,0) |- (r2.west);
\draw[decision] (d2.east) -- ++(1.4,0) |- (r3.west);

\end{tikzpicture}}
    \caption{Illustration of the process to determine the rounds that will be rejected based on prior detection events. The rounds that are not rejected are only measured using $\{\tilde{F},\id - \tilde{F}\}$, which determines the detect vs no-detect outcome. They will have the rest of the measurement, including basis choice, completed at a later stage. The rounds that are rejected have their entire measurements completed in order to determine the detection event.}
\label{fig:correlated}
\end{figure}
\end{proof}

\subsection{Reduction to perfect correlated detectors} \label{subsubsec:reductiontoperfect}

As mentioned earlier, the applicability of \cref{lem:corrFilter} crucially relies on the fact that the detectors are perfect threshold detectors -- except for correlations that one detect event might trigger other detect events within the correlation length. However, as argued in the rest of this paper, this is not a practical assumption. To resolve this issue, we use the results of Ref.~\cite{nahar2025imperfectdetectorsadversarialtasks}. 

Ref.~\cite{nahar2025imperfectdetectorsadversarialtasks} constructs a basis-choice independent reduction from a detection setup with detectors with different dark counts and loss, to a detection setup with perfect detectors. It shows that the imperfect detection setup can be treated as a basis-choice independent channel (similar to the basis-independent filter $\tilde{F}$) followed by the perfect detection setup. To do this, it crucially relies on the flag-state squasher \cite{zhangSecurity2021}. The intuition is that the flag-state squasher makes it possible to `give' Eve full information about multi-photon signals by introducing a classical flag space. Crucially, this flag space can also then be used to transfer imperfections such as loss and dark counts to Eve. This is argued formally in \cite[Theorem 1 and 2]{nahar2025imperfectdetectorsadversarialtasks}.

The main open task in our approach outlined in this section is the use of the flag-state squasher in an EUR-based security proof. In the typical usage of the flag-state squasher, one considers a single round state, and the weight of the state in the flag space is bounded through suitable methods to prevent trivial results (since the entire flagged state is revealed to Eve). In the EUR context, there no single round state. Instead, the natural alternative is to instead bound the \textit{number} of rounds in the flag space. We believe that the usual methods of bounding the weight of the state along with suitable concentration inequalities will suffice for this task. However, we emphasize that in this work, we provide only a sketch of the proof for handling correlated detectors, leaving a detailed analysis to future work --- a non-trivial task. In the next section, we shift our focus to detector side-channels, and in particular, we will see how our methods naturally address certain detector side-channels as a by-product.

\section{Detector Side-Channels} \label{sec:detectorside} 		

The analysis presented so far assumes that the detector behavior, while possibly varying between rounds or exhibiting correlations, is described by a single mode characterized by bounded loss and dark count rates $\eta_{b_i}$ and $d_{b_i}$, as specified in \cref{eq:etadcmodel}.
Thus, we have presented an analysis of Case~\ref{casetwo} from \cref{sec:introduction}. This analysis allows us to drastically reduce the requirements on device characterization: the proof technique is now robust to imperfect characterization.
However, physical implementations of QKD protocols are vulnerable to side-channel attacks where Eve can control, to a limited extent, the POVMs used. For example, by controlling the frequency, spatial mode \cite{rau2014spatial,sajeedSecurity2015} or arrival time \cite{qiTimeshift2007} of the light, Eve can partially choose the detector efficiencies and induce a suitable basis-efficiency mismatch. This is the scenario described by Case~\ref{casethree} from \cref{sec:introduction}.

While our proof technique advances the theory to the point where this case can be handled in principle, a complete analysis first requires the physical modeling of multi-mode detectors, which remains an open problem. In this section, we outline how the results of this work can be applied to a simple multi-mode model.
 
We expand our detector model (and Bob's Hilbert space) to account for spatio-temporal modes \cite{zhangSecurity2021} as 
\begin{equation} \label{eq:povmmodes}
    \Gamma^{\text{multi}}_{(b_A,b_B),(k)} = \bigoplus_{\mode} \Gamma_{(b_A,b_B),(k)} (\{{\eta_{b_i}(\mode),  d_{b_i}(\mode)}\}),
\end{equation}
where $\mode$ denotes the spatio-temporal mode, and $ \Gamma_{(b_A,b_B),(k)} (\{{\eta_{b_i}(\mode),  d_{b_i}(\mode)}\})$ denotes the single-mode POVM element corresponding to that mode, and is given by \cref{eq:alicebobpovmmodel}. The multi-mode detector has loss $\eta_{b_i}(\mode)$ for this mode, and a dark count rate of $d_{b_i}(\mode)$. 

The block-diagonal structure with respect to $\mode$ in the above equation reflects the fact that our model assumes no interference between any pair of spatio-temporal modes during the measurement process.
In particular, it captures the possibility that an adversary may exploit different times-of-arrival, frequencies, or angles of incidence to attack the system, provided that each instance corresponds to a definite spatio-temporal mode and no coherent superpositions across modes, or multi-excitation states that simultaneously occupy several modes are used. Even with these limitations, the model protects against a wide range of known classical side-channel attacks.
For instance, the time-shift attack \cite{makarovEffects2006,qiTimeshift2007} is fully captured within this model, as it simply corresponds to Eve selecting different times-of-arrival to exploit the time-dependent efficiency mismatch of the gated detectors.
Thus, the block-diagonal model represents a first step toward a more complete analysis of realistic side-channels.
This perspective also captures other potential attack strategies, such as modifying the temperature of the detection setup.
   
\begin{remark} \label{remark:sidechannelmodel}
   We stress that our results in this subsection should be interpreted within the context of this model, and may not accurately describe the physical reality of multi-mode detectors. Nevertheless, while we only consider models of the above form in this work, our proof provides a framework to accommodate more complicated models of multi-mode detectors with off-diagonal blocks, as long as one can suitably bound $\deltaone,\deltatwo$.  In general, this would require a model of the detectors, and characterization of the detectors over all the modes. For examples of such attempts to experimentally characterize all the modes, see Ref.~\cite{rau2014spatial,sajeedSecurity2015}. 
\end{remark}

Due to the block-diagonal structure of the above POVM element \cref{eq:povmmodes}, and the fact that $\deltaone, \deltatwo$ are $\infty$-norms which can be computed on each block-diagonal part separately, it is straightforward to see that our computation of $\deltaone,\deltatwo$  is directly applicable to the above scenario.  To see this,  note that our metrics are obtained by first constructing POVMs corresponding to a  multi-step measurement process, as outlined in \cref{subsec:recipe}. This construction preserves the block-diagonal structure of \cref{eq:povmmodes}.  Thus, if $\deltaone(\{{\eta_{b_i}(\mode),  d_{b_i}(\mode)}\}), \deltatwo(\{{\eta_{b_i}(\mode),  d_{b_i}(\mode)}\}) $ are the values of these metrics computed according to \cref{eq:deltaCostsRecipe}, for the appropriate single-mode POVMs,  then the metrics for the multi-mode case are given by 
    \begin{equation} \label{eq:deltamultimode}
        \begin{aligned}
        \deltaone^\text{multi} =& \max_{\mode}  \deltaone(\{{\eta_{b_i}(\mode),  d_{b_i}(\mode)}\}), \\
            \deltatwo^\text{multi} =& \max_{\mode}  \deltatwo(\{{\eta_{b_i}(\mode),  d_{b_i}(\mode)}\}).
        \end{aligned}
    \end{equation}
    If the values of $\{{\eta_{b_i}(\mode),  d_{b_i}(\mode)}\}$ are characterized and satisfy  \cref{eq:etadcmodel} for all $\mode$, then 	\cref{eq:deltamultimode} is exactly the same as the computation as in Step (5) of \cref{subsec:recipe} (which corresponds to computing $\deltaone,\deltatwo$ for  Case~\ref{casetwo} from \cref{sec:introduction}).		

This means that the recipe from \cref{subsec:recipe}, and the computed key rates from \cref{subsec:plots} are valid for the scenario where Eve can choose the value of $\eta_{b_i}, d_{b_i}$ in the specified ranges (\cref{eq:etadcmodel}), via some extra spatio-temporal modes. Most importantly, our analysis does not depend on the \textit{number} of such spatio-temporal modes. Thus, we are able to address scenarios where Eve has an arbitrary number of spatio-temporal modes, to induce (a bounded amount of) basis-efficiency mismatch in the detector.

			\begin{remark} \label{remark:allowingevedarkcounts}
				As discussed above, our methods are such that allowing Eve to choose the detector parameters within the characterized range yields the same key rate as having fixed detector parameters that are characterized within the same range. However, this observation need not be fundamental, and may be a consequence of the proof technique used in this work. This is because intuitively, we expect scenarios where Eve cannot choose the detector parameters (from within their respective ranges), to lead to higher key rates than scenarios where she can, since she is strictly stronger in the latter scenario.
				Nevertheless, while we do not know of a physical mechanism by which Eve can choose dark count rates, we allow Eve to choose them along with the detection efficiency. 
			\end{remark}

We have picked this model for its theoretical simplicity. However, more realistic models such as the one introduced in \cite[Section 3]{fungSecurity2009a} can also be analysed with the results in this work. In that case, the computation of $\deltaone$ and $\deltatwo$ would constitute a more involved version of our current computations described in \cref{app:detModelCalc}. Specifically, \cref{eq:triangleIneqDiffOfFilters} would need to be modified with a different choice of operator $P$.
We note that the basis dependent filters for this model are still block-diagonal in the total number of photons $n$ across all modes. Moreover, as $n$ increases, the filtering operators approach the identity operator (since the probability of detect approaches $1$). Thus, we expect $\deltaone$ and $\deltatwo$ to depend on the $n\leq1$ blocks. If this monotonicity can be rigorously proven, then the $n\leq1$ block contributions to $\deltaone$ and $\deltatwo$ can even be computed numerically.

Finally we note that this work does not apply to \textit{all} detector side-channels. For instance, our model does not fit Trojan horse attacks \cite{Gisin_2006}. Moreover, some blinding attacks on detectors \cite{Gerhardt_2011} lead to complete  knowledge of Bob's detection events to Eve. In this case, our methods naturally lead to trivial key rates, since no key generation is possible.

		\section{Summary and Discussion} \label{sec:conclusion} 
		In this work, we presented a finite-size security proof of the decoy-state BB84 protocol in the presence of imperfectly characterized and (bounded) adversary controlled basis-efficiency mismatch. Thus, we addressed a longstanding assumption made in security proofs for such protocols  within the EUR and phase error correction frameworks. Before this work, proofs within these frameworks were not stable, and would be invalidated by infinitesimal amounts of basis-efficiency mismatch (This problem does not arise in MDI-QKD, and is not resolved by this work for entanglement-based protocols). Since our methods permit (bounded) adversarial control over the efficiency mismatch, we also develop a framework to address an important class of detector side-channels, which has remained unresolved in existing security proof approaches for standard QKD.
        
        We also fixed several technical issues in the security analysis of decoy-state QKD within the EUR framework. We applied our results to the decoy-state BB84 protocol, demonstrating practical key rates in the finite-size regime even in the presence of basis-efficiency mismatch. We also investigate quantitatively, the effect of methods such as random swapping of detectors, to reduce efficiency mismatch. Taken together, these results are a significant step towards protocol security of the EUR proof technique, and the implementation security of QKD using trusted detection setups. 
       
        Moreover, although the rigorous results we obtain for multi-mode detectors depend on a specific model, we expect this framework to be adapted to more realistic models with subsequent work.    For computing key rates based on our results, suitable bounds on $\deltaone,\deltatwo$ are required. While our current analysis relies on some simplifications to obtain these bounds, they can likely be improved. Finally, examining a wider spectrum of detector imperfections -- well characterized by state-of-the-art experimental methods -- would further broaden the applicability of our results. Such an endeavor would require close collaboration with experimentalists to refine the characterization of imperfections, as well as theoretical advancements to extend the framework to encompass a wider class of side channels.
		
  We note that results from this work have already been used for subsequent work on QKD security analysis. For instance, in this work, we only sketch a possible approach to handling correlated detectors - an open problem on which there has been little progress so far.  Ref.~\cite{wang2025phaseerrorestimationpassive} obtains rigorous results for correlated detector effects within phase error based frameworks. It follows the same essential idea presented here, but incorporates some modifications necessary for a fully rigorous analysis. Another natural extension is to integrate our methods with established methods for addressing source imperfections \cite{pereiraModified2023,tamakiLosstolerant2014,curras-lorenzoSecurity2024}. Such a combination would lead to a security proof robust to both source and detector imperfections. Such a result has also been recently obtained in Ref.~\cite{curraslorenzo2025securityquantumkeydistribution}. Another avenue is applying these methods to passive detection setups, where the basis-efficiency mismatch assumption translates to an assumption of a perfectly balanced beam splitter and identical detectors. Such a result has also been recently obtained in Refs.~\cite{wang2025phaseerrorestimationpassive,Mizutani_2026}.

	\bibliographystyle{quantum}  
		\bibliography{bibliographyEUR}

		\section*{Author Contributions} \label{sec:authorcont}
		This project was formulated and led by DT. DT is responsible for the contributions in \cref{app:varlengthproof,sec:perfectdetectors,sec:decoy,app:variablelengthdecoy}. DT and SN are responsible for the contributions in \cref{sec:imperfectdetectors,sec:results,sec:detectorside,sec:correlated}. The proofs of the lemmas in \cref{subsec:pulkitlemmas} are due to PS. NL contributed to supervision and general directions for the project.

\section*{Code Availablility}
The code used in this paper is available at \href{https://openqkdsecurity.wordpress.com/repositories-for-publications/}{https://openqkdsecurity.wordpress.com/repositories-for-publications/}.
        
		\section*{Acknowledgements}
		 We thank Victor Zapatero, Margarida Pereira, and Guillermo Currás-Lorenzo for useful discussions on the phase error correction framework and the EUR approach to QKD security proofs, and for providing helpful comments on this manuscript.  We thank Ernest Tan for useful discussions that led to the variable-length security proof in this work. We thank Ashutosh Marwah for some helpful comments on our random sampling arguments. We thank Masato Koashi, and Antia Lamas-Linares for helpful discussions on the detector model used in this work. We thank John Burniston for help with code. We thank Federico Grasselli for detailed feedback on early versions of this work.
		This work was funded by the NSERC Discovery Grant, Micro Net DND Grant, and the Alliance Quint Grant, and was conducted
		at the Institute for Quantum Computing, University of Waterloo, which is funded by the
		Government of Canada through ISED. DT and PS are partially funded by the Mike and Ophelia
		Lazaridis Fellowship.
		
		\appendix

		\section{Technical Statements}
		\label{appendix:technical}
		We use $S_\circ(Q)$ to denote the set of normalized states on $Q$. We use $S_\bullet(Q)$ to denote the set of all sub-normalized states on $Q$. We use $\text{Pos}(Q)$ to denote the set of positive semi-definite operators on $Q$. The smoothing on the min and max entropies is with respect to the purified distance \cite[Definition 3.8]{tomamichelQuantum2016}.

		\begin{lemma}(\cite[Lemma 7]{tomamichelLargely2017}) \label{lemma:smalleventstate}
			Let $\rho_{CQ} \in S_\bullet (CQ)$ be classical in $C$, and let $\Omega$ be any event on $C$ such that $\Pr(\Omega)_\rho \leq \varepsilon$. Then there exists a sub-normalized state $\tilde{\rho}_{CQ} \in S_\bullet(CQ)$ with $\Pr(\Omega)_{\tilde{\rho}} =0$, and $P(\rho,\tilde{\rho}) \leq \sqrt{\varepsilon}$, where $P$ denotes the purified distance.
		\end{lemma}
		
		We use the above lemma in the proof of the following statement. The following statement allows us to replace the smooth max entropy term in the EUR statement with our bound on the phase error rate. The proof is basically the same as the proof of \cite[Proposition 8]{tomamichelLargely2017}.
		\begin{lemma} \label{lemma:smoothmaxerror}
			Let $\rho \in S_\bullet(XY)$ where $X,Y$ store $n$-bit strings, and let $\bm{e}_{XY}$ denote the error rate in these strings. Let $\Omega$ be any event such that $\bm{e}_{XY} > e_\text{max}$, and let $\Pr(\Omega)_{\rho} \leq \kappa$. For any $e_\text{max}<1/2$, we have
			\begin{equation}
				\smoothmax{\sqrt{\kappa}}(X|Y)_\rho \leq n h (e_\text{max})
			\end{equation}
			
		\end{lemma}
		\begin{proof}
			By \cref{lemma:smalleventstate}, there exists a state $\tilde{\rho}_{XY}$ such that $\Pr(\Omega)_{\tilde{\rho}}=0$ and $P(\rho,\tilde{\rho}) \leq \sqrt{\kappa}$. Therefore we have
			\begin{equation}
				\begin{aligned}
					\smoothmax{\sqrt{\kappa}} (X | Y)_{\rho} &\leq H_\text{max}(X|Y)_{\tilde{\rho}} \\
					&= \log(   \sum_{y \in \{0,1\}^n} \Pr(Y=y)_{\tilde{\rho}} 2^{H_\text{max}(X|Y)_{\tilde{\rho} | Y=y}  }         ) \\
					&\leq \max_{y \in \{0,1\}^n } H_\text{max}(X|Y)_{\tilde{\rho } | Y=y}	\\
					& = \max_{y \in \{0,1\}^n } \log  \bigg \lvert \left\{ x \in \{0,1\}^n : \Pr(X=x \land Y=y)_{\tilde{\rho}} > 0 \right\}  \bigg  \rvert  \\
					&\leq \log ( \sum_{k=0}^{n e_\text{max}}  {n\choose k} ), \\
					&\leq \log(2^{ nh(e_\text{max})})
				\end{aligned}
			\end{equation}
			where we used the definition of the smooth max entropy in the first inequality, and \cite[Sec. 4.3.2]{tomamichelFramework2013} for the second equality. The third inequality and the fourth equality follow from the definitions. The fifth inequality follows from the fact that the state $\tilde{\rho}$ is guaranteed to have $\leq n e_\text{max}$ errors, while the final inequality follows from the suitable bound on the sum of binomial coefficients.
		\end{proof}

		\twosteplemma*
		\begin{proof}
			Observe that $\{\tilde{F}_i | i \in \mathcal{P}_\mathcal{A}\}$ is a valid set of POVMs by construction. Moreover, $\{ G_{k} | k \in \mathcal{A}_i\}$ is a valid set of POVMs for each $i$, also by construction.  Thus we only need to show that $\rho_\text{final} = \rho^\prime_\text{final}$. Using the cyclicity of trace in \cref{eq:twostepsecondstate}, it suffices to prove
			\begin{equation}
				\sqrt{\tilde{F}_i} {G}_{k} \sqrt{\tilde{F}_i} = \Gamma_{k} \quad \forall i \in \mathcal{P}_\mathcal{A}, \forall k \in \mathcal{A}_i.
			\end{equation}
			Substituting the expression for $F_k$ into the above equation, we obtain 
			\begin{equation}
				\begin{aligned}
					\sqrt{\tilde{F}_i} G_{k} \sqrt{\tilde{F}_i} &= \sqrt{\tilde{F}_i} \left(  \sqrt{\tilde{F}}^+_i \Gamma_{k} \sqrt{\tilde{F}}^+_i + P_{k} \right)  \sqrt{\tilde{F}_i}  \\
					&= \sqrt{\tilde{F}_i} \left(  \sqrt{\tilde{F}}^+_i \Gamma_{k} \sqrt{\tilde{F}}^+_i \right)  \sqrt{\tilde{F}_i} \\
					&=\mathrm{\Pi}_{\tilde{F}_i} 			\Gamma_{k} \mathrm{\Pi}_{\tilde{F}_i} \\
					&=	\Gamma_{k},
				\end{aligned}
			\end{equation}
			where the second equality follows from the fact that $P_k$ and $\tilde{F}_i$ have orthogonal supports, and the final equality uses the fact that the support for $\tilde{F}_i$ is larger than the support for $\Gamma_k$ for $k\in \mathcal{A}_i$.
			This concludes the proof.
		\end{proof}

		\section{Variable-length security proof} \label{app:varlengthproof}

		In this appendix we will prove the following theorem regarding variable-length security of the protocol from \cref{sec:protocol}.

		\variablelengthproof*

		\begin{proof}
			Our proof will consist of three parts. In the first part, we will discuss the security definition for variable-length QKD protocols. In the second part we will use entropic uncertainty relations and \cref{eq:req} to obtain a suitable lower bound on the smooth min entropy of the raw key register in the QKD protocol. In the third part, we use this to prove variable-length security.
			
			\subsubsection{Variable-length security definition}
			In order to prove the $\epsSec$-security of a variable-length QKD protocol \cite{portmannSecurity2022,Benor2004}, one must show that for all attacks by the adversary, the following statement is true :
		\begin{equation} \label{eq:securitydef}
			\sum_{\len=1}^{\infty} \Pr(\Omega_{\lenrv = \len}) \frac{1}{2} \norm{\rho_{K_A K_B \Cfull E^n |\Omega_{\lenrv = \len} } - \sum_{k \in \{0,1\}^l} \frac{1 }{2^l} \ketbra{kk}_{K_AK_B} \otimes  \rho_{ \Cfull  E^n |\Omega_{\lenrv = \len} } }_1 \leq \epsSec
		\end{equation}
		where $\Omega_{\lenrv = \len} $ denotes the event that a key of length $\len$ bits is produced. The error-verification step of the protocol guarantees the $\epsEV$-correctness of the protocol \cite{tupkarySecurity2023}. The fact that $(2 \epsAT+\epsPA)$-secrecy (\cref{eq:secrecyclaim}) and $\epsEV$-correctness implies $\epsSec = (2 \epsAT+\epsPA+\epsEV)$-security of the QKD protocol (\cref{eq:securitydef}) has already been shown in many prior works \cite{tupkarySecurity2023, Benor2004}  and we do not repeat it here. Therefore, in this work we focus on proving the $(2 \epsAT+\epsPA)$-secrecy of the QKD protocol. This requires us to show 
		\begin{equation} \label{eq:secrecyclaim}
			\sum_{\len=1}^{\infty} \Pr(\Omega_{\lenrv = \len}) \frac{1}{2} \norm{\rho_{K_A \Cfull E^n |\Omega_{\lenrv = \len} } - \sum_{k \in \{0,1\}^l} \frac{1}{2^l} \ketbra{k}_{K_A} \otimes  \rho_{ \Cfull  E^n |\Omega_{\lenrv = \len} } }_1 \leq 2 \epsAT+\epsPA
		\end{equation}
		which is essentially the same statement as \cref{eq:securitydef}, but with Bob's key register omitted.
		
		\subsubsection{Bounding the smooth min entropy}
		Let us turn our attention to the phase error rate estimate. Let
		\begin{equation} \label{eq:defserfbound}
			\serfbound(\params) \coloneq 	\Pr(\ephrv \geq \Boundbasicdelta(\eX ,\nX ,\nK))_{ | \event{\params} } 
		\end{equation}
		where $\serfbound(\params)$ denotes the probability that our computed bound $\Boundbasicdelta(\eX,\nX,\nK)$ fails, conditioned on the event $\event{\params}$. We will not be able to directly bound $\serfbound(\params)$ (see footnote.~\footnote{In fact, it is easy to see that if Eve implements an intercept-resend attack, it is impossible to obtain any non-trivial bounds on $\kappa(\params)$. This is because during an intercept-resend attack, even if the observations indicate a low error rate (due to an unlucky protocol run), the phase error rate is still equal to 1/2}). However note that \cref{eq:req} trivially implies
		\begin{equation} \label{eq:req2}
			\begin{aligned}
				\sum_{\nX,\nK,\eX,\eZ}  \Pr(\event{\nX,\nK,\eX\eZ} )\serfbound(\params) &= 
				\Pr(\ephrv \geq \Boundbasicdelta(\eXrv,\nXrv,\nKrv) ) \\
				&\leq \epsAT^2,
			\end{aligned}
		\end{equation}
		where the sum is over all possible values of $\nX,\nK,\eX,\eZ$. 
		We will utilize \cref{eq:req2}, which follows from \cref{eq:req}, in bounding the smooth min entropy of the raw key register.
		
	To do so, focus on the state $\rho_{A^{\nK}B^{\nK} E^n C^n | \event{\params} }$, which is the state on the detected key generation rounds. This state can be obtained by transforming Bob's measurement  procedure to consist of two steps, and then only implementing the first step measurement which determines the detect vs no-detect outcome. Such a state can be rigorously obtained using \cref{lemma:twostep} from \cref{sec:perfectdetectors}. For the purposes of this proof, we only need the fact that it is well defined.  We will obtain a bound on the smooth min entropy of the key generated from this state.
			
			Suppose Alice measures her $\nK$ systems in the $Z$ basis. Let the post-measurement state be given by \\  $\rho_{Z_A^{\nK}B^{\nK} E^n C^n  | \event{\params} }$. Suppose she measures it in the $X$ basis, and let the post-measurement state be given by $\rho^\text{virt}_{{X_A}^{\nK}B^{\nK} E^n C^n  | \event{\params} }$. This $X$ measurement is not actually done in the protocol, and is only required for the theoretical proof. Using the entropic uncertainty relation \cite{tomamichelUncertainty2011}, we can relate the smooth min and max entropies $\left(\text{with smoothing parameter }\sqrt{\kappa(\params)}\right)$ of the two states obtained via $Z$ and $X$ measurements as 
			\begin{equation}
				\begin{aligned}
					& \smoothmin{\sqrt{\serfbound(\params)}}(Z_A^{\nK} | C^n  E^n)_{\rho | \event{\params} } \\
					+& \smoothmax{\sqrt{\serfbound(\params)}}(X_A^{\nK} | B^{\nK})_{\rho^\text{virt} | \event{\params}} \geq \nK \qualityfactor.
				\end{aligned}
			\end{equation}
			where $ \qualityfactor \coloneq \log(\frac{1}{ \max_{i,j} \norm{\AlicePOVM{X}{i} \AlicePOVM{Z}{j}}^2_\infty})$. We have deliberately chosen an appropriate smoothing parameter (see \cref{eq:defserfbound} for $\serfbound$) in the above equation. This choice will play a role at a later stage in the proof.

			\begin{remark}
				Notice that the value of $\qualityfactor$ \textit{only} depends on the POVM's used by Alice, after using the source-replacement scheme, and is equal to $1$ in this work. Thus, we set $\qualityfactor=1$ in the remainder of this work. Moreover, directly using the EUR in this context requires Alice to implement an \textit{active} basis choice measurement, which requires perfect signal state preparation. However, as stated earlier, several techniques of dealing with imperfect source preparation exist.
			\end{remark}

			We can make Bob measure his systems $B$ in the $X$ basis to obtain the classical outcome $X_B$. Then, using  data processing \cite[Theorem 6.2]{tomamichelQuantum2016}, we obtain
			\begin{equation} \label{eq:eurafterdpi}
				\begin{aligned}
					& \smoothmin{\sqrt{\serfbound(\params)}}(Z_A^{\nK} | C^n  E^n)_{\rho | \event{\params} } \\
					+& \smoothmax{\sqrt{\serfbound(\params)}}(X_A^{\nK} | X_B^{\nK})_{\rho^\text{virt} | \event{\params}} \geq \nK .
				\end{aligned}
			\end{equation}

			Recall that we have a probabilistic upper bound on $\ephrv$ (the error rate in $X_A^{\nK},X_B^{\nK}$) conditioned on the event $\event{\params}$. This bound fails with probability $\serfbound(\params)$ (see \cref{eq:defserfbound}). Thus,  using \cref{lemma:smoothmaxerror} (see \cref{appendix:technical}) along with this fact, we obtain:
			\begin{equation}
				\smoothmax{\sqrt{\serfbound(\params)}}(X_A^{\nK} | X_B^{\nK})_{\rho^\text{virt} | \event{\params} } \leq \nK h \left( \Boundbasicdelta(\eX,\nX,\nK) \right),
			\end{equation}
			which along with \cref{eq:eurafterdpi} gives us
			\begin{equation} \label{eq:eurbound}
				\begin{aligned}
					\smoothmin{\sqrt{ \serfbound(\params)}}(Z_A^{\nK} | C^n  E^n)_{\rho | \event{\params} }  \geq \nK ( 1 - h\left( \Boundbasicdelta(\eX ,\nX,\nK)\right)).
				\end{aligned}
			\end{equation}
			This is the required bound on the smooth min entropy of the raw key. We will now use \cref{eq:eurbound} to prove the $(2\epsAT+\epsPA)$-secrecy of the QKD protocol. 
			\subsubsection{Proving variable-length security}

			To obtain $(2 \epsAT+\epsPA)$-secrecy, we must show that \cref{eq:secrecyclaim} is true. Note that \cref{eq:secrecyclaim} groups together terms with the same length of the output key. However, different events $\fullevent$ may correspond to the same length of the output key.  Nevertheless, $\fullevent$ is a deterministic function of the classical announcements $C^n$. Thus, the states conditioned on different $\fullevent$ have orthogonal supports. Therefore, it is enough to show that
			\begin{equation} \label{eq:actualsecrecyclaim} 
				\begin{aligned}
					&\Delta \coloneq	\half \sum_{\nX,\nK,\eX,\eZ} \Pr(\fullevent \wedge \EVevent) \multiNorm {\rho_{K_A \Cfull E^n | \event{\params} \wedge \EVevent}  \\
						&  -  \sum_{k \in \{0,1\}^{l(\params)}} \frac{\ketbra{k}_{K_A}}{2^{l (\params)} } \otimes \rho_{ \Cfull E^n | \event{\params} \wedge \EVevent} }_1 \leq2 \epsAT+\epsPA,
				\end{aligned}
			\end{equation}
			since we can group together terms with the same output key to obtain \cref{eq:secrecyclaim} from \cref{eq:actualsecrecyclaim}. We will now prove \cref{eq:actualsecrecyclaim}.

			Now, note that without loss of generality, we can assume that we are summing over events that lead to a non-trivial length of the key (since events where the protocol aborts do not contribute to $\Delta$). Let $\accset = \{ (\params) | l(\params) > 0\}$ be the set of parameters that produce a non-trivial length of the key. Then, $\Delta$ can bounded using the following chain of expressions, which we explain below:
			\begin{equation} 
				\begin{aligned}
					\Delta
					& \leq \sum_{(\params) } \Pr(\fullevent )   \bigg(  2\sqrt{ \serfbound(\params)} \\
					&+ \half 2^{-\half \left( \smoothmin{\sqrt{ \serfbound(\params) }}(Z^{\nK} | E^n C^n C_E )_{(\rho | \fullevent) \wedge \EVevent } - l(\params)  \right)}   \bigg) \\
					& \leq \sum_{(\params) } \Pr(\fullevent )   \bigg(  2\sqrt{ \serfbound(\params)} \\
					&+ \half 2^{-\half \left( \smoothmin{\sqrt{ \serfbound(\params) }}(Z^{\nK} | E^n C^n C_E )_{\rho | \fullevent } - l(\params)  \right)}   \bigg) \\
					& \leq \sum_{(\params) \in \accset} \Pr(\fullevent)  \bigg( 2 \sqrt{  \serfbound(\params)} \\
					&+  \half 2^{-\half \left(  \smoothmin{\sqrt{ \serfbound(\params) }}(Z^{\nK} | E^n C^n  )_{\rho | \fullevent}- l(\params) -\leak(\params) - \ECost \right)       }   \bigg) \\
					& \leq \sum_{(\params) \in \accset} \Pr(\fullevent) \bigg( 2 \sqrt{ \serfbound(\params)} \\
					&+ \half 2^{-\half \left(  \nK \left(  1 - h\left( \Boundbasicdelta(\eX ,\nX,\nK) \right) \right)- l(\params) -\leak(\params) - \ECost \right)}   \bigg) \\
					& = \sum_{(\params) \in \accset} \Pr(\fullevent) \left( \epsPA + 2 \sqrt{ \serfbound(\params)} \right) \\
					&\leq \epsPA +  2\sqrt{\sum_{\params} \Pr( \event{\params} )  \serfbound(\params)}  \\ 
					&\leq \epsPA + 2\epsAT.
				\end{aligned}
			\end{equation}
			Here, we used the leftover-hashing lemma \cite[Proposition 9]{tomamichelLargely2017} with the appropriate smoothing parameter on the sub-normalized state $(\rho_{|  \event{\params}} )_{\wedge \EVevent}$ for the first inequality. Since we use sub-normalized conditioning on $\EVevent$, it only appears in the smooth min entropy term and not inside the probability. Next, we use \cite[Lemma 10]{tomamichelLargely2017} to get rid of the sub-normalized conditioning ($\wedge \EVevent$) in the smooth min entropy term in the second inequality. We used \cite[Lemma 6.8]{tomamichelQuantum2016} to split off the error-correction information ($\leak(\params)$) and error-verification information ($\log(2/\epsEV)$) in the third inequality.   We used the bound on the smooth min entropy from \cref{eq:eurbound} for the fourth inequality, and the values of $l(\params)$ and $\leak (\params)$ from \cref{eq:lvalue} for the fifth equality.
			We used concavity of the square root function and Jensen's inequality to pull the sum over probabilities inside the square root for the sixth inequality, and \cref{eq:req2} for the final inequality.

		\end{proof}

		\begin{remark}
			Note that the critical step here was using the concavity of the square root function to see that a bound on the \textit{average} failure probability of the phase error estimation procedure is enough to prove security. This is the same fundamental trick used by  Ref.~\cite{hayashiConcise2012,kawakamiSecurity,curras-lorenzoSecurity2024}. Our presentation here is in the EUR framework, where this manifests in the deliberate choice of our smoothing parameter in the first part of the proof. 
		\end{remark}

		\begin{remark}
			Notice that in the variable-length protocol for which we proved security, the number of bits on which privacy amplification is applied is variable. It depends on the number of key generation rounds obtained in each protocol run. Some subtle issues regarding two-universal hashing on a variable-length \textit{input} register were pointed out and addressed in \cite[Section VII]{tupkarySecurity2023}. In particular, it was noted that first looking at the number of bits in the raw key, and then choosing an appropriate two-universal hashing procedure for that many input bits, \textit{does not produce a valid two-universal hashing procedure on the input space of variable-length bit strings}. Due to this, the leftover-hashing lemma cannot be straightforwardly applied to such a scenario. However, this issue was addressed by showing that when the locations of the discard rounds are publicly announced,  the theoretical analyses of scenarios where the rounds are actually discarded, vs mapped to special symbols such as $0$ or $\bot$ (where leftover-hashing lemma can be applied), are equivalent \cite[Lemmas 4, 5]{tupkarySecurity2023}. Due to this equivalence, the PA procedure described above can be applied in QKD protocols. 
			
			It is interesting to note that these issues are completely avoided by the above proof, in a very \textit{different} manner than \cite{tupkarySecurity2023}. This is because in this proof, we always apply the leftover-hashing lemma on a state conditioned on the specific length of the raw key register (\cref{eq:eurbound}). Therefore, the leftover-hashing lemma can be applied in a  straightforward manner, and there are no issues is choosing the hashing family based on the specific length of the raw key register. In other words, the PA procedure described above is valid for this proof. In a similar sense, the variable-length security proof from \cite{tupkarySecurity2023} critically relied on a technical lemma (Lemma 9), that necessitated the use of R\'enyi entropies instead of smooth min entropy in that work. However, the variable-length proof presented above takes a different approach, and does \textit{not} impose the same requirements on the behavior of smooth min entropy. Again, this is due to the use of \cref{eq:eurbound}. 
		\end{remark}
	
	\subsection{Comparing to fixed-length protocols}

We will now compare the results obtain above with the key rate obtained for fixed-length protocols. Consider a fixed-length protocol that is identical to the steps in \cref{sec:protocol}, except that it accepts if and only if $\nX \geq \nX^L$, $\nK \geq \nK^L$, $\eX \leq e_X^U$, and $\eZ \leq e_Z^U$. Then, the key length obtained for such a protocol can be shown to be the same expressions as \cref{eq:lvalue}, with $\nX,\nK,\eX,\eZ$ replaced with the acceptance thresholds $\nK^L,\nK^L,e_X^U,e_Z^U$. Since $l(\nX,\nK,\eX,\eZ)$ is an increasing function in $\nX,\nK$ and a decreasing function of $\eX,\eZ$, the variable-length protocol \textit{always produces at least as much key as} the the fixed-length protocol, for any possible observations $(\params)$ during a protocol run (for the above proof technique).

Furthermore,theoretical works with such acceptance conditions typically require Alice and Bob to pick a uniformly random sample of $\nX^L$ rounds for testing, and $\nK^L$ rounds for key generation (in the highly-likely event that they obtain extra rounds) \cite{limConcise2014,tomamichelTight2012,pfisterSifting2016}. This requires additional randomness in the protocol implementation. The variable-length version does not require this step, and therefore had a reduced requirement on local randomness.

		\section{Sampling}\label{appendix:sampling}
		In this section, we prove the technical statements needed to prove our sampling bounds.
		\subsection{Random Sampling} \label{appsubsec:randomsampling}
		We start with the usual Serfling \cite{serflingProbability1974}  statement in the following lemma. The following lemma is obtained from \cite[Eq. 74, Lemma 6]{tomamichelLargely2017}
		\begin{lemma}[Serfling] \label{lemma:serfling}
			Let $\X_1\dots \X_{m+n}$ be bit-valued random variables. Let $\bm{J}_m$ denote the choice of a uniformly random subset of $m$ positions, out of $m+n$ positions.  Then,
			\begin{equation}
				\begin{aligned}
					\Pr(\sum_{i \notin \bm{J}_m} \frac{\X_i}{n} \geq \sum_{i \in \bm{J}_m} \frac{\X_i}{m} + \gamma_\text{serf}) &\leq e^{-2 \gamma_\text{serf}^2 \fserf(m,n)}, \\
					\fserf(m,n) \coloneq \frac{nm^2}{(n+m)(m+1)}.
				\end{aligned}
			\end{equation}
		\end{lemma}
		Serfling basically states that if one chooses a random set of positions, then the fraction of $1$s in those positions gives us a good estimate of the fraction of $1$s in the remaining positions.
		However, observe that the sampling procedure in the protocol from \cref{subsec:samplingperfect,subsec:sampling} does not actually choose a random subset of fixed-length for testing. Instead, the protocol decides to map each conclusive round to test or key in an IID manner. Therefore, the application of the Serfling bound is not straightforward. In the following lemma, we show how the Serfling bound can still be rigorously used.

		\serflinglemma*
		
		\begin{proof}
			Since the sampling procedure randomly assigns each bit to test or key (or does nothing with them if $p_t+p_k <1$), \cref{lemma:serfling} cannot be directly applied. However, consider what happens if we condition on the event $\event{\nX,\nK}$. Then, for a given set of positions that form the $\nX+\nK$ positions selected for test or key, it is the case that each set of $\nX$ positions is equally likely. Therefore, the above sampling procedure is exactly equivalent to:
			\begin{enumerate}
				\item First determining the event $\event{\nX,\nK}$ by sampling from \textit{some} probability distribution. 
				\item Pick some $\nX+\nK$ positions at random.
				\item Then determining the exact positions of the $\nX$ test rounds, by choosing a random subset of fixed-size $\nX$ out of these $\nX+\nK$ positions.	\end{enumerate}
			The necessary claim follows by applying \cref{lemma:serfling} for step 3 of the above procedure.
		\end{proof}

		\subsection{Sampling with imperfect detectors} \label{subsec:pulkitlemmas}
		We now turn our attention to proving \cref{lemma:pulkitgeneric}, which is the main statement utilized in extending the EUR approach to imperfect detectors in \cref{sec:imperfectdetectors}. We start by proving \cref{lemma:orderingonPOVMs,lemma:smallPOVM} which we use later in the proof of \cref{lemma:pulkit}.
		
		Recall our notation: If  $\rho_{Q^n} \in S_\circ(Q^{\otimes n})$ is an arbitrary state, and $\{P_1,P_2,\dots,P_m\}$ is a set of POVM elements, then we let $\num_{P_i}$ denote the classical random variable corresponding to the number of measurement outcomes corresponding to $P_i$ when the state $\rho_{Q^n}$ is measured. Moreover, let $ \bm{D}_P$ denote the classical random variable that describes the measurement outcomes when each subsystem of $\rho$ is measured using $\{P_1,P_2,\dots,P_m\}$.  We use $S \sim \bm{D}_P$ to denote the statement S is sampled from $\bm{D}_P$.
		\begin{lemma} \label{lemma:orderingonPOVMs}
			Let $\rho_{Q^n} \in S_\circ(Q^{\otimes n})$ be an arbitrary state. Let $\{P,I - P\}$ and $\{P^\prime,I-P^\prime\}$ be two sets of POVM elements such that $P \leq P^\prime $. Then, for any $e$, it is the case that
			\begin{equation}
				\Pr(\frac{\num_{P}}{n} \geq e ) \leq \Pr(\frac{\num_{P^\prime}}{n} \geq e) 
			\end{equation}
		\end{lemma}
		\begin{proof}
			We will describe a procedure to generate random strings $S,S^\prime$ such that  $S \sim \bm{D}_P, S^\prime \sim \bm{D}_P^\prime$. Consider the POVM $\{P,P^\prime-P,I-P^\prime \}$, and let $T$ be the classical string taking values in $\{0,1,2\}^n$ which stores the measurement outcomes when measured using this POVM. Then, $S,S^\prime$ can be obtained by first obtaining $T$, followed by the following remapping
			\[(S_i,S^\prime_i)=\begin{cases}
				(1,1) &\text{$T_i=0$}\\
				(0,1) &\text{$T_i=1$}\\
				(0,0) &\text{$T_i=2$}
			\end{cases} \] 
			where $i$ denotes the position in the string. The required claim follows from the observation that the above procedure maps more $S^\prime_i$ to $1$ than $S_i$. Thus, 
			\begin{equation}
				\Pr(w( \bm{S}^\prime) \geq w( \bm{S})) \geq 0 \implies \Pr(w( \bm{S}^\prime) > ne) \geq \Pr(w(\bm{S}) \geq ne)
			\end{equation}
			where $w$ denotes the hamming weight of the string (sum of each element of the string). The necessary statement follows after noting that $w(S) \sim \num_P$ and $w(S^\prime) \sim \num_{P^\prime}$ (which can be argued rigorously using the two-step measurement \cref{lemma:twostep}).  
		\end{proof}
		\begin{remark}
			Note a conceptual subtlety in the above proof: The procedure used to generate $S, S^\prime$ in the above proof has some joint probability distribution associated to it. This means that $(S,S^\prime)$ is a well-defined random variable. However, one cannot talk about the joint probability distribution of two different sets of measurement on the same quantum state. This subtle issue is avoided by noting that we are only interested in making statements on the marginal probability distribution of $S$ and $S^\prime$, and how they relate to one another. And it is indeed true that these distributions satisfy $S \sim \bm{D}_P$ and $S^\prime \sim \bm{D}_{P^\prime}$, which is enough to prove our claim. The fact that $S,S^\prime$ has some joint distribution associated with it is immaterial.
		\end{remark}
		
		\smallPOVM*
		
		\begin{proof}
			Since $\norm{P}_\infty \leq \delta$, we have $P \leq \delta I$. By \cref{lemma:orderingonPOVMs}, we have
			\begin{equation}
				\Pr(\frac{\num_{P}}{n} \geq \delta+c ) \leq \Pr(\frac{\num_{\delta I}}{n} \geq \delta+c) 
			\end{equation}
			Observe that measurement using $\{\delta I, (1-\delta) I\}$ is equivalent to Bernoulli sampling. Thus $\num_{\delta I}$ obeys the binomial distribution. Therefore, 
			\begin{equation}
				\begin{aligned}
					\Pr(\frac{\num_{\delta I}}{n} \geq \delta+c) &\leq \sum_{i = n (\delta+c)}^{n} {n \choose i} \delta^ i (1-\delta)^{n-i}.
				\end{aligned}
			\end{equation}
		\end{proof}

		\begin{lemma}
			\label{lemma:pulkit} Let $\rho_{Q^n} \in S_\circ(Q^{\otimes n})$ be an arbitrary state. Let $\{P,I - P\}$ and $\{P^\prime,I-P^\prime\}$ be two sets of POVM elements. Suppose there exists a $0 \leq \tilde{P} \leq I$ such that $P\leq \tilde{P}, P^\prime \leq \tilde{P} $, and $ \norm{\tilde{P}-{P}}_\infty  \leq \delta$. Then
			\begin{equation} \label{eq:pulkitequation}
				\Pr(\frac{\num_{P^\prime}}{n} \geq e +  (\delta+c)) \leq \Pr(\frac{\num_P}{n} \geq e) + \binfunction{n}{\delta}{c} ,
			\end{equation}
			where $\binfunction{n}{\delta}{c}$ was defined in \cref{lemma:smallPOVM}.
		\end{lemma}
		\begin{proof}
			We will describe a process to generate $S \sim \bm{D}_P$ and $\tilde{S} \sim  \bm{D}_{\tilde{P}}$, in a similar manner as in the proof of \cref{lemma:orderingonPOVMs}. In particular, let $T$ be the random variable taking values in $\{0,1,2\}^n$ that stores the measurement outcomes of $\{P,\tilde{P}-P,I-\tilde{P}\}$ measurements. We generate $S,\tilde{S}$ by first obtaining $T$, followed by the following remapping
			\[(S_i,\tilde{S}_i)=\begin{cases}
				(1,1) &\text{$T_i=0$}\\
				(0,1) &\text{$T_i=1$}\\
				(0,0) &\text{$T_i=2$}.
			\end{cases} \]
			Then,
			\begin{equation} \label{eq:pulkitwithordering} 
				\begin{aligned} 
					\Pr(\num_{\tilde{P}} \geq  ne +n (\delta+c)) &= 	\Pr(w(\bm{\tilde{S} })\geq  ne +n (\delta+c) ) \\
					&= 	\Pr(w( \bm{\tilde{S}} )\geq  ne +n (\delta+c)  \cap w(\bm{S}) \geq ne) \\
					&+ 	\Pr(w(\bm{\tilde{S} })\geq  ne +n (\delta+c) \cap w(\bm{S})  < ne) \\
					& \leq  \Pr( w(\bm{S})  \geq ne ) + \Pr(w(\bm{\tilde{S}}) - w(\bm{S}) \geq n(\delta+c)) \\
					& = \Pr(\num_P \geq ne) + \Pr(\num_{\tilde{P} - P} \geq n (\delta+c)) \\
					& \leq \Pr(\num_P \geq ne) +  \binfunction{n}{\delta}{c} . 
				\end{aligned}
			\end{equation}
			where we used \cref{lemma:smallPOVM} in the final inequality, the fact that $w(S) \sim \num_P$ and $w(\tilde{S}) - w(S) = w(\tilde{S}-S) \sim \num_{\tilde{P}-P}$ ($\tilde{S}_i - S_i = 1$ if and only if $T_i=1$) for the penultimate inequality, and basic properties of probabilities for the remaining steps. Next, we replace the $\tilde{P}$ with $P^\prime$ using \cref{lemma:orderingonPOVMs} and $\tilde{P}\geq P^\prime$, and obtain
			\begin{equation}\label{eq:pulkittemptwo} 
				\Pr(\num_{P^\prime} \geq ne + n(\delta+c)) \leq \Pr(\num_{\tilde{P}} \geq ne + n(\delta+c)).
			\end{equation}
			The proof follows after noting that \cref{eq:pulkitwithordering,eq:pulkittemptwo} $\implies$ \cref{eq:pulkitequation}.
		\end{proof}
		
		\cref{lemma:pulkit} above  requires an explicit construction of a $\tilde{P}$ satisfying the necessary requirements. However, this requirement can be removed, and we obtain a sightly worse result with greater generality below.

		\pulkitlemmageneric*

		\begin{proof}
			Let $G^\prime = (1-\delta) P^\prime $, and $G = (1-\delta) P$. 
			Using $0\leq G\leq P$ and \cref{lemma:orderingonPOVMs}, we obtain
			\begin{equation}
				\label{eq:pulkitgenerictempone}
				\Pr(\num_{G} \geq ne) \leq \Pr(\num_{P} \geq ne).
			\end{equation}
			Using $0 \leq G^\prime + \delta I \leq I$, $G^\prime+\delta I \geq G$,  $\norm{G^\prime+\delta I - G}_\infty \leq   \delta+\delta(1-\delta) \leq 2\delta $, and \cref{eq:pulkitwithordering}, we obtain
			\begin{equation} \label{eq:pulkitgenerictemptwo}
				\Pr(\num_{G^\prime+\delta I} \geq ne + n(2 \delta + c)) \leq \Pr(\num_{G} \geq ne) + \binfunction{n}{2\delta}{c},
			\end{equation}
			Finally, using $G^\prime + \delta I \geq P^\prime$,  and \cref{lemma:orderingonPOVMs}, we obtain
			\begin{equation} \label{eq:pulkitgenerictemp3}
				\Pr(\num_{P^\prime} \geq ne + n(2 \delta+c)) \leq \Pr(\num_{G^\prime + \delta I} \geq n e + n(2\delta+c)).
			\end{equation}
			The proof follows from the observation that \cref{eq:pulkitgenerictempone,eq:pulkitgenerictemptwo,eq:pulkitgenerictemp3} $\implies$ \cref{eq:pulkitequationgeneric}.
		\end{proof}

		\subsection{Sampling with independent imperfect detectors} \label{app:pulkitindependent}
		The statements above are written for a measurement procedure where the same POVM is used to measure each round of the state. However the proofs do not actually use this fact. The same proofs are valid even if the measurement for each round is done using a different POVM element (as long it satisfies the required bounds on the $\infty$-norm). Thus we write a generalized version of \cref{lemma:pulkit,lemma:pulkitgeneric} below. They can be proved by simply redoing the proofs in the earlier section.
		
		\begin{lemma}
			\label{lemma:pulkitind} Let $\rho_{Q^n} \in S_\circ(Q^{\otimes n})$ be an arbitrary state. Suppose the $i$th round is measured using POVM $\{P(i),I - P(i)\}$ and let $\num_P$ be the number of outcomees corresponding to the first POVM element. Similarly, suppose the $i$th round is measured using POVM $\{P^\prime(i), \id - P^\prime(i)\}$  and let $\num_{P^\prime}$ be the number of outcomes corresponding to the first POVM element. Suppose for all $i$, there exists a $0 \leq \tilde{P}(i) \leq I$ such that $P(i) \leq \tilde{P} (i) , P^\prime (i) \leq \tilde{P} (i) $, and $ \norm{\tilde{P}(i) -{P}(i)}_\infty  \leq \delta$. Then
			\begin{equation} 
				\Pr(\frac{\num_{P^\prime}}{n} \geq e +  (\delta+c)) \leq \Pr(\frac{\num_P}{n} \geq e) + \binfunction{n}{\delta}{c} ,
			\end{equation}
			where $\binfunction{n}{\delta}{c}$ was defined in \cref{lemma:smallPOVM}.
			
		\end{lemma}

		\begin{lemma}
			\label{lemma:pulkitgenericind} Let $\rho_{Q^n} \in S_\circ(Q^{\otimes n})$ be an arbitrary state. Suppose the $i$th round is measured using POVM $\{P(i),I - P(i)\}$ and let $\num_P$ be the number of outcomees corresponding to the first POVM element. Similarly, suppose the $i$th round is measured using POVM $\{P^\prime(i), \id - P^\prime(i)\}$  and let $\num_{P^\prime}$ be the number of outcomes corresponding to the first POVM element. Suppose  $ \norm{{P}^\prime(i) -{P}(i)}_\infty  \leq \delta \quad \forall i$. Then
			\begin{equation} 
				\Pr(\frac{\num_{P^\prime}}{n} \geq e +  2\delta+c) \leq \Pr(\frac{\num_P}{n} \geq e) + \binfunction{n}{2\delta}{c} ,
			\end{equation}
			where $\binfunction{n}{\delta}{c}$ was defined in \cref{lemma:smallPOVM}.
			
		\end{lemma}

		\section{Combining bounds} \label{appendix:combining}
		In this section, we will combine \cref{eq:boundone,eq:boundtwo,eq:boundthree,eq:boundfour} and obtain \cref{eq:finalbound}. This process is simply some cumbersome algebra and the use of the union bound for probabilities.
		
		Combining \cref{eq:boundone,eq:boundtwo}, we obtain
		
		\begin{equation}  \label{eq:bound12}
			\begin{aligned}
				&\Pr(\ephwierdrv \geq \eXrv + \gamma^{\epsATa}_\text{serf}   (\nXperf,\nKperf)   )_{ | \event{\nXperf,\nKperf}} \\
				&=   	\Pr(  \left(\ephwierdrv \geq \eXrv + \gamma^{\epsATa}_\text{serf}   (\nXperf,\nKperf)  \right) \bigcap \left(  \eXperfrv \geq \eXrv  \right)  )_{ | \event{\nXperf,\nKperf}}  \\
				&+	\Pr( \left( \ephwierdrv \geq \eXrv + \gamma^{\epsATa}_\text{serf}   (\nXperf,\nKperf)   \right)  \bigcap \left(   \eXperfrv < \eXrv  \right) )_{ | \event{\nXperf,\nKperf}}   \\
				&\leq  \Pr(\eXperfrv \geq \eXrv )_{ | \event{\nXperf,\nKperf}}  + \Pr(\ephwierdrv \geq \eXperfrv + \gamma^{\epsATa}_\text{serf}   (\nXperf,\nKperf)  )_{ | \event{\nXperf,\nKperf}} \\
				&\leq \epsATa^2.
			\end{aligned}
		\end{equation}
	We will now combine  \cref{eq:bound12,eq:boundthree}. To do so we will additionally need to condition on $\eX$. However, note that  \cref{eq:boundthree} remains true with this additional conditioning (because $\eX$ is observed on a different set of rounds). Thus we obtain

		\begin{equation} \label{eq:bound123}
			\begin{aligned}
				&	\Pr(\ephperfrv \geq  \eXrv + \gamma^{\epsATa}_\text{serf}   (\nXperf,\nKperf) + \deltaone + \gamma^{\epsATb}_{\text{bin}}(\nKperf,\deltaone) )_{ | \event{\nXperf,\nKperf}} \\
				&=\sum_{\eX} \Pr(\event{\eX} | \event{\nXperf,\nKperf})	\Pr(\ephperfrv \geq  \eX + \gamma^{\epsATa}_\text{serf}   (\nXperf,\nKperf) + \deltaone + \gamma^{\epsATb}_{\text{bin}}(\nKperf,\deltaone) )_{ | \event{\nXperf,\nKperf,\eX}} \\
				&\leq \sum_{\eX} \Pr(\event{\eX} | \event{\nXperf,\nKperf})	 \Pr(\ephwierdrv \geq  \eX + \gamma^{\epsATa}_\text{serf}   (\nXperf,\nKperf))_{| \event{\nXperf,\nKperf,\eX}}  +\frac{\epsATb^2}{2} \\
						&= \Pr(\ephwierdrv \geq  \eXrv + \gamma^{\epsATa}_\text{serf}   (\nXperf,\nKperf))_{| \event{\nXperf,\nKperf}}  +\frac{\epsATb^2}{2} \\
				&\leq \epsATa^2+\epsATb^2.
			\end{aligned}
		\end{equation}
		where the first equality follows from the definition of conditional probability. The second inequality is obtained by setting $e=\eX +  \gamma^{\epsATa}_\text{serf}   (\nXperf,\nKperf)$ in \cref{eq:boundthree}, the third equality follows from the definition of probability and the final inequality follows from \cref{eq:bound12}

		Combing \cref{eq:bound123,eq:boundfour}, we obtain
		
		\begin{equation}
			\begin{aligned}
				&\Pr (   \ephrv \geq \frac{\eXrv+  \gamma^{\epsATa}_\text{serf}   (\nXperf,\nKperf) + \deltaone + \gamma^{\epsATb}_{\text{bin}}(\nKperf,\deltaone)}{ (1-\deltatwo- \gamma^{\epsATc}_{\text{bin}}(\nKperf,\deltatwo) )}    )_{ | \event{\nXperf,\nKperf}} \\
				= &\Pr (\left(   \ephrv \geq \frac{\eXrv +  \gamma^{\epsATa}_\text{serf}   (\nXperf,\nKperf) + \deltaone + \gamma^{\epsATb}_{\text{bin}}(\nKperf,\deltaone) }{ (1-\deltatwo- \gamma^{\epsATc}_{\text{bin}}(\nKperf,\deltatwo) )} \right)  \bigcap \left(  \ephperfrv \leq \ephrv (1-\deltatwo- \gamma^{\epsATc}_{\text{bin}}(\nKperf,\deltatwo) ) \right) )_{ | \event{\nXperf,\nKperf}} \\
				+& \Pr (\left(   \ephrv \geq \frac{\eXrv+  \gamma^{\epsATa}_\text{serf}   (\nXperf,\nKperf) + \deltaone + \gamma^{\epsATb}_{\text{bin}}(\nKperf,\deltaone) }{ (1-\deltatwo- \gamma^{\epsATc}_{\text{bin}}(\nKperf,\deltatwo) )}  \right)  \bigcap \left(  \ephperfrv > \ephrv (1-\deltatwo- \gamma^{\epsATc}_{\text{bin}}(\nKperf,\deltatwo) ) \right) )_{ | \event{\nXperf,\nKperf}} \\
				\leq &\Pr( \ephperfrv \leq \ephrv (1-\deltatwo- \gamma^{\epsATc}_{\text{bin}}(\nKperf,\deltatwo) ) )_{ | \event{\nXperf,\nKperf}} \\
				+ & \Pr(\ephperfrv \geq  \eXrv+  \gamma^{\epsATa}_\text{serf}   (\nXperf,\nKperf) + \deltaone + \gamma^{\epsATb}_{\text{bin}}(\nKperf,\deltaone)) 
				\leq \epsATa^2 + \epsATb^2 + \epsATc^2,
			\end{aligned}
		\end{equation}
		which is the required result.

		\section{Decoy Analysis} \label{app:decoyanalysis}
		In this section, we will rigorously justify the application of Hoeffdings concentration inequality \cite{hoeffdingProbability1963} in the decoy analysis of this work. To do so, we will first state the following general lemma.

		\begin{lemma} \label{lemma:hoeffdings}
			Let $\X_1 \dots \X_n$ be random variables. Let $X_i$ be a specific value taken by the random variable $\X_i$. For each $i$, a new random variable $\Y_i$ is generated from $X_i$ via the probability distribution $\Pr(\Y_i | X_i)$. Then
			\begin{equation}
				\Pr( \abs{ \sum_i (\Y_i )_{| \event{X_1 \dots X_{\nO}}}  - \mathrm{E}\left(\sum_i (\Y_i )_{| \event{X_1 \dots X_{\nO}}} \right) } \geq t ) \leq 2 \exp{\frac{-2t^2}{\sum_i (b_i-a_i)^2}}
			\end{equation}
			where $[a_i,b_i]$ denotes the range of $\Y_i$, and $\mathrm{E}$ denotes the expectation value. (Note that we do not require the $\X_i$s to be independent random variables, nor do we require the $\Y_i$s to be independent random variables).
		\end{lemma} 
		\begin{proof}
			Fix a specific sequence $X_1 \dots X_n$ of values taken by the random variables $\X_i$s.  The variables $\Y_i$, conditioned on this specific input $X_1 \dots X_n$, are then \textit{independent} random variables (since they are generated by $\Pr(\Y_i | X_i)$). Thus, Hoeffding's inequality applies.
			
		\end{proof}
		The above lemma is utilized to perform decoy analysis in the following lemma.
		\begin{lemma} \label{lemma:decoy}
			In the decoy-state QKD protocol of \cref{sec:decoy}, fix an outcome $O$, and intensity  $\mu_k$. 
			Then, we have
			\begin{equation}
				\Pr ( \abs{\nOmurv{k} - \sum_{m=0}^{\infty} p_{\mu_k | m} \nOphrv{m} }  \geq \sqrt{ \frac{\nOrv}{2} \ln( \frac{2}{\epsdecoy^2}) } ) \leq \epsdecoy^2.
			\end{equation}
		\end{lemma}
		
		\begin{proof}
			Consider all the rounds where $O$ is observed.  Condition on the event that $\nO$ such rounds are observed. Let $X_1 \dots X_{\nO}$ be the sequence of photon numbers of Alice's signals corresponding to these rounds. Condition further on the event that a specific sequence $X_1 \dots X_{\nO}$ is observed.

			Fix an intensity $\mu_k$ of interest. Define $\Y_i$ as
			\begin{equation}
				\Y_i \coloneq 
				\left\{
				\begin{array}{ll}
					1  & \mbox{if intensity $\mu_k$ is assigned to the $i$th round}   \\
					0 & \mbox{if intensity $\mu_k$ is not assigned to the $i$th round} .
				\end{array}
				\right.
			\end{equation}
			Since the intensity of each round is chosen from a probability distribution that only depends on the photon number of each round, each $\Y_i$ is generated independently via $\Pr(\Y_i | X_i)$. 
			By the construction of $\Y_i$s, $\sum_i \Y_i = \nOmurv{k}$. By the construction of $X_i$s, $ |\{  i | X_i = m\} | = \nOph{m}$. Then $ \mathrm{E}\left(\sum_i (\Y_i | X_i) \right)  = \sum_{m=0}^\infty p_{\mu_k | m} \nOph{m}$. Applying \cref{lemma:hoeffdings}, we directly obtain
			\begin{equation}
				\Pr( \abs{ \nOmurv{k}   - \sum_{m=0}^\infty p_{\mu_k | m} \nOph{m}  } \geq t )_{| \Omega(X_1 \dots X_{\nO},\nO) } \leq 2 \exp{\frac{-2t^2}{\nO}}.
			\end{equation}
			
			The above statement is valid for all $X_1 \dots X_{\nO}$ compatible with $\nO,\nOph{\vec{m}}$.  We now obtain a statement that only conditions on $\nO$ via
			
			\begin{equation}
				\begin{aligned}
					&\Pr( \abs{ \nOmurv{k}   - \sum_{m=0}^\infty p_{\mu_k | m} \nOphrv{m}  } \geq t )_{| \Omega(\nO) }  \hspace{-1em} \\ &= \sum_{X_1 \dots X_{\nO}} \Pr(\event{X_1 \dots X_{\nO}} | \event{\nO}) \Pr( \abs{ \nOmurv{k}   - \sum_{m=0}^\infty p_{\mu_k | m} \nOph{m}  } \geq t )_{| \Omega(X_1 \dots X_{\nO},\nO) }  \\
					&\leq \sum_{X_1 \dots X_{\nO}} \Pr(\event{X_1 \dots X_{\nO}} | \event{\nO})  2 \exp{\frac{-2t^2}{\nO}} \\
					&=   2 \exp{\frac{-2t^2}{\nO}} .
				\end{aligned}
			\end{equation}

			Setting $t=\sqrt{ \frac{\nO}{2} \ln( \frac{2}{\epsdecoy^2}) }$, we obtain
			\begin{equation}
				\Pr ( \abs{\nOmurv{k} - \sum_{m=0}^{\infty} p_{\mu_k | m} \nOphrv{m} }  \geq \sqrt{ \frac{\nO}{2} \ln( \frac{2}{\epsdecoy^2}) } )_{| \event{\nO}} \leq \epsdecoy^2,
			\end{equation}
			which directly implies the required statement.
			
		\end{proof} 
		
		\section{Variable-length security proof for decoy-state BB84} 	\label{app:variablelengthdecoy}
		In this appendix, we will prove the following theorem regarding the variable-length security of the decoy-state BB84 protocol.
		
		\variablelengthproofdecoy*
		\begin{proof}
			
				Again, the proof is similar to that of \cref{thm:variablesecurity} with some differences. The first part of the proof is identical to that of \cref{thm:variablesecurity} and we do not repeat it here. Thus, this proof consists of two parts. In the first part, we use \cref{eq:decoyreq} along with the entropic uncertainty relations to bound the smooth min-entropy of the raw key register. Here the main difference is the use of entropic chain rules to isolate the single-photon component of the raw key, before applying the EUR statement.  In the second part, we use the obtained bound to prove the variable-length security statement. Here the main difference is due to to the presence of the $\nKph{1}$ in the event $\event{\decoyparams,\nKph{0},\nKph{1}} =\tempevent $, which is handled differently. Let us first focus on the event $\event{\decoyparams,\nKph{0},\nKph{1}} =\tempevent $. (In the remainder of this proof, we identify $\tempevent = \event{\decoyparams,  \nKph{1} }$ for brevity.)
				
				\subsubsection{Bounding the smooth min entropy}
	
			Similar to the proof of \cref{thm:variablesecurity}, we will use $\kappa(\tempevent)$, to denote the probability of our computed bounds failing conditioned on the event $\tempevent$. We will see that the events that we will need to condition on are given by $\event{\decoyparams,  \nKph{1} }$. Even though we do not have access to the value $\nKph{1}$, we will see that this is the ``right" event to condition on.  We do not generate key from the zero-photon component $\nKph{0}$, since it has little impact on key rates for typical use cases. To see how one may do so, see \cref{remark:zerophotonkey}.   As in \cref{app:varlengthproof}, we define 
		\begin{equation} \label{eq:kappadecoydefined}
			\begin{aligned}
				&\kappa(\decoyparams,\nKph{1}) = \kappa(\tempevent) \coloneq 	\\
				&\Pr(\ephphrv{1} \geq \Bound_e(\eXmu{\allk}, \nXmu{\allk}, \nKmu{\allk} ) \quad  \lor \quad \nKphrv{1} \leq \Bound_1(\nKmu{\allk} )  )_{\event{\decoyparams,\nKph{1}} },
			\end{aligned}
		\end{equation}
		which when combined with \cref{eq:decoyreq} implies 
		\begin{equation} \label{eq:decoyreq2}
			\sum_{\decoyparams, \nKph{1} } \Pr(\decoyparams, \nKph{1} ) \kappa(\decoyparams,\nKph{1} ) \leq \epsAT^2.
		\end{equation}					
		Thus, the average probability of \textit{either} of our bounds failing (for any attack undertaken by Eve) is small.

			The state prior to the final (third step) measurements by Alice and Bob, is given by $\rho_{A^{\nK}B^{\nK} E^n C^n  | \tempevent } $. Suppose Alice measures her $\nK$ systems in the $Z$ basis. Let the post-measurement state be given by  $\rho_{Z^{\nK}B^{\nK} E^n C^n  | \tempevent }$.  This state actually exists in the protocol, and we want to compute the smooth min entropy on this state.
			Instead of $Z$ measurements, suppose Alice measures (virtually) in the $X$ basis. In this case, let the post-measurement state be given by  $\rho^\text{virt}_{X_A^{\nK}B^{\nK} E^n C^n  | \tempevent }$. 
			Since we condition on a specific value of $\nKph{1}$ in $\tempevent$, we can split up $Z^{\nK}$ as $Z^{\nKph{1} } Z^\text{rest}$. The registers before measurements ($A^{\nK}$), and the $X$ basis measurement registers ($X^{\nK}_A$) can also be split up in the same way. 
			
			Then, the required bound can be obtained via the following inequalities
			\begin{equation} \label{eq:EURdecoybound}
				\begin{aligned}
					\smoothmin{\sqrt{\kappa(\tempevent)}} (Z^{\nK} | C^n E^n)_{\rho | \tempevent}& \geq 			\smoothmin{\sqrt{\kappa(\tempevent)}} (Z^{\nKph{1}} | C^n E^n)_{\rho | \tempevent } \\
					&\geq \nKph{1} - 	\smoothmax{\sqrt{\kappa(\tempevent)}} (X_A^{\nKph{1}} | B^{\nK})_{\rho | \tempevent } \\
					&\geq \nKph{1} \left(1 - \Bound_e \left( h(\decoyparams) \right) \right)
				\end{aligned}
			\end{equation}
			where we used \cite[Lemma 6.7]{tomamichelQuantum2016} to isolate the single-photon contribution to the key in the first inequality, and the entropic uncertainty relations \cite{tomamichelUncertainty2011}  followed by the same series of steps as in the proof of \cref{thm:variablesecurity} to replace the smooth max entropy term with the bound on the phase error rate. This is the required bound on the smooth min entropy of the raw key.
			\begin{remark} \label{remark:zerophotonkey}
				If one wishes to obtain key from multi-photon or zero-photon events, one can use a suitable chain rule on the smooth min entropy at this stage (as is done in \cite{limConcise2014}). However, note that one must first fix the number of pulses corresponding to these events in order to have the registers be well-defined. For instance, one cannot apply the above analysis for a state $\rho$ \textit{without} conditioning on a $\tempevent$, since a fixed value of $\nKph{1}$ is required to meaningfully define the register $Z^{\nKph{1}}$. This subtlety is missing in \cite{limConcise2014}. While one may choose to define these registers to store strings of variable length, this then complicates the application of EUR in the proof (since the number of rounds on which the EUR is applied is now variable). Our analysis is one rigorous way to avoid these problems.
			\end{remark}

			We will now use \cref{eq:EURdecoybound} to prove the $(2 \epsAT+\epsPA)$-secrecy of the QKD protocol. To obtain $(2 \epsAT+\epsPA)$-secrecy, we must show that \cref{eq:secrecyclaim} is true. Again, as in the proof of \cref{thm:variablesecurity}, the states conditioned on $\tempevent = \Omega(\decoyparams , \nKph{1}) \wedge \EVevent$ have orthogonal supports. Therefore, it is enough to show that
			\begin{equation} \label{eq:actualsecrecyclaimdecoy}
				\begin{aligned} 
					&\Delta := 	\half \sum_{\decoyparams, \nKph{1} }  \Pr(\tempevent \wedge \EVevent) \dist(\decoyparams,\nKph{1}) \leq2 \epsAT+\epsPA, \\
					&\dist(\decoyparams,\nKph{1}) \coloneq   \multiNorm {\rho_{K_A \Cfull E^n | \tempevent \land \EVevent }  -  \sum_{k \in \{0,1\}^{l (\dots)}} \frac{\ketbra{k}_{K_A}}{2^{l (\dots)}} \otimes \rho_{ \Cfull E^n | \tempevent \land \EVevent} }_1
				\end{aligned}
			\end{equation}
			since we can group together terms with the same output key to obtain \cref{eq:secrecyclaim} from \cref{eq:actualsecrecyclaimdecoy} (and $l(\dots) = l(\decoyparams)$ for brevity). We will now prove \cref{eq:actualsecrecyclaimdecoy}. 			
			First, we split the sum over $ \nKph{1}$ into two parts, depending on whether it satisfies our estimates from \cref{eq:decoyreq}. Thus, we obtain							
			\begin{equation} \label{eq:Deltasplit}
				\begin{aligned}
					\Delta & = 	\half \sum_{ \substack{\decoyparams \\ \nKph{1} > \Bound_1(\nKmu{\allk})}}  \Pr(\tempevent \wedge \EVevent) \dist(\decoyparams,\nKph{1}) \\
					& +	\half \sum_{ \substack{\decoyparams \\ \nKph{1} \leq \Bound_1(\nKmu{\allk})}}  \Pr(\tempevent \wedge \EVevent) \dist(\decoyparams,\nKph{1}) \\
					&=\Delta_1 + \Delta_2
				\end{aligned}
			\end{equation} 
			Using the fact that $\dist(.) \leq 2$, the second term ($\Delta_2$) can be upper bounded via
			\begin{equation} \label{eq:Delta2bound}
				\Delta_2 \leq 		 \sum_{ \substack{\decoyparams \\ \nKph{1} \leq \Bound_1(\nKmu{\allk})}}  \Pr(\tempevent)  = \Pr(\nKphrv{1} \leq \Bound_1(\nKmurv{\allk}))  \\
			\end{equation}
			The first term ($\Delta_1$) can be bounded in an \textit{identical} manner
			as in the proof of \cref{thm:variablesecurity}, we shown below. Again, we can assume that we are summing over events that lead to a non-trivial length of the key (since events where the protocol aborts do not contribute to $\Delta$). Let $\accset = \{ (\decoyparams) | l(\decoyparams) > 0\}$ be the set of parameters that produce a non-trivial length of the key. Then, we obtain the following set of inequalities, which we explain below:
			\begin{equation}  \label{eq:Delta1Bound}
				\begin{aligned}
					\Delta_1
					& \leq \sum_{ \substack{\decoyparams \in \mathcal{F} \\ \nKph{1} > \Bound_1(\nKmu{\allk})}}  \Pr(\tempevent )  \bigg(  2\sqrt{ \serfbound(\tempevent)} + \half2^{-\half \left( \smoothmin{\sqrt{ \serfbound(\tempevent) }}(Z^{\nK} | E^n C^n C_E )_{(\rho | \tempevent) \wedge \EVevent } - l(\decoyparams)  \right)}   \bigg) \\
					& \leq \sum_{ \substack{\decoyparams \in \mathcal{F} \\ \nKph{1} > \Bound_1(\nKmu{\allk})}}  \Pr(\tempevent )  \bigg(  2\sqrt{ \serfbound(\tempevent)} + \half2^{-\half \left( \smoothmin{\sqrt{ \serfbound(\tempevent) }}(Z^{\nK} | E^n C^n C_E )_{\rho | \tempevent } - l(\decoyparams)  \right)}   \bigg) \\
					& \leq \sum_{ \substack{\decoyparams  \in \mathcal{F}\\ \nKph{1} > \Bound_1(\nKmu{\allk})}}  \Pr(\tempevent  )  \bigg(  2\sqrt{ \serfbound(\tempevent)} \\
					&+ \half2^{-\half \left( \smoothmin{\sqrt{ \serfbound(\tempevent) }}(Z^{\nK} | E^n C^n  )_{\rho | \tempevent} - l(\decoyparams - \leak(\decoyparams) - \log(2/\epsEV))  \right)}   \bigg) \\  
					& \leq \sum_{ \substack{\decoyparams  \in \mathcal{F}\\ \nKph{1} > \Bound_1(\nKmu{\allk})}}  \Pr(\tempevent )  \bigg(  2\sqrt{ \serfbound(\tempevent)} \\
					&+ \half2^{-\half \left(  \nKph{1} \left(1 - h\left(\Bound_e  \left(\decoyparams\right) \right) \right)   - l(\decoyparams) - \leak(\decoyparams) - \log(2/\epsEV))  \right)}   \bigg) \\  
					& = \sum_{ \substack{\decoyparams   \in \mathcal{F} \\ \nKph{1} > \Bound_1(\nKmu{\allk})}}  \Pr(\tempevent )  \bigg(  2\sqrt{ \serfbound(\tempevent)} +  \epsPA    \bigg) \\  
					& \leq  \epsPA + 2 \sqrt{ \sum_{ \substack{\decoyparams  \\ \nKph{1} > \Bound_1(\nKmu{\allk})}}  \Pr(\tempevent )  \serfbound(\tempevent)}  \\
					&= \epsPA + 2 \sqrt{   \Pr(\ephphrv{1} \geq \Bound_e(\eXmurv{\allk}, \nXmurv{\allk}, \nKmurv{\allk} ) \quad \land \quad \nKphrv{1} > \Bound_1(\nKmurv{\allk} )  )   }.  \\
				\end{aligned}
			\end{equation}
			Here, we used the leftover-hashing lemma \cite[Proposition 9]{tomamichelLargely2017} on the sub-normalized state $(\rho_{ | \tempevent} )_{\wedge \EVevent}$ for the first inequality, and \cite[Lemma 10]{tomamichelLargely2017} to get rid of the sub-normalized conditioning ($\wedge \EVevent$) in the smooth min entropy term in the second inequality. We used \cite[Lemma 6.8]{tomamichelQuantum2016} to split off the error-correction information ($\leak(\decoyparams)$) and error-verification information ($\log(2/\epsEV)$) in the third inequality. We used  the bound on the smooth min entropy from \cref{eq:EURdecoybound} for the fourth inequality. The fifth equality follows by replacing the values of $l(\params)$ and $\leak (\params)$ from \cref{eq:lvaluedecoy}. We use the concavity of the square root function and Jensen's inequality for the sixth inequality to pull the sum over the events and the probability inside the square root, while the seventh equality follows simply from the definition of conditional probabilities and $\kappa$ (\cref{eq:kappadecoydefined}).
			
			Combining  our bounds on $\Delta_1$ and $\Delta_2$ from \cref{eq:Delta1Bound,eq:Delta2bound}, we obtain
			\begin{equation} \label{eq:DeltaCombined}
				\begin{aligned}
					\Delta & \leq \epsPA + 2 \sqrt{   \Pr(\ephphrv{1} \geq \Bound_e(\eXmurv{\allk}, \nXmurv{\allk}, \nKmurv{\allk} ) \quad \land \quad \nKphrv{1} \geq \Bound_1(\nKmurv{\allk} )  )   }   + \Pr(\nKphrv{1} \leq \Bound_1(\nKmurv{\allk})) \\
					&\leq \epsPA +  2 \sqrt{   \Pr(\ephphrv{1} \geq \Bound_e(\eXmurv{\allk}, \nXmurv{\allk}, \nKmurv{\allk} ) \quad \land \quad \nKphrv{1} \geq \Bound_1(\nKmurv{\allk} )  )  + \Pr(\nKphrv{1} \leq \Bound_1(\nKmurv{\allk}))  }  \\
					&= \epsPA  + 2\sqrt{ \Pr(\ephphrv{1} \geq \Bound_e(\eXmurv{\allk}, \nXmurv{\allk}, \nKmurv{\allk} ) \quad \lor \quad \nKphrv{1} \leq \Bound_1(\nKmurv{\allk} )  )    } \\
					&\leq \epsPA + 2\epsAT.
				\end{aligned}
			\end{equation}
			Here we use the fact that  $2\sqrt{a} + b \leq 2\sqrt{a+b}$ for $ 0\leq a,b,a+b\leq 1$ in the second inequality. We use $\Pr(\event{1} \land \event{2}^c) + \Pr(\event{2}) = \Pr(\event{1} \lor \event{2})$ (where $\Omega^c$ denotes the complement of $\Omega$) for the third equality, and \cref{eq:decoyreq} for the final inequality.

		\end{proof}

		Notice that our proofs of variable-length security (\cref{thm:variablesecurity,thm:decoyvariablesecurity}) rely on obtaining a suitable bound on the smooth min entropy of the raw key register, with a suitable smoothing parameter, for suitable events. In particular, the smoothing parameter averaged over all events satisfies certain bounds. However, the theorem statements so far have been specific to BB84 and decoy-state BB84. We state the following technical theorem regarding variable-length security which is applicable to generic protocols, as long as suitable bounds on the smooth min entropy can be obtained.

		\begin{theorem}
			In a QKD protocol, let $\event{i,j}$ denote well-defined events (we use $i$ for observed events and $j$ for unobserved events) that can take place. Let $\vec{Z}$ denote the raw key register, let $\vec{E}$ denote Eve's quantum system, and let $\vec{C}$ denote public announcements (excluding error-correction and error-verification). Let the protocol be such that it produces a key of length $l(i)$ bits, using $\leak(i)$ bits for error-correction and $\log(2/\epsEV)$ bits for error-verification, upon the observed event $\event{i} \wedge \EVevent$. For each $i$, let $S_i$ denote a subset of possible values of $j$, and let $\kappa_{(i,j)}\geq 0$ be a set of values such that
			\begin{equation} \label{eq:genericvariable}
				\begin{aligned}
				&	\smoothmin{\sqrt{\kappa_{(i,j)}}}(\vec{Z} | \vec{E}  \vec{C})_{\rho | \Omega_{(i,j)}} \geq \beta_i \qquad \qquad  \forall i, \forall j \in S_i\\
				&	\sum_{i} \sum_{j \in S_i} \Pr(\event{i,j}) \kappa_{(i,j)}  + \sum_{i} \sum_{j\notin S_i} \Pr(\event{i,j})\leq \epsAT^2 \\
				& l(i) \coloneqq \max\big\{\beta_i - \leak(i) - 2\log(1/2\epsPA) - \log(2/\epsPA), 0\big\}.
					\end{aligned}
			\end{equation}
			Then the QKD protocol is $(2\epsAT+ \epsPA + \epsEV)$-secure.
		\end{theorem}
\textit{Proof Sketch.}			The proof follows an identical series of steps as the proof of \cref{thm:decoyvariablesecurity}, and we do not repeat it here. This can be seen by identifying $i$ with observed events in the protocol (analogous to $(\decoyparams)$), and $j$ with events that are not directly observed in the protocol (analogous to $\nKph{1}$). One identifies the set $S_i$ with values of $j$ that satisfy certain bounds with high probability.  
			
			Intuitively the conditions in \cref{eq:genericvariable} split all possible combinations of $(i,j)$ into two sets, depending on whether $j \in S_i$ or $j \notin S_i$. If $j \in S_i$, then a suitable bound $\beta_i$ on the min entropy is known. This bound is utilized in the leftover hash lemma, and the trace distance for QKD security is bounded in the same manner as \cref{eq:Delta1Bound}. The bound obtained in this case is given by $ \epsPA + 2 \sqrt{\sum_{i,j \in S_i} \Pr(\event{i,j}) \kappa_{(i,j)}}$.   If $j \notin S_i$, then the trace distance for QKD security is bounded in the same manner as  \cref{eq:Delta2bound}. The bound obtained is given by $\sum_{i,j\notin S_i} \Pr(\event{i,j})$. These two bounds can be combined as in \cref{eq:DeltaCombined}.

			Note that if $j$ is set to be a trivial value, then given the suitable bound on the min entropy, we recover  \cref{thm:variablesecurity}.

	In general, when dealing with events, one must ensure that all events considered are well-defined, i.e, there exists (in theory) a classical register that determines whether the event occured or did not occur \cite[Section 2]{tomamichelLargely2017}.

		\section{Detector Model Calculations} \label{app:detModelCalc} 
	
		In this section we will compute upper bounds on the $\deltaone,\deltatwo$. To do so, we follow the recipe from \cref{subsec:recipe}.
		\subsection{Computing POVMs} \label{app:povmcalcs}
		
		Recall that Alice and Bob's joint POVM $\{ \AliceBobPOVM{b_A,b_B}{\neq}, \AliceBobPOVM{b_A,b_B}{=}, \AliceBobPOVM{b_A,b_B}{\bot} \}$ is given in \cref{eq:alicebobpovmmodel}.
		We can choose 
		\begin{equation}
			\begin{aligned}
				\tilde{F} &= \id_A\otimes \sum_{N_0, N_1=0}^\infty (1-(1-\dMax)^2(1-\etaMax)^{N_0+N_1}) \ketbra{N_0,N_1}_{Z} \\
				&= \id_A\otimes \sum_{N_0, N_1=0}^\infty (1-(1-\dMax)^2(1-\etaMax)^{N_0+N_1}) \ketbra{N_0,N_1}_{X},  \quad \text{where} \\
				\etaMax &= \max_{b\in\{X,Z\}}\{\eta_{b_0},\eta_{b_1}\}, \quad \text{ and } \dMin = \min_{b\in\{X,Z\}}\{d_{b_0},d_{b_1}\}, \\
				\etaMin &= \min_{b\in\{X,Z\}}\{\eta_{b_0},\eta_{b_1}\}, \quad \text{ and } \dMax = \max_{b\in\{X,Z\}}\{d_{b_0},d_{b_1}\}.
			\end{aligned}
		\end{equation}
	
		It is straightforward to verify that $\tilde{F}$ satisfies our requirement $\tilde{F}\geq \AliceBobPOVM{b_A,b_B}{\neq}+\AliceBobPOVM{b_A,b_B}{=} = \id -\AliceBobPOVM{b_A,b_B}{\bot}$ for all choices of $b_A,b_B$. Then, we calculate the POVM elements from \cref{eq:recipeone} as
		\begin{equation} \label{eq:tempE3}
			\begin{aligned}
				\AliceBobPOVMprimeFilter{b}{\con} &= \id_A\otimes \sum_{N_0, N_1=0}^\infty \frac{1-(1-d_{b_0})(1-d_{b_1})(1-\eta_{b_0})^{N_0} (1-\eta_{b_1})^{N_1}}{(1-(1-\dMax)^2(1-\etaMax)^{N_0+N_1})} \ketbra{N_0,N_1}_b\\
				\AliceBobPOVMprimeFilter{b}{\bot} &= \id_A\otimes \sum_{N_0, N_1=0}^\infty \left(1-\frac{1-(1-d_{b_0})(1-d_{b_1})(1-\eta_{b_0})^{N_0} (1-\eta_{b_1})^{N_1}}{(1-(1-\dMax)^2(1-\etaMax)^{N_0+N_1})}\right) \ketbra{N_0,N_1}_b\\
				\AliceBobPOVMsecond{b}{\neq} =& \sum_{N_0, N_1 =0}^\infty\ketbra{0}_b \otimes \frac{(1+(1-d_{b_0})(1-\eta_{b_0})^{N_0}) (1-(1-d_{b_1})(1-\eta_{b_1})^{N_1})}{2(1-(1-d_{b_0})(1-d_{b_1})(1-\eta_{b_0})^{N_0} (1-\eta_{b_1})^{N_1})} \ketbra{N_0,N_1}_b\\
				&\qquad \qquad  + \ketbra{1}_b \otimes \frac{(1+(1-d_{b_1})(1-\eta_{b_1})^{N_1})(1-(1-d_{b_0})(1-\eta_{b_0})^{N_0})}{2(1-(1-d_{b_0})(1-d_{b_1})(1-\eta_{b_0})^{N_0} (1-\eta_{b_1})^{N_1})} \ketbra{N_0,N_1}_b\\
				\AliceBobPOVMsecond{b}{=} =& \sum_{N_0, N_1 =0}^\infty\ketbra{0}_b \otimes \frac{(1+(1-d_{b_1})(1-\eta_{b_1})^{N_1})(1-(1-d_{b_0})(1-\eta_{b_0})^{N_0})}{2(1-(1-d_{b_0})(1-d_{b_1})(1-\eta_{b_0})^{N_0} (1-\eta_{b_1})^{N_1})} \ketbra{N_0,N_1}_b\\
				&\qquad \qquad  + \ketbra{1}_b \otimes \frac{(1+(1-d_{b_0})(1-\eta_{b_0})^{N_0}) (1-(1-d_{b_1})(1-\eta_{b_1})^{N_1})}{ 2 (1-(1-d_{b_0})(1-d_{b_1})(1-\eta_{b_0})^{N_0} (1-\eta_{b_1})^{N_1})} \ketbra{N_0,N_1}_b.
			\end{aligned}
		\end{equation}
	
		\subsection{Computing $\delta_1$} \label{subsec:deltaonecalcs}

		To compute the costs in \cref{eq:deltaCostsRecipe}, we first note that all POVMs appearing in this work are block-diagonal in the total photon number $N_0+N_1$. We wish to compute the infinity norm, i.e
		
		\begin{align}
			\nonumber \delta_1 &= 2  \norm{\sqrt{\AliceBobPOVMprimeFilter{Z}{\con}}\AliceBobPOVMsecond{X}{\neq} \sqrt{\AliceBobPOVMprimeFilter{Z}{\con}} -  \sqrt{\AliceBobPOVMprimeFilter{X}{\con}}\AliceBobPOVMsecond{X}{\neq} \sqrt{\AliceBobPOVMprimeFilter{X}{\con}} }_\infty.
		\end{align}
		
		Due to block-diagonal structure, the operator appearing inside the $\norm{.}_\infty$ in the above expressions are also block-diagonal in photon number $N_0+N_1$. From the properties of the norm, we have that 		
		\begin{equation} \label{eq:deltaN}
			\deltaone = \max_N \deltaone^{(N)},
		\end{equation} where  $\deltaone^{(N)}$ is the infinity norm for the block corresponding to total photon number $N=N_0+N_1$. We now focus on computing $\deltaone^{(N)}$.
		
		First, we compute $\deltaone^{(0)}$ directly as
		\begin{align} 
			\nonumber \frac{\deltaone^{(0)}}{2} =& \norm{\id_A\otimes\ketbra{0,0}\left(\sqrt{\AliceBobPOVMprimeFilter{Z}{\con}}\AliceBobPOVMsecond{X}{\neq} \sqrt{\AliceBobPOVMprimeFilter{Z}{\con}} -  \sqrt{\AliceBobPOVMprimeFilter{X}{\con}}\AliceBobPOVMsecond{X}{\neq} \sqrt{\AliceBobPOVMprimeFilter{X}{\con}}\right) \id_A\otimes\ketbra{0,0}}_\infty\\
			\nonumber =& \abs{\frac{1-(1-d_{Z_0})(1-d_{Z_1})}{1-(1-\dMax)^2}-\frac{1-(1-d_{X_0})(1-d_{X_1})}{1-(1-\dMax)^2}}  \\
            \nonumber \times & \max \left\{\frac{d_{X_1}(2-d_{X_0})}{2(1-(1-d_{X_0})(1-d_{X_1}))},\; \frac{d_{X_0}(2-d_{X_1})}{2(1-(1-d_{X_0})(1-d_{X_1}))}\right\}\\
			\nonumber =&  \abs{\frac{(1-d_{X_0})(1-d_{X_1})-(1-d_{Z_0})(1-d_{Z_1})}{1-(1-\dMax)^2}} \max \left\{\frac{d_{X_1}(2-d_{X_0})}{2(1-(1-d_{X_0})(1-d_{X_1}))},\; \frac{d_{X_0}(2-d_{X_1})}{2(1-(1-d_{X_0})(1-d_{X_1}))}\right\}\\ 
			\leq& \left(1-\frac{1-(1-\dMin)^2}{1-(1-\dMax)^2}\right)\frac{\dMax(2-\dMin)}{2(1-(1-\dMin)^2)}. \label{eq:deltaoneZeroPart}
		\end{align}
		To bound $\deltaone^{(N)}$ for $N\neq 0$, we write 
		\begin{align}
			\nonumber \frac{\deltaone}{2} =&  \norm{\sqrt{\AliceBobPOVMprimeFilter{Z}{\con}}\AliceBobPOVMsecond{X}{\neq} \sqrt{\AliceBobPOVMprimeFilter{Z}{\con}} -  \sqrt{\AliceBobPOVMprimeFilter{X}{\con}}\AliceBobPOVMsecond{X}{\neq} \sqrt{\AliceBobPOVMprimeFilter{X}{\con}} }_\infty\\
			\nonumber \leq&  \norm{\sqrt{\AliceBobPOVMprimeFilter{Z}{\con}}\AliceBobPOVMsecond{X}{\neq} \sqrt{\AliceBobPOVMprimeFilter{Z}{\con}} - \sqrt{\AliceBobPOVMprimeFilter{Z}{\con}}\AliceBobPOVMsecond{X}{\neq}\sqrt{\AliceBobPOVMprimeFilter{X}{\con}}}_\infty\\
			& \label{eq:triangleIneqDelta1} + \norm{\sqrt{\AliceBobPOVMprimeFilter{Z}{\con}}\AliceBobPOVMsecond{X}{\neq}\sqrt{\AliceBobPOVMprimeFilter{X}{\con}} - \sqrt{\AliceBobPOVMprimeFilter{X}{\con}}\AliceBobPOVMsecond{X}{\neq} \sqrt{\AliceBobPOVMprimeFilter{X}{\con}}}_\infty\\
			\nonumber\leq& \norm{\sqrt{\AliceBobPOVMprimeFilter{Z}{\con}}\AliceBobPOVMsecond{X}{\neq}}_\infty \norm{\sqrt{\AliceBobPOVMprimeFilter{Z}{\con}}-\sqrt{\AliceBobPOVMprimeFilter{X}{\con}}}_\infty\\
			\label{eq:subMultDelta1} & + \norm{\sqrt{\AliceBobPOVMprimeFilter{Z}{\con}}-\sqrt{\AliceBobPOVMprimeFilter{X}{\con}}}_\infty \norm{\AliceBobPOVMsecond{X}{\neq} \sqrt{\AliceBobPOVMprimeFilter{X}{\con}}}_\infty,
		\end{align}
		where \cref{eq:triangleIneqDelta1} follows from the triangle inequality, and \cref{eq:subMultDelta1} follows from the submultiplicativity of the $\infty$-norm.
		We can then compute each term individually.
		Note that due to the submultiplicativity of the $\infty$-norm, we have
		\begin{align}
			\label{eq:FGnormDelta1Calc}\norm{\sqrt{\AliceBobPOVMprimeFilter{b}{\con}}\AliceBobPOVMsecond{X}{\neq}}_\infty \leq& \norm{\sqrt{\AliceBobPOVMprimeFilter{b}{\con}}}_\infty\norm{\AliceBobPOVMsecond{X}{\neq}}_\infty \leq 1.
		\end{align}
		
		Thus, we only need to bound  $\norm{\sqrt{\AliceBobPOVMprimeFilter{Z}{\con}}-\sqrt{\AliceBobPOVMprimeFilter{X}{\con}}}_\infty$. To do so, we define 
		
		\begin{equation}
			\begin{aligned}
				\intPOVM &\coloneq \id_A\otimes\sum_{N=0}^\infty\sum_{N_0+N_1 = N} \frac{\sqrt{1-(1-\dMax)^2(1-\etaMax)^N}+\sqrt{1-(1-\dMin)^2(1-\etaMin)^N}}{2\sqrt{1-(1-\dMax)^2(1-\etaMax)^N}} \ketbra{N_0,N_1}_X \\
				&= \id_A\otimes\sum_{N=0}^\infty\sum_{N_0+N_1 = N} \frac{\sqrt{1-(1-\dMax)^2(1-\etaMax)^N}+\sqrt{1-(1-\dMin)^2(1-\etaMin)^N}}{2\sqrt{1-(1-\dMax)^2(1-\etaMax)^N}} \ketbra{N_0,N_1}_Z,
			\end{aligned}
		\end{equation}
		and obtain
		\begin{align}
			\label{eq:triangleIneqDiffOfFilters} \norm{\sqrt{\AliceBobPOVMprimeFilter{Z}{\con}}-\sqrt{\AliceBobPOVMprimeFilter{X}{\con}}}_\infty \leq & \norm{\sqrt{\AliceBobPOVMprimeFilter{Z}{\con}}-\intPOVM}_\infty + \norm{\intPOVM-\sqrt{\AliceBobPOVMprimeFilter{X}{\con}}}_\infty,
		\end{align}
		where \cref{eq:triangleIneqDiffOfFilters} is a consequence of the triangle inequality.	Let us focus on a fixed value of $N=N_0+N_1 \neq 0$. In this case, combining \cref{eq:subMultDelta1,eq:FGnormDelta1Calc,eq:triangleIneqDiffOfFilters} allows us to obtain
		\begin{equation} \label{eq:generalDiffOfFiltersBound}
			\begin{aligned}
				\deltaone^{(N)} &\leq 4 \sum_{b =X,Z}  \Biggl| \sqrt{\frac{1-(1-d_{b_0})(1-d_{b_1})(1-\eta_{b_0})^{N_0} (1-\eta_{b_1})^{N_1}}{1-(1-\dMax)^2(1-\etaMax)^{N_0+N_1}}} \\ &-\frac{\sqrt{1-(1-\dMax)^2(1-\etaMax)^{N_0+N_1}}+\sqrt{1-(1-\dMin)^2(1-\etaMin)^{N_0+N_1}}}{2\sqrt{1-(1-\dMax)^2(1-\etaMax)^{N_0+N_1}}}  \Biggl|  \\
				&\leq 4 \abs{ 1- \sqrt{ \frac{1-(1-\dMin)^2(1-\etaMin)^N}{1-(1-\dMax)^2(1-\etaMax)^N}  }},
			\end{aligned}
		\end{equation}
		where the final inequality above follows from the monotonicity of the expression inside the modulus with respect to $\eta_{b_0},\eta_{b_1},d_{b_0},d_{b_1}$. However, as described in Ref.~\cite[Section III C]{zhangEntanglement2017} we can renormalize detection efficiencies and treat the common loss in detectors as part of the channel. Thus, without loss of generality, we consider one of the detectors to be lossless. This can be thought of as setting $\etaMax \rightarrow 1$ and $\etaMin \rightarrow \etaMin / \etaMax$. This significantly simplifies the calculations. Combining  \cref{eq:triangleIneqDelta1,eq:FGnormDelta1Calc,eq:triangleIneqDiffOfFilters,eq:generalDiffOfFiltersBound,eq:deltaN}, we obtain
		
			\begin{equation} \label{eq:deltaOneOtherBound}
			\begin{aligned}
				\deltaone^{(N)} &\leq  4 \abs{ 1- \sqrt{ 1-(1-\dMin)^2(1-\etaRenorm)^N}  }, \qquad (N \geq 1)
			\end{aligned}
		\end{equation}
		Combining \cref{eq:deltaOneOtherBound,eq:deltaoneZeroPart}, we obtain
		\begin{equation}
			\begin{aligned}
				\deltaone  &\leq \max \left\{ \left(1-\frac{1-(1-\dMin)^2}{1-(1-\dMax)^2}\right)\frac{\dMax(2-\dMin)}{1-(1-\dMin)^2},   4 \abs{ 1- \sqrt{ 1-(1-\dMin)^2(1-\etaRenorm)}} \right\},
			\end{aligned}
		\end{equation}
		where $\etaRenorm = \etaMin/\etaMax$.

		\subsection{Computing $\deltatwo$} \label{subsec:deltatwocalcs}
		We can now also compute $\deltatwo =	\norm{\id - \AliceBobPOVMprimeFilter{Z}{\con} }_\infty $ in a similar way, using \cref{eq:tempE3,eq:deltaCostsRecipe}. Again, we have $\deltatwo = \max_N \deltatwo^{(N)}$,  where 	
		\begin{align}
			\nonumber \deltatwo^{(N)} 
			=& \max_{N_0+N_1=N} \left\{1- \frac{1-(1-d_{Z_0})(1-d_{Z_1})(1-\eta_{Z_0})^{N_0} (1-\eta_{Z_1})^{N_1}}{1-(1-\dMax)^2(1-\etaMax)^{N}}\right\}\\
			\leq &\max_{N_0+N_1=N} \left\{1- \frac{1-(1-\dMin)^2(1-\etaMin)^{N} }{1-(1-\dMax)^2(1-\etaMax)^{N}}\right\},
			\label{eq:generalDelta2Bound}
		\end{align}
		where the second inequality follows from the fact that the term inside is monotonous with respect to $\eta_{Z_0},\eta_{Z_1},d_{Z_0},d_{Z_1}$.  As in the computation for $\deltaone$, we pull out common loss by setting $\etaMax \rightarrow 1$, $\etaMin \rightarrow \etaMin / \etaMax$. 
		Using \cref{eq:generalDelta2Bound}, $\deltatwo$ can be bounded  as
		
		\begin{equation}
			\begin{aligned}
				\deltatwo \leq \max \left\{1-\frac{1-(1-\dMin)^2}{1-(1-\dMax)^2},(1-\dMin)^2 (1-\etaRenorm) \right\},
			\end{aligned}
		\end{equation}
		where $\etaRenorm = \etaMin / \etaMax$.

		\section{Random Swapping} \label{app:deltarandomswaps}
We compute $\deltaone, \deltatwo$ for the scenario where random swapping is implemented with probability $p=1/2$. Bob's POVM elements are given as in \cref{eq:bobswappovmmodel}.
		
		Note in particular that the zero-photon  and single-photon component of $\BobPOVMswap{b}{\bot}$ is independent of the basis choice $b$. Thus, we can choose
		\begin{equation}
			\begin{aligned}
				\tilde{F} &= \id_A \otimes (1-(1-\dMultAvg)^2)\ketbra{0,0}\\
				&+ \id_A \otimes \left(1-(1-\dMultAvg)^2(1-\etaAvg)\right)\left(\ketbra{0,1}_Z+\ketbra{1,0}_Z\right)\\        
				&+ \id_A\otimes \sum_{\substack{N_0, N_1=0\\ N_0+N_1>1}}^\infty (1-(1-\dMultAvg)^2(1-\etaMultAvg)^{N_0+N_1}) \ketbra{N_0,N_1}_{Z},  \quad \text{where} \\
				\etaMultAvg &= 1-\sqrt{(1-\eta_0)(1-\eta_1)}, \quad \text{ and } \dMultAvg = 1-\sqrt{(1-d_0)(1-d_1)}, \\
				\etaAvg &= \frac{\eta_0+\eta_1}{2}, \quad \text{ and } \etaMin = \min \{\eta_0, \eta_1\},
			\end{aligned}
		\end{equation}
		Note that replacing $Z$ with $X$ does not change the above expressions. This results in
	
		\begin{equation} \label{eq:swaptempE3}
			\begin{aligned}
				\AliceBobPOVMprimeFilter{b}{\con} &= \id_A\otimes\ketbra{0,0}\\
				&+ \id_A\otimes\left(\ketbra{0,1}_b+\ketbra{1,0}_b\right)\\
				&+\id_A\otimes \sum_{\substack{N_0, N_1=0\\ N_0+N_1>1}}^\infty \frac{2- (1-\dMultAvg)^2\left((1-\eta_{b_0})^{N_0}(1-\eta_{b_1})^{N_1}+(1-\eta_{b_1})^{N_0}(1-\eta_{b_0})^{N_1}\right)}{2(1-(1-\dMultAvg)^2(1-\etaMultAvg)^{N_0+N_1})} \ketbra{N_0,N_1}_b.
			\end{aligned}
		\end{equation}
		As a consequence of the way in which we bound the metrics $\deltaone$ and $\deltatwo$, these are the only POVM elements we need to explicitly compute.
		
		\subsection{Computing $\deltaone$} \label{subsec:deltaonecalcsrandom}
		
		Similar to the analysis performed in \cref{subsec:deltaonecalcs}, we reduce the problem to bounding $\norm{\sqrt{\AliceBobPOVMprimeFilter{Z}{\con}}-\sqrt{\AliceBobPOVMprimeFilter{X}{\con}}}_\infty$ through \cref{eq:subMultDelta1,eq:FGnormDelta1Calc}. Once again we exploit the block-diagonal structure in the POVM elements to compute the infinity norm $\deltaone^{(N)}$ of each block with total photon-number $N$ separately as done in \cref{eq:deltaN}. First, note that $\deltaone^{(0)} = \deltaone^{(1)} = 0$.
		To compute $\delta^{(N)}$ for $N\geq 2$, we define 	
		\begin{equation}
			\begin{aligned}
				\intPOVM &\coloneq \id_A\otimes\sum_{N=2}^\infty\sum_{N_0+N_1 = N} \frac{\sqrt{1-(1-\dMultAvg)^2(1-\etaMultAvg)^N}+\sqrt{1-(1-\dMultAvg)^2\frac{(1-\eta_0)^N+(1-\eta_1)^N}{2}}}{2\sqrt{1-(1-\dMultAvg)^2(1-\etaMultAvg)^N}} \ketbra{N_0,N_1}_X \\
				&= \id_A\otimes\sum_{N=2}^\infty\sum_{N_0+N_1 = N} \frac{\sqrt{1-(1-\dMultAvg)^2(1-\etaMultAvg)^N}+\sqrt{1-(1-\dMultAvg)^2\frac{(1-\eta_0)^N+(1-\eta_1)^N}{2}}}{2\sqrt{1-(1-\dMultAvg)^2(1-\etaMultAvg)^N}} \ketbra{N_0,N_1}_Z,
			\end{aligned}
		\end{equation}
		and obtain
		\begin{align}
			\norm{\sqrt{\AliceBobPOVMprimeFilter{Z}{\con}}-\sqrt{\AliceBobPOVMprimeFilter{X}{\con}}}_\infty \leq & \norm{\sqrt{\AliceBobPOVMprimeFilter{Z}{\con}}-\intPOVM}_\infty + \norm{\intPOVM-\sqrt{\AliceBobPOVMprimeFilter{X}{\con}}}_\infty,
		\end{align}
		identically to \cref{eq:triangleIneqDiffOfFilters}.
		
		Following a similar calculation as in \cref{eq:generalDiffOfFiltersBound}, we obtain
		\begin{equation} \label{eq:swapgeneralDiffOfFiltersBound}
			\begin{aligned}
				\deltaone^{(N)} \leq 4 \left(1-\sqrt{\frac{1-(1-\dMultAvg)^2\frac{(1-\eta_0)^N+(1-\eta_1)^N}{2}}{1-(1-\dMultAvg)^2(1-\etaMultAvg)^N}}\right),
			\end{aligned}
		\end{equation}
		for all $N \geq 2$.	Once again we use the argument in \cite[Section III C]{zhangEntanglement2017} to treat the common loss in detectors as part of the channel. This is equivalent to setting $\etaMax \rightarrow 1$ and $\etaMin \rightarrow \etaRenorm = \etaMin/\etaMax$. (Actually this sets $\etaMultAvg \rightarrow 1$ and one of $\eta_i$ to $1$ and the other to $\etaRenorm$). Thus we obtain
		
			\begin{equation} 
			\begin{aligned}
				\deltaone^{(N)} \leq 4 \left(1-\sqrt{1-(1-\dMultAvg)^2\frac{(1-\etaRenorm)^N}{2}}\right).
			\end{aligned}
		\end{equation}
		The above expression is monotonic in $N$. Therefore, we obtain
		\begin{equation}
			\deltaone \leq 4 \left(1-\sqrt{1-(1-\dMultAvg)^2\frac{(1-\etaRenorm)^2}{2}}\right).
		\end{equation}
		
		\subsection{Computing $\deltatwo$} 
		\label{subsec:deltatwocalcsrandom}

		We can also compute $\deltatwo =	\norm{\id - \AliceBobPOVMprimeFilter{Z}{\con} }_\infty $ similarly to the computation in \cref{subsec:deltatwocalcs}. Again, we have $\deltatwo = \max_N \deltatwo^{(N)}$. Similar to \cref{subsec:deltaonecalcsrandom}  we obtain $\deltatwo^{(0)} = \deltatwo^{(1)} = 0$. A straightforward computation for the other blocks results in
		\begin{equation}
			\deltatwo \leq (1-\dMultAvg)^2\frac{(1-\etaRenorm)^2}{2}.
		\end{equation}
		
	\end{document}